\documentclass[11pt]{article}
\usepackage[letterpaper,margin=1in]{geometry}
\usepackage[T1]{fontenc}
\usepackage[utf8]{inputenc}
\usepackage{amsmath,amssymb}
\usepackage{graphicx}
\usepackage{booktabs,longtable,array,calc}
\usepackage{enumitem}
\usepackage{hyperref}
\hypersetup{colorlinks=true,linkcolor=blue,citecolor=blue,urlcolor=blue,hypertexnames=false}
\begin{document}

\textbf{Identifying Neural State Changes due to Gain versus Off-Manifold
Displacement}

Sam McKenzie\textsuperscript{1}

\begin{enumerate}
\def\labelenumi{\arabic{enumi}.}
\item
  University of New Mexico Health Sciences Center, Department of
  Neuroscience, Albuquerque, NM 87131
\end{enumerate}

\textbf{Summary}

Memory segmentation is thought to be caused by rapid decorrelation in
neural activity, commonly quantified by Euclidean distance and cosine
angle. While these metrics are useful for detecting that a transition
has occurred, they cannot reveal the nature of that change relative to
the repertoire of pre-existing states -- a space known as the neural
manifold. This distinction matters, since neuromodulators that are
thought to drive state transitions also influence neuronal excitability,
and there is ongoing controversy over whether learning involves
repurposing old representations or creating new ones. Here, I introduce
a decomposition that separates changes in neural activity into those
that can be understood by gain modulation of a nearby manifold point
versus genuine off-manifold displacement. The approach exploits the
privileged role of the radial axis in neural population activity by
partitioning the normal space of a local region of the neural manifold.
A central challenge is identifiability: from a static reference manifold
and a single test state, the state from which a perturbation began, its
gain magnitude, and, in general, the mechanistic decomposition of the
observed displacement cannot be uniquely recovered. I therefore
formulate identifiability as a cascade of geometric gates that specify
the conditions under which each component can be interpreted. These
gates distinguish structural failures, such as the absence of a local
manifold chart or incorrect intrinsic dimensionality, from estimation
error and systematic bias arising from the manifold reference, gain-axis
misalignment, and poor ratio conditioning. I derive exact expressions
for how reference error propagates into the reported components and
through simulations on idealized manifolds show that several
factors---neighborhood size, curvature, sampling density, ambient
dimension, and noise---affect the decomposition through a small set of
geometric quantities. This framework therefore does not simply assign a
neural state change to gain or novelty; it specifies when those
assignments are identifiable, how they become biased, and which
diagnostics can reveal the relevant failure regime. By quantifying the
nature of the neural state change, not just its magnitude, this
framework provides a more interpretable description of population
dynamics and a principled basis for evaluating candidate mechanisms
underlying neural state transitions.

\hypertarget{introduction}{%
\section{Introduction}\label{introduction}}

Though the external world arrives to our senses as a continuous stream,
it is recorded in episodic memory in discretized chunks bounded by
salient moments of change\textsuperscript{1-4}. This segmentation is
thought to be driven by rapid neuronal decorrelation facilitated by
phasic release of neuromodulators, such as
norepinephrine\textsuperscript{5,6}. When neural states change in
response to novelty, it is not always the case that a never-before-seen
stimulus produces a never-before-seen brain state\textsuperscript{7,8}.
We understand the future through the lens of the past, and the brain
maintains powerful neural attractors to categorize the new by the
old\textsuperscript{9,10}. Moreover, the same neuromodulators that
catalyze state transitions increase neuronal excitability, and indeed
such gain modulation is thought to be a driving force behind state
switching\textsuperscript{5}. Therefore, when neuronal activity shows
rapid transitions, it is important to know if these changes reflect
movements within the pre-existing state space, genuine creation of a
novel representation, or mere amplification of a prior state.

Cosine and Euclidean distances are standard metrics used to quantify
neuronal state similarity. While useful in detecting whether a state
change occurred, these metrics cannot resolve how the displaced neural
state relates to a reference manifold. Differential geometry provides a
mathematical language for describing the local state space through its
tangent and normal bundles. What is missing is the privileged role of
gain amplification, and more precisely whether one should consider a
familiar gain-modulated state to be on or off manifold. In recognition
of the central role that multiplicative gain modulation plays in
neuronal coding, the proposed decomposition further partitions the
normal space into a radial gain direction and its orthogonal complement.

The goal then is to apportion a neural state change into that which is
on the manifold of existing states, that which is attributable to
multiplicative gain of a pre-existing state, and a residual that is
off-manifold, in directions the sampled reference does not span. Such an
analysis raises important questions of identifiability, since gain is
defined only relative to a reference state, and that reference is not
generally recoverable from a single observation. Here I provide a
systematic account of how error in the reference sample, the intrinsic
dimension, the tangent frame, the anchor and the gain axis propagates
into the reported components, and of which of those conditions can be
checked from the data in hand.

Table 1. Symbols used throughout.

\begin{longtable}[]{@{}
  >{\raggedright\arraybackslash}p{(\columnwidth - 4\tabcolsep) * \real{0.1720}}
  >{\raggedright\arraybackslash}p{(\columnwidth - 4\tabcolsep) * \real{0.6559}}
  >{\raggedright\arraybackslash}p{(\columnwidth - 4\tabcolsep) * \real{0.1720}}@{}}
\toprule\noalign{}
\begin{minipage}[b]{\linewidth}\raggedright
\textbf{Symbol}
\end{minipage} & \begin{minipage}[b]{\linewidth}\raggedright
\textbf{Meaning}
\end{minipage} & \begin{minipage}[b]{\linewidth}\raggedright
\textbf{First used}
\end{minipage} \\
\midrule\noalign{}
\endhead
\bottomrule\noalign{}
\endlastfoot
\multicolumn{3}{@{}l@{}}{%
\textbf{Ambient space and manifold}} \\
\emph{F} & dimension of the ambient space, \ensuremath{\mathbb{R}}\^{}F & \S{}2.1 \\
\emph{\ensuremath{\mathcal{M}}} & reference manifold & \S{}2.1 \\
\emph{Z\textsubscript{A}} & reference sample, a point set in \ensuremath{\mathbb{R}}\^{}F
drawn from \ensuremath{\mathcal{M}} & \S{}2.1 \\
\emph{p} & a point on the reference manifold & \S{}2.1 \\
\emph{z\textsubscript{B}} & test point & \S{}2.1 \\
\emph{d} & assumed tangent dimension used by the estimator & \S{}2.1 \\
\emph{d}\textsubscript{true} & true intrinsic dimension of the manifold
& \S{}5.2 \\
\multicolumn{3}{@{}l@{}}{%
\textbf{Local geometry at a point}} \\
\emph{T\textsubscript{p}\ensuremath{\mathcal{M}}} & tangent space at p & Eq. (1) \\
\emph{N\textsubscript{p}\ensuremath{\mathcal{M}}} & normal space at p & Eq. (1) \\
\emph{P\textsubscript{T}}(p) & projector onto T\ensuremath{_p}\ensuremath{\mathcal{M}} & Eq. (2) \\
\emph{P\textsubscript{N}}(p) & projector onto N\ensuremath{_p}\ensuremath{\mathcal{M}}, I -- P\ensuremath{_t}(p) & Eq.
(2) \\
\ensuremath{\mathbb{I}} & second fundamental form, the symmetric bilinear map T\ensuremath{_p}\ensuremath{\mathcal{M}} \ensuremath{\times} T\ensuremath{_p}\ensuremath{\mathcal{M}} \ensuremath{\to} N\ensuremath{_p}\ensuremath{\mathcal{M}}
& \S{}5.4 \\
\emph{\ensuremath{\vec{H}}} & mean curvature vector & \S{}5.4 \\
\multicolumn{3}{@{}l@{}}{%
\textbf{Radial gain axis}} \\
\emph{\ensuremath{\hat{\rho}}}(p) & unit radial direction at p, p/\ensuremath{\lVert}p\ensuremath{\lVert} & Eq. (3) \\
\emph{a}(p) & alignment, \ensuremath{\lVert}P\ensuremath{_N}(p) \ensuremath{\hat{\rho}}(p)\ensuremath{\lVert} \ensuremath{\in} {[}0,1{]} & Eq. (3) \\
\emph{\ensuremath{\hat{n}}\textsubscript{g}}(p) & gain axis: the normalized radial
direction with its tangential part removed & Eq. (5) \\
\emph{N}\textsubscript{res}(p) & residual normal space: the orthogonal
complement of the gain axis within N\ensuremath{_p}\ensuremath{\mathcal{M}} & Eq. (4) \\
\emph{P}\textsubscript{res}(p) & projector onto N\ensuremath{_r}\ensuremath{_e}\ensuremath{_s}(p) & Eq. (6) \\
\multicolumn{3}{@{}l@{}}{%
\textbf{Displacement and energy fractions}} \\
\emph{r\textsubscript{p}} & displacement from the reference point, z\ensuremath{_B} --
p & \S{}2.1 \\
\emph{G}, \emph{T}, \emph{O} & gain, on-manifold (tangential), and
off-manifold energy fractions; G + T + O \ensuremath{\equiv} 1 & Eq. (7) \\
\emph{r}\textsubscript{tan} & on-manifold (tangential) component of r\ensuremath{_p} &
Eq. (6) \\
\emph{r}\textsubscript{novel} & off-manifold component of r\ensuremath{_p} & Eq.
(6) \\
\multicolumn{3}{@{}l@{}}{%
\textbf{Estimator (data-based substitutes)}} \\
\emph{N\textsubscript{A}} & local neighborhood: the k nearest reference
points to z\ensuremath{_B} & \S{}2.1 \\
\emph{k} & number of nearest neighbors defining the local neighborhood &
\S{}2.1 \\
\emph{n}\textsubscript{ref} & number of reference points & \S{}5.4.2 \\
\emph{\ensuremath{\mu}} & anchor: centroid of N\_A & \S{}2.1 \\
\emph{X\textsubscript{c}} & mean-centered neighbor matrix, N\_A -- \ensuremath{\mu} &
\S{}2.1 \\
\emph{V\textsubscript{d}} & estimated tangent basis: leading d right
singular vectors of X\_c & \S{}2.1 \\
\emph{\ensuremath{\hat{\rho}}\textsubscript{\ensuremath{\mu}}} & estimated radial direction, \ensuremath{\mu}/\ensuremath{\lVert}\ensuremath{\mu}\ensuremath{\lVert} & \S{}2.1 \\
\emph{\ensuremath{\ell}} & tangential extent of the local neighborhood & \S{}5.4 \\
\emph{r} & displacement from the anchor, z\textsubscript{\ensuremath{_B}} -- \ensuremath{\mu} & \S{}3 \\
\multicolumn{3}{@{}l@{}}{%
\textbf{Identifiability and anchor drift}} \\
\emph{m} & true manifold state from which the change proceeded & \S{}3 \\
\(\lambda\) & gain magnitude, z\textsubscript{\ensuremath{_B}} \ensuremath{\approx} \(\lambda\)m & \S{}3 \\
\multicolumn{3}{@{}l@{}}{%
\textbf{Anchor-frame decomposition (Proposition 1)}} \\
\emph{\ensuremath{\delta}} & signed anchor displacement along the gain direction,
\ensuremath{\hat{n}}\textsubscript{g}\ensuremath{\top}(m -- \ensuremath{\mu}) & Eq. (10) \\
\emph{\ensuremath{\vec{\eta}}}, \emph{\ensuremath{\eta}} & residual-normal component of the anchor
displacement, and its magnitude & Eq. (10) \\
\emph{\ensuremath{\tau}}\textsuperscript{2} & squared tangential component of the anchor
displacement & Eq. (10) \\
\emph{\ensuremath{\gamma}} & imposed displacement along the gain direction & \S{}3.2 \\
\emph{\ensuremath{\beta}} & imposed residual-normal displacement & \S{}3.2 \\
\emph{\ensuremath{\hat{b}}} & unit vector in the residual normal space (direction of the
imposed novelty) & \S{}3.2 \\
\emph{\ensuremath{\psi}} & angle between \ensuremath{\hat{b}} and \(\widehat{\eta}\) & Eq. (11) \\
\(O_{*}\) & ideal (veridical) off-manifold novelty, \ensuremath{\beta}\ensuremath{^2}/(\ensuremath{\gamma}\ensuremath{^2} + \ensuremath{\beta}\ensuremath{^2}) &
\S{}3.2(i) \\
\multicolumn{3}{@{}l@{}}{%
\textbf{Gate 1 --- dimension excess}} \\
\emph{q} & dimension excess, d -- d\textsubscript{true} & \S{}5.2 \\
\emph{L}(\ensuremath{\hat{b}}) & leakage, \ensuremath{\lVert}V\textsubscript{d}\textsuperscript{\ensuremath{\top}} \ensuremath{\hat{b}}\ensuremath{\lVert}\ensuremath{^2} & Eq.
(16) \\
\multicolumn{3}{@{}l@{}}{%
\textbf{Gate 5 --- where the displacement is booked}} \\
\emph{c} & second alignment, \ensuremath{\hat{n}}g\textsuperscript{\ensuremath{\top}}/
\ensuremath{\lVert}\(\overrightarrow{H}\)\ensuremath{\lVert} & Eq. (23) \\
\multicolumn{3}{@{}l@{}}{%
\textbf{Gate 6 --- ratio conditioning}} \\
\emph{C}\textsubscript{6} & observable conditioning index, k\ensuremath{\lVert}r\ensuremath{\lVert}\ensuremath{^2}/\ensuremath{\ell}\ensuremath{^2} &
Eq. (25) \\
\multicolumn{3}{@{}l@{}}{%
\textbf{Other}} \\
\emph{R} & radius of the validation sphere & \S{}5 \\
\emph{\ensuremath{\sigma}}, \emph{\ensuremath{\sigma}\textsubscript{w}} & noise scale, and noise scale along
a specified direction & \S{}5.5.1 \\
\end{longtable}

\hypertarget{the-decomposition}{%
\section{The decomposition}\label{the-decomposition}}

\hypertarget{the-tangent-and-normal-bundles}{%
\subsection{\texorpdfstring{\emph{The tangent and normal bundles:}
}{The tangent and normal bundles: }}\label{the-tangent-and-normal-bundles}}

Differential geometry describes an embedded manifold through its tangent
and normal bundles (Figure~1A), which at each point \emph{p} partition
the ambient space into directions tangent and orthogonal to the
manifold:

\(\mathbb{R}^{F} = T_{p}\mathcal{M \oplus}N_{p}\mathcal{M,}\quad\quad\)
(1)

the tangent space \(T_{p}\mathcal{M}\) contains the locally admissible
tangent directions of the manifold at point p, and the normal space
\(N_{p}\mathcal{M}\) contains directions that are orthogonal to those
tangent directions. Let \(\mathcal{M}\) be a reference manifold, sampled
as a point set \(Z_{A} \subset \mathbb{R}^{F}\), and let \(z_{B}\) be a
test point. We wish to know whether \(z_{B}\) is on-manifold, a
gain-modulated pre-existing point, or truly off-manifold. From static
geometry we cannot know which point was gain-modulated, and we are left
with a choice of where to anchor the gain estimate. A natural choice is
the nearest on-manifold point, and it is from this anchor that the
manuscript explores what is knowable and where the biases lie. Let p \ensuremath{\in} M
be a reference point on the manifold with p \ensuremath{\ne} 0. The projectors onto the
tangent and normal spaces at p are

\(P_{T}(p) = proj\ onto\ T_{p}M,\ \ \ \ P_{N}(p) = I - P_{T}(p)\quad\quad\)
(2)

Neuroscience accords privileged status to the radial axis, and to the
origin as a meaningful zero, since multiplicative gain is not usually
assumed to change what is being coded but rather the intensity of that
code. The unit radial axis and gain within the normal bundle are defined
as

\({\widehat{\rho}}_{p} = p/ \parallel p \parallel\),
\(a(p) = \left\| P_{N}(p){\widehat{\rho}}_{p} \right\|\  \in \lbrack 0,1\rbrack\)
(3)

Accounting for this gain axis, the decomposition further partitions the
normal space into a gain-modulated component and the off-manifold
residual,

\(N_{p}\mathcal{M} = span\left( P_{N}(p){\widehat{\rho}}_{p} \right) \oplus N_{res}(p)\quad\quad\)
(4)

The gain direction as it appears within the normal space is

\({\widehat{n}}_{g}(p) = \frac{P_{N}(p){\widehat{\rho}}_{p}}{a(p)},\quad defined\ iff\ a(p) > 0\quad\)
(5)

Defining the gain axis this way, rather than with \(\widehat{\rho}\)
alone, matters since some of that energy may otherwise be conflated with
tangential movement along the manifold (Figure~1B). Since
\({\widehat{n}}_{g}(p)\  \in \ N_{p}\mathcal{M}\) by construction, the
three subspaces \(T_{p}\mathcal{M}\), \(span({\widehat{n}}_{g}(p))\),
and \(N_{res}(p)\) are mutually orthogonal and the decomposition does
not depend on the order in which the projections are applied. Writing
\(r_{p} = z_{B} - p\) for the displacement from the manifold reference
point, and
\(P_{res}(p)\  = \ P_{N}(p)\  - \ {\widehat{n}}_{g}(p){{\widehat{n}}_{g}(p)}^{\top}\)
for the projector onto \(N_{res}(p)\), the decomposition can be written,

\(r_{p} = \underset{\text{on-manifold pattern}}{\underbrace{P_{T}(p)r_{p}}} + \underset{\text{gain}}{\underbrace{\left( {\widehat{n}}_{g}(p)^{\top}r_{p} \right)\,{\widehat{n}}_{g}(p)}} + \underset{\text{off-manifold}}{\underbrace{P_{res}(p)r_{p}}}\quad\quad\)
(6)

We report the three energy fractions

\(T = \frac{\parallel P_{T}(p)r_{p} \parallel^{2}}{\parallel r_{p} \parallel^{2}},\quad\quad G = \frac{\left( {\widehat{n}}_{g}(p)^{\top}r_{p} \right)^{2}}{\parallel r_{p} \parallel^{2}},\quad\quad O = \frac{\parallel P_{res}(p)r_{p} \parallel^{2}}{\parallel r_{p} \parallel^{2}},\quad\quad G + T + O \equiv 1\quad\quad\)
(7)

The novel contribution is the partition of the normal space into a
radial direction and its complement, which is what makes gain separable
from novelty. When the radial direction lies completely within the
tangent space (as on a cone through the origin), it has no normal
component and the split between gain and off-manifold displacement is
undefined (Figure 1C).

Equation 6 is stated for a manifold point \(p\) with its true tangent
space and radial direction. In an experimental setting, none of those is
observed and estimators are needed. To identify the anchor, find the
k-nearest neighbors of \(z_{B}\) in \(Z_{A}\), giving a local
neighborhood \(N_{A}\) with centroid \(\mu\) and mean-centered matrix
\(X_{c} = N_{A} - \mu\). Let \(V_{d}\) be the leading \(d\) right
singular vectors of \(X_{c}\), an estimate of the local tangent space.
The tangential extent of the neighborhood is
\(\mathcal{l}^{2}\  = \ \Sigma_{j}{\| V_{d}^{\top}(x_{j}\  - \ \mu)\|}^{2}/(k\  - \ 1)\);
normalizing by \(k\) instead biases \(\mathcal{l}^{2}\) low by
\((k\  - \ 1)/k\). Therefore, the estimator substitutes: a reference
sample \(Z_{A}\) for \(\mathcal{M}\), the \(k\)-nearest-neighbor
centroid \(\mu\) for \(p\), the leading \(d\) right singular vectors
\(V_{d}\) for \(T_{p}\mathcal{M}\), and
\({\widehat{\rho}}_{\mu} = \mu/ \parallel \mu \parallel\) for the radial
direction at \(p\).
Sections~\protect\hyperlink{what-is-identifiable-from-static-manifold-geometry}{3}
and~\protect\hyperlink{the-gates}{5} are about the biases introduced by
these substitutions: which of the four substitutions is consequential,
in what order the errors propagate, and which of them can be checked
from the data at hand. Reference sampling and the full estimator
construction are detailed in Supplement Section 1.

\hypertarget{what-is-identifiable-from-static-manifold-geometry}{%
\section{What is identifiable from static manifold
geometry}\label{what-is-identifiable-from-static-manifold-geometry}}

Static geometry provides information about the displacement from the
observable local anchor \(\mu\), but not generally about the
displacement from the state \(m\) at which the perturbation began. We
therefore distinguish anchor-relative geometric components from
perturbation-relative mechanistic components. Often one wants the total
between two states, \(z_{B} - m\), written here with
\(r\  = \ z_{B}\  - \ \mu\) the displacement from the anchor:

\(z_{B} - m = \underset{\text{anchor drift}}{\underbrace{(\mu - m)}} + \underset{\text{residual tangent}}{\underbrace{r_{\tan}}} + \underset{\text{residual gain}}{\underbrace{\left( {\widehat{n}}_{g}^{\top}r \right){\widehat{n}}_{g}}} + \underset{\text{residual novelty}}{\underbrace{r_{novel}}}\quad\quad\)
(8)

Since in the current construction \(m\) is unavailable, the present
decomposition reports what may be learned about deviations from the
anchor \(\mu\), and the relevance of that anchor in determining the
validity of off-manifold energy is a modelling choice. We do not
estimate a gain magnitude. The scalar \(\lambda\) for which
\(z_{B} \approx \lambda m\), and the manifold state \(m\) from which the
change proceeded, are not identifiable from static geometry. A change
may mix radial, tangential and off-manifold components, so \(m\) cannot
be read off \(z_{B}\); and even for a purely radial change, where the
same ray meets the manifold twice (\(m\) and \(\lambda m\) both on it),
the state \(\lambda m\) at unit gain and the state \(m\) amplified by
\(\lambda\) produce the same observation (Figure 3B).

\hypertarget{the-anchor-is-selected-by-the-test-point-not-by-the-base-point}{%
\subsection{\texorpdfstring{\emph{The anchor is selected by the test
point, not by the base
point}}{The anchor is selected by the test point, not by the base point}}\label{the-anchor-is-selected-by-the-test-point-not-by-the-base-point}}

Equation~(8) writes the anchor drift as \(\mu - m\), where the anchor
\(\mu\) is the centroid of the \(k\) nearest neighbors of \(z_{B}\).
Changing the imposed displacement can therefore change which reference
points are selected, and the estimator is a nonlinear function of the
very quantity it is being used to decompose (Figure 2A). Write
\(p = \pi_{\mathcal{M}}\left( z_{B} \right)\) for the projection of the
test point onto the local manifold and split the drift accordingly:

\(\mu - m = \underset{\text{selection-referred anchor bias}}{\underbrace{(\mu - p)}} + \underset{\text{change-dependent projection shift}}{\underbrace{(p - m)}}\quad\quad\)
(9)

The two terms have different characters. The first is a property of the
neighborhood geometry around the point the probe actually sits above,
and it is what Proposition~2 of
Section~\protect\hyperlink{gate-3-where-is-the-anchor}{5.4} derives. The
second is a property of the change itself: it is nonzero exactly when
the imposed displacement has a component along the manifold, and it is
therefore the same phenomenon as the anchor tracking described at the
start of Section~\protect\hyperlink{sec:whatasked}{5}; in the present
construction such tangential movement is absorbed through the selection
of a closer anchor.

\hypertarget{when-the-second-term-vanishes.}{%
\subsubsection{\texorpdfstring{\emph{When the second term
vanishes.}}{When the second term vanishes.}}\label{when-the-second-term-vanishes.}}

Two cases make \(p = m\) exactly, each sufficient (Figure 2B). Case 1:
If the displacement lies orthogonal to the span of the reference
manifold. Case 2: If the displacement is radial and the reference set
lies on an origin-centered sphere of radius \(R\), every reference point
has the same norm, so for \(z_{B}\  = \ \lambda m\) with
\(\lambda\  > \ 0\) the squared distance
\((\lambda^{2}\  + \ 1)R^{2}\  - \ 2\lambda\ m^{\top}x\) to a reference
point \(x\) decreases monotonically in \(m^{\top}x\) and the ordering is
unchanged. The radial and ambient validation probes of Section 5 use
exactly these two cases, so their neighborhoods are invariant to both
gain and novelty and \(p = m\). Those results therefore isolate
\(\mu - p\) with the second term in (9) exactly zero. The on-manifold
probes of Section 5 instead refer the anchor displacement to the probe's
own projection, which removes the second term by construction. Outside
these constructions the projection shift is not characterized and
requires additional information about the system.

\hypertarget{how-anchor-misspecification-contributes-bias}{%
\subsection{\texorpdfstring{\emph{How anchor misspecification
contributes
bias}}{How anchor misspecification contributes bias}}\label{how-anchor-misspecification-contributes-bias}}

Equation~(8) identifies anchor drift but does not show how it changes
the reported fractions. We can account for that effect by working in the
frame used by the estimator: the unit gain direction within the normal
space \({\widehat{n}}_{g}\), the tangent basis \(V_{d}\), and the
off-manifold residual normal space orthogonal to both.

Let \(m\) be the true base manifold point and \(\mu\) the estimated
anchor. Decompose the anchor displacement as

\(m - \mu = \delta\,{\widehat{n}}_{g} + \overrightarrow{\eta} + P_{T}(m - \mu),\quad\quad\overrightarrow{\eta} = P_{res}(m - \mu),\quad\eta = \parallel \overrightarrow{\eta} \parallel ,\quad\tau^{2} = \parallel P_{T}(m - \mu) \parallel^{2}\quad\quad\)
(10)

Here \(\delta = {\widehat{n}}_{g}^{\top}(m - \mu)\) is the signed anchor
displacement along the normal-space gain direction, \(\eta\) is its
residual-normal component, and \(\tau^{2}\) is the squared magnitude of
its tangential component (Figure 2C). Write the imposed change as
\(z_{B} - m = \gamma{\widehat{n}}_{g} + \beta\widehat{b}\), where
\(\gamma\) is the imposed displacement along the normal-space gain
direction, \(\beta\) the imposed residual-normal component, and
\(\widehat{b}\) a unit vector in the residual normal space. Here
\({\widehat{n}}_{g}\) and \(\widehat{b}\) are taken in the anchor frame
at \(\mu\). The resulting residual \(r = z_{B} - \mu\) is then resolved
exactly into gain, residual-normal and tangent components.

\hypertarget{proposition-1-exact-decomposition-of-the-fractions.}{%
\subsubsection{\texorpdfstring{\emph{Proposition 1 (exact decomposition
of the
fractions).}}{Proposition 1 (exact decomposition of the fractions).}}\label{proposition-1-exact-decomposition-of-the-fractions.}}

\(\parallel r \parallel^{2} = (\gamma + \delta)^{2} + \beta^{2} + \eta^{2} + 2\beta\eta\cos\psi + \tau^{2},\quad\quad\cos\psi = \widehat{b} \cdot \widehat{\eta}\quad\quad\)
(11)

\(G = \frac{(\gamma + \delta)^{2}}{\parallel r \parallel^{2}},\quad\quad O = \frac{\beta^{2} + \eta^{2} + 2\beta\eta\cos\psi}{\parallel r \parallel^{2}},\quad\quad T = \frac{\tau^{2}}{\parallel r \parallel^{2}}\quad\quad\)
(12)

This identity follows only from the mutual orthogonality of the gain,
residual-normal and tangent subspaces. It therefore provides an exact
accounting of how anchor drift changes the reported fractions,
independently of the approximations used below. The proof is given in
Supplement Section 1, and Table S1 verifies the identity numerically at
machine precision. Proposition 1 assumes that the imposed change has no
component in the anchor frame's tangent space; in general the tangential
term is \({\| P_{T}(z_{B}\  - \ m)\  + \ P_{T}(m\  - \ \mu)\|}^{2}\) in
place of \(\tau^{2}\), and Table S1 checks this general form.

Several useful consequences follow; (ii)--(iv) are derived in the
Supplement.

\emph{(i) The ideal case.} Where the anchor lies on the correct manifold
point, \(\delta = \eta = \tau = 0\) and the veridical off-manifold
novelty
is\(\ O_{\star} = \beta^{2}/\left( \gamma^{2} + \beta^{2} \right)\).

\emph{(ii) First order.} Ordering terms in the neighborhood extent ---
\(\delta\) and \(\eta\) are \(O(\mathcal{l}^{2})\), \(\tau^{2}\) is
\(O(\mathcal{l}^{2}/k)\), and \(\delta^{2}\) and \(\eta^{2}\) are
\(O(\mathcal{l}^{4})\) and dropped --- the bias in the off-manifold
estimate is,

\(O - O_{\star} \simeq \frac{2O_{\star}\gamma}{\gamma^{2} + \beta^{2}}\left\lbrack \frac{\gamma\eta\cos\psi}{\beta} - \delta \right\rbrack - \frac{O_{\star}\tau^{2}}{\gamma^{2} + \beta^{2}}\quad\quad\)
(13)

When \(\gamma\delta\  > \ 0\), anchor drift along the gain direction
increases the apparent gain and reduces \(O\) (Section 5.7 treats the
reversal). Residual-normal anchor drift enters Eq. (13) only through
\(cos\ \psi\), so its first-order sign is set by the angle between
\(\widehat{b}\) and \(\overrightarrow{\eta}\); its positive-definite
contribution is the second-order \(\eta^{2}\), seen directly at
\(\beta\  = \ 0\) in Eq. (14). The final term is the contribution from
tangential displacement.

\emph{(iii) Zero imposed novelty.} At \(\beta = 0\) the first-order form
(13) diverges, since the \(\eta\) term carries \(1/\beta\), but the
exact form does not:

\(O = \frac{\eta^{2}}{(\gamma + \delta)^{2} + \eta^{2} + \tau^{2}}\quad\quad\)
(14)

With no imposed novelty, any measured \(O\) arises from the
residual-normal component of anchor drift.

\emph{(iv) A random novel direction carries no bias.} For
\(\widehat{b}\) uniformly oriented in the residual normal space,
\(\mathbb{E}\left\lbrack \cos\psi \right\rbrack = 0\). The cross term in
(12) therefore contributes variance but no first-order mean bias; what
\(\eta\) leaves on average is the \(\eta^{2}\) term, which appears as a
floor in sweeps that randomize the imposed novel direction.

\hypertarget{the-cascade}{%
\section{The cascade}\label{the-cascade}}

The preceding sections establish the construction and introduce the
associated limits in identifiability. Here, I establish which quantities
can be interpreted and the conditions in which each may be estimated.
\hypertarget{tab:cascade}{}Table~\protect\hyperlink{tab:cascade}{2} gives this dependency order
which is validated in Section~\protect\hyperlink{the-gates}{5}. The
ordering is a dependency of the computation, not an ordering of damage:
for example, over-estimating intrinsic dimension can bypass the
intermediate gates and directly corrupt the gain axis
(Section~\protect\hyperlink{gate-1-is-mathbfd-right}{5.2}).

The results also differ in evidentiary status. Some are exact
identities, including \(G + T + O = 1\) and the accounting in
Proposition~1. Others are local asymptotic results, such as
Propositions~2 and~3, whose validity depends on the stated sampling
regime and approximation order. A third class comprises model-specific
scalings, whose coefficients depend on the manifold and sampling
distribution. The remainder are empirical results from the synthetic
sweeps.

\begin{longtable}[]{@{}
  >{\raggedright\arraybackslash}p{(\columnwidth - 8\tabcolsep) * \real{0.0681}}
  >{\raggedright\arraybackslash}p{(\columnwidth - 8\tabcolsep) * \real{0.1471}}
  >{\raggedright\arraybackslash}p{(\columnwidth - 8\tabcolsep) * \real{0.2845}}
  >{\raggedright\arraybackslash}p{(\columnwidth - 8\tabcolsep) * \real{0.1275}}
  >{\raggedright\arraybackslash}p{(\columnwidth - 8\tabcolsep) * \real{0.3728}}@{}}
\toprule\noalign{}
\begin{minipage}[b]{\linewidth}\raggedright
\textbf{Gate}
\end{minipage} & \begin{minipage}[b]{\linewidth}\raggedright
\textbf{Identifiability question}
\end{minipage} & \begin{minipage}[b]{\linewidth}\raggedright
\textbf{Diagnostic quantity}
\end{minipage} & \begin{minipage}[b]{\linewidth}\raggedright
\textbf{Directly\\
evaluable?}\strut
\end{minipage} & \begin{minipage}[b]{\linewidth}\raggedright
\textbf{Cost of failure / interpretation}
\end{minipage} \\
\midrule\noalign{}
\endhead
\bottomrule\noalign{}
\endlastfoot
0 & Does a local chart exist? & Local sheet separation relative to
neighborhood extent and noise & No & \(\widehat{\ensuremath{\mu}}\) and
\({\widehat{V}}_{d}\) become unreliable; mixed sheets create artifacts
that mimic gain or novelty. \\
1 & Is d right? & \(q = d - d_{true}\) stability of the estimated
tangent space as d varies & No & Over-estimating d removes directions
from the normal complement, corrupting O and potentially destroying the
gain axis. \\
2 & Given a valid chart and d, is the estimate good? &
\(\left\| {\widehat{V}}_{d} - V_{d} \right\|\) and gain-axis-direction
error & No & Frame-estimation error rotates the tangent/normal
decomposition. In the tested regime it is bounded by simulation and is
small at \(d = d_{true}\). \\
3 & Where is the anchor? & Anchor bias: (\ensuremath{\ell}\ensuremath{^2}/2) \(\overrightarrow{H}\);
anchor jitter: \ensuremath{\ell}/\ensuremath{\sqrt{}}k & \emph{Partial} & \ensuremath{\ell} and k are measurable, but
\(\overrightarrow{H}\) is not estimated here. Curvature-induced anchor
bias is the dominant systematic error. \\
4 & Does the gain axis exist and remain usable? &
\(a = \left\| P_{N}\widehat{\rho} \right\|\) & \textbf{Yes} & \(a = 0\)
: gain axis is undefined. 0 \textless{} \(a\) \ensuremath{\ll} 1: defined but
ill-conditioned. Decreasing \(a\) can convert G into O. \\
5 & Where is the anchor displacement booked? &
\(c = \ {\widehat{n}}_{g}^{\top}\ \overrightarrow{H}\ /\ \left\| \overrightarrow{H} \right\|\)
& No & Requires the mean-curvature direction \(\overrightarrow{H}\ \).
\textbar c\textbar{} = 1 routes anchor displacement to G; c = 0 routes
it to spurious O. \\
6 & Is the G/O ratio well conditioned? & C\ensuremath{_6} = k\ensuremath{\lVert}r\ensuremath{\lVert}\ensuremath{^2}/\ensuremath{\ell}\ensuremath{^2}, computed per
probe; C\ensuremath{_6} -- 1 \ensuremath{\gg} 1 & \textbf{Yes} & When the leading denominator is
comparable to anchor jitter, the reported O becomes unstable or
uninformative. Table S17 gives the recovery error by range of C\ensuremath{_6}. \\
\end{longtable}

Table 2. The cascade. Each gate is conditional on those above it.
``Directly evaluable'' means that the relevant quantity can be estimated
from the observed reference/test data without an external manifold model
or synthetic calibration. Partial indicates that only part of the gate
is directly measurable. The decomposition is therefore conditional:
Gates 0--2 establish whether a local geometric estimate is trustworthy;
Gate 3 quantifies anchor bias; Gate 4 tests whether the gain axis is
defined and sufficiently aligned; Gate 5 determines how anchor bias is
apportioned between G and O; and Gate 6 determines whether the resulting
ratio is numerically informative.

\hypertarget{the-gates}{%
\hypertarget{sec:whatasked}{}\section{The gates}\label{the-gates}}

Validation is constrained by the same anchor-drift problem that limits
the decomposition itself. Because the anchor is defined from the test
point's own neighbors, it moves to absorb any tangential component of
the imposed displacement, so a na\"{i}ve test would confound estimator error
with the anchor-drift term of Eq. (8). Every construction below
therefore either leaves neighbor selection invariant to the imposed
displacement, or refers the anchor to the probe's own projection; the
two families and the manifolds each is used on are detailed in
Supplement Section 1.2.

\hypertarget{gate-0-does-a-local-chart-exist}{%
\subsection{\texorpdfstring{\emph{Gate 0: does a local chart
exist?}}{Gate 0: does a local chart exist?}}\label{gate-0-does-a-local-chart-exist}}

When two sheets of a manifold lie within the neighborhood radius, the
neighborhood mixes different local geometries, so the estimated tangent
space belongs to neither sheet (Figure 3A). This can occur at a branch
point or between nearby turns of a Swiss roll. The resulting artifact is
largest when sheet separation is comparable to neighborhood extent, and
it cannot be identified from \(G,T,O\) alone because it can mimic
genuine gain or novelty (radial stacking reads as gain, ambient stacking
as novelty; Table S2). Cross-sheet neighbor counts are likewise
insufficient: the relevant quantity is the ratio of sheet separation to
neighborhood extent (Table S2); under anisotropic noise the noise scale
along the separation direction should also enter, which is not tested
here. In practice, neighborhoods should therefore be chosen small enough
that distinct sheets remain separated at the relevant noise scale.

Large gain can introduce a second failure: if radial scaling carries the
test point into a region where the manifold intersects the same ray
again, the gain representation is no longer injective and the nearest
anchor can switch to another state (Figure 3B). Gain magnitude must
therefore remain within a range over which the radial map is injective.

\hypertarget{gate-1-is-mathbfd-right}{%
\subsection{\texorpdfstring{\emph{Gate 1: is} \(\mathbf{d}\)
\emph{right?}}{Gate 1: is \textbackslash mathbf\{d\} right?}}\label{gate-1-is-mathbfd-right}}

Assessing whether a vector lies off-manifold requires knowledge of the
dimensionality of the manifold tangent space within an ambient space of
dimension \emph{F}. The off-manifold normal space is defined as the
orthogonal complement of the assumed tangent space: over-estimating
\(d\) removes directions from the space in which \(O\) and the gain axis
are defined (Figure 4A). The expected survival law is obtained when the
\(q = d - d_{true}\) excess directions are uniformly oriented within the
\(F - d_{true}\)-dimensional complement. Rotational invariance then
gives a Beta distribution for the fraction of an off-manifold direction
that survives the projection, with mean

\(\mathbb{E}\left\lbrack \text{survival} \right\rbrack = 1 - \frac{q}{F - d_{true}},\quad\quad\)
(15)

and its median is the appropriate comparator when survival is summarized
by medians (Table S3; Supplement Section 1). We test this directly using
a purely off-manifold displacement \(\beta\widehat{b}\), with
\(\widehat{b}\) the unit residual-normal direction of
Section~\protect\hyperlink{how-anchor-misspecification-contributes-bias}{3.2}.
Define leakage as

\(\mathcal{L}\left( \widehat{b} \right) = \parallel V_{d}^{\top}\widehat{b} \parallel^{2},\quad\quad\)
(16)

the fraction of \(\widehat{b}\) incorrectly absorbed by the estimated
tangent space. Leakage isolates the cost of dimension over-estimation.

The Beta prediction holds when the excess directions are effectively
random (Table~S3), but curvature creates an important exception. A
curved manifold has a genuine normal direction associated with its local
sagitta. If the variance that curvature contributes exceeds the noise
floor, that specific direction has more variance than a generic ambient
direction would, so it wins the \((d_{true}\  + \ 1)\)-th singular
vector, and the excess dimension absorbs the normal direction and with
it the gain axis (Figure 4B). At \(q = 1\) this produces two opposing
effects: leakage is lower than the random-direction prediction because
the discarded direction is not the imposed novelty direction, but the
gain axis is destroyed, because that axis is constructed from the normal
space. Consequently Gate~4 does not degrade smoothly with dimension
over-estimation: its alignment can collapse at the first excess
dimension and remain low thereafter.

When curvature is below the directional noise scale, the excess
directions become effectively random and the Beta law is recovered, with
alignment declining more gradually as the dimensional mismatch \(q\)
increases. Table S3 locates the crossover by measuring the squared
overlap between the local neighborhood\textquotesingle s
\(d_{true} + 1\)-th singular vector and the true normal direction ---
its \emph{capture} of the excess dimension --- which is nearly complete
at low noise and falls to the chance level once the noise scale exceeds
the curvature\textquotesingle s spread along the normal. On an
origin-centered manifold the gain axis is the normal, so the alignment
at \(q\  = \ 1\) satisfies \(a^{2}\  \approx \ 1\  - \ capture\). Under
anisotropic noise the relevant comparison is directional: whether the
normal appears as the first excess direction depends on noise variance
along the normal, not on overall noise magnitude. The expected loss of
off-manifold energy is only part of the problem; depending on curvature
and directional noise, the excess dimension can remove the normal
direction itself and thereby invalidate the gain axis.

\hypertarget{gate-2-given-a-chart-and-the-right-mathbfd-is-the-local-tangent-estimate-correct}{%
\subsection{\texorpdfstring{\emph{Gate 2: given a chart and the right}
\(\mathbf{d}\)\emph{, is the local tangent estimate
correct?}}{Gate 2: given a chart and the right \textbackslash mathbf\{d\}, is the local tangent estimate correct?}}\label{gate-2-given-a-chart-and-the-right-mathbfd-is-the-local-tangent-estimate-correct}}

Next it is necessary to confirm whether the normal and tangent axes may
be estimated. For each validation probe we decompose \(r = z_{B} - \mu\)
using both the estimated local geometry (\(V_{d}\),
\({\widehat{n}}_{g},\ \widehat{b}\)) and the exact geometry of the
generating manifold at the projection of \(\mu\) onto it. Their
difference isolates estimation error from anchor-driven effects.

For the uniformly sampled 2-D sphere used here, the leading
finite-sample error is a random tilt of the fitted plane (Figure 5).
Under symmetric sampling the tilt has zero mean, but its variance is set
by the sample covariance between tangential position and normal offset.
For a two-dimensional neighborhood sampled uniformly, with normal
curvature \(\kappa\) (\(\kappa\  = \ 1/R\) on a sphere of radius \(R\)),

\(E{\| V_{d}^{\top}\widehat{n}\|}^{2}\  = \ \frac{\mathcal{l}^{2}\kappa^{2}}{3k}\lbrack 1\  + \ O(1/k)\rbrack\)
(17)

which on a uniformly sampled 2-sphere, where \(\kappa\  = \ 1/R\) and
\(\mathcal{l}^{2}\  \approx \ 2kR^{2}/n_{ref}\), is
\(2/(3n_{ref})\lbrack 1\  + \ O(1/k)\rbrack\). Tables S4 and S5 confirm
the prefactor, the \(1/n_{ref}\) scaling and the independence from \(k\)
and \(R\), up to an \(O(1/k)\) excess that is largest at small \(k\)
(derivation in the Supplement). The rotation is approximately
exponentially distributed across probes, so its median lies below its
mean.

\hypertarget{two-mechanisms-by-which-the-frame-rotates}{%
\subsubsection{\texorpdfstring{\emph{Two mechanisms by which the frame
rotates}}{Two mechanisms by which the frame rotates}}\label{two-mechanisms-by-which-the-frame-rotates}}

Splitting the local coordinates into a tangential part \(u\) and a
normal part \(w\),

\(M = \begin{pmatrix}
\mathbb{E}\left\lbrack uu^{\top} \right\rbrack & \mathbb{E}\left\lbrack uw^{\top} \right\rbrack \\
\mathbb{E}\left\lbrack wu^{\top} \right\rbrack & \mathbb{E}\left\lbrack ww^{\top} \right\rbrack
\end{pmatrix},\quad\quad\) (18)

the estimated tangent space can rotate through two mechanisms.
Variance-driven rotation occurs when normal variance becomes large
enough to compete with tangent variance, as with excess intrinsic
dimension or anisotropic noise; this error does not vanish with
additional reference data. Covariance-driven rotation occurs when
\(\mathbb{E}\left\lbrack uw^{\top} \right\rbrack \neq 0\), tilting the
leading eigenspace even when normal variance is small.

Under symmetric sampling on a homogeneous manifold the covariance term
vanishes in expectation, because the leading normal displacement is
quadratic in tangent coordinates and therefore even in \(u\). Finite
sampling breaks this symmetry, producing the floor of Eq. (17).
Systematic asymmetry, however, produces persistent covariance-driven
rotation.

\protect\hypertarget{sec:covroute}{}{}Locally, normal displacement is
determined by curvature, and curvature gradients introduce higher-order
terms that can break the symmetry between opposite tangent directions.
Expanding the normal displacement along a single tangent direction gives
a quadratic term set by the local normal curvature and a cubic term set
by its derivative along that direction (derivation in the Supplemental
proofs). The quadratic term is even in the tangent coordinate and so
cannot covary with it; the cubic term is odd and can.

Beyond that floor, two forms of asymmetry break the symmetric case of
Figure 6A and generate systematic covariance-driven rotation: a manifold
whose curvature is not constant (Figure 6B), and a probe close to the
edge of the sampled region (Figure 6C). In either case the symmetry
argument that made the frame rotation k-independent no longer holds.
Once the finite-sample floor is subtracted, the rotation grows as
\(\mathcal{l}^{4}\  \propto \ k^{2}\) in the presence of a curvature
gradient, as predicted (Table S5). Near a boundary the rotation leaves
the floor once the probe is within about one neighborhood extent of the
edge, and grows close to \(\mathcal{l}^{2}\) (Table S6). Therefore,
large \emph{k} influences bias through both anchor bias and, near an
edge or a curvature gradient, through a frame tilt that grows faster
than the anchor bias does. A density gradient produces a third route,
although its scaling is not isolated here.

Neighborhood size therefore acts differently in symmetric and asymmetric
regions. In symmetric interior regions, larger neighborhoods primarily
increase anchor bias while leaving the frame stable. Near boundaries, or
where sampling density changes, larger neighborhoods also increase frame
rotation. Frame tilt and excess intrinsic dimension both rotate normal
directions into \(V_{d}\), and so both enter the leakage of Eq. (16);
only an excess dimension also removes a direction from the complement,
which is what the survival law of Eq. (15) counts.

\hypertarget{gate-3-where-is-the-anchor}{%
\subsection{\texorpdfstring{\emph{Gate 3: where is the
anchor?}}{Gate 3: where is the anchor?}}\label{gate-3-where-is-the-anchor}}

Equation~(8) isolates the anchor drift but leaves it unquantified.
Because \(\mu\) is a centroid of points sampled from a curved set, its
displacement from the manifold has a leading term fixed by the local
second-order geometry. Let \(\mathbb{I\ }\)be the second fundamental
form, the symmetric bilinear map
\(T_{p}\mathcal{M \times}T_{p}\mathcal{M \rightarrow}N_{p}\mathcal{M}\)
whose value \(\mathbb{I}(u,u)\) is twice the leading normal displacement
of a point at tangent coordinate \(u\); its component along a single
normal direction is the local normal curvature used in
Section~\protect\hyperlink{sec:covroute}{5.3.1}. Let
\(\overrightarrow{H} = d^{- 1}\sum_{i}^{}\mathbb{I}\left( e_{i},e_{i} \right)\)
be the mean curvature vector, and let the \(k\) neighbors be drawn from
a locally isotropic measure with
\(\mathcal{l}^{2}\mathbb{= E \parallel}u \parallel^{2}\) the squared
tangential extent of the neighborhood. Because the \(k\)-th neighbor
lies on the boundary of the selection ball, \(\mathcal{l}^{2}\) exceeds
the areal value \(kArea/(2\pi n_{ref})\) by \((k\  + \ 1)/k\); the
centered estimator of Section 2.1 is unbiased for it.

\hypertarget{proposition-2-anchor-displacement.}{%
\paragraph{Proposition 2 (anchor
displacement).}\label{proposition-2-anchor-displacement.}}

With \(p = \pi_{\mathcal{M}}\left( z_{B} \right)\) the manifold point
the probe sits above, the expected bias in the anchor placement relative
to that projected point can be described by,

\(\mathbb{E}\lbrack\mu\rbrack - p = \frac{\mathcal{l}^{2}}{2}\,\overrightarrow{H} + O\left( \mathcal{l}^{3} \right)\quad\quad\)
(19)

conditional on a neighborhood-selection distribution centered at
\emph{p} and locally isotropic in tangent coordinates.

\hypertarget{proposition-3-anchor-jitter.}{%
\paragraph{Proposition 3 (anchor
jitter).}\label{proposition-3-anchor-jitter.}}

The tangential fluctuation of the anchor has root-mean-square magnitude

\(\mathbb{(E \parallel}\bar{u} \parallel^{2})^{1/2}\mathcal{= l/}\sqrt{k}\quad\quad\)
(20)

Both are proved in the Supplement. The displacement is a normal-space
vector, so at leading order the anchor is not biased along the manifold.
It is governed by the \emph{trace} of \(\mathbb{I}\) rather than by
curvature magnitude (Figure 7A), so at a point of zero mean curvature
(for example a saddle with equal and opposite principal curvatures) the
anchor is unbiased at leading order, however sharply the surface curves
(Figure 7B). And its direction is that of \(\overrightarrow{H}\), which
need not be the gain axis --- the topic of Gate 5. Enlarging the
neighborhood increases the curvature bias, while its effect on anchor
jitter depends on intrinsic dimension. Under locally uniform sampling
\(\mathcal{l \propto}k^{1/d}\), so the jitter scales as
\(\mathcal{l/}\sqrt{k}\ \  \propto k^{1/d\  - 1/2}\ \). Thus, anchor
jitter decreases in \emph{k} only for \emph{d\textgreater2}, k-invariant
for d=2, and increasing for d=1.

For a sphere of radius \(R\), \(\overrightarrow{H} = - \widehat{n}/R\)
and the displacement is \(\mathcal{l}^{2}/2R\) directed inward,
independent of \(d\) and \(F\); when the sphere is origin-centered this
is collinear with the gain axis. With \(\mathcal{l}\) measured in the
tangent plane the next order is known,
\(\delta\  = \ (\mathcal{l}^{2}/2R)(1\  + \ \mathcal{l}^{2}/3R^{2})\)
(Supplement). Table~S7 tests Propositions~2--3 on quadric patches, where
the anisotropic, biaxial and zero-mean-curvature rows are the
informative tests, and Table~S8 repeats the test on a torus of
revolution, where curvature varies over the surface and, for a
sufficiently fat torus, the mean curvature changes sign --- the measured
displacement reverses with it, a directional test the sphere cannot
supply.

\hypertarget{what-the-free-parameters-actually-control}{%
\subsubsection{\texorpdfstring{\emph{What the free parameters actually
control}}{What the free parameters actually control}}\label{what-the-free-parameters-actually-control}}

Proposition~2 expresses the anchor bias through neighborhood size
\(\mathcal{l}\), which is not itself free but a consequence of several
parameters that are. For a \(d\)-dimensional manifold of sampled volume
\(A\) and \(n_{ref}\) uniformly distributed reference points, \(k\)
neighbors span a region of area \(kA/n_{ref}\), so
\(\mathcal{l \propto}\left( kA/n_{ref} \right)^{1/d}\). For the 2-sphere
of radius \(R\) this closes, to leading order in \(\mathcal{l/}R\) and
including the boundary factor \((k\  + \ 1)/k\):

\(\mathcal{l}^{2}\  = \ \frac{2(k\  + \ 1)R^{2}}{n_{ref}},\ \ \delta\  = \ \frac{\mathcal{l}^{2}}{2R}\  = \ \frac{(k\  + \ 1)R}{n_{ref}},\ \ \frac{\mathcal{l}}{\sqrt{k}}\  = \ R\sqrt{\frac{2(k\  + \ 1)}{k\ n_{ref}}}\)
(21)

On a closed sphere at fixed \(n_{ref}\) and \(k\) the whole
configuration scales with \(R\), so \(\delta\  \propto \ R\) is
dimensional: \(\delta/R\) is fixed (Table S9). A flatter reference
manifold is worse only relative to a fixed absolute displacement or
noise scale; on a patch of fixed area, \(\mathcal{l}\) does not grow
with \(R\) and \(\delta\) falls. \(\delta\) is reduced by raising
\(n_{ref}\) or lowering \(k\). Bias and jitter respond differently to
\(k\) and \(n_{ref}\), though both scale with \(R\): the bias grows as
\(k\) while the jitter is \(k\)-invariant at \(d\  = \ 2\), and the bias
falls as \(n_{ref}^{- 1}\) while the jitter falls only as
\(n_{ref}^{- 1/2}\). At high sampling density the residual error is
therefore dominated by fluctuation rather than by bias, although both
vanish (Table S9).

Nearest-neighbor selection favors points whose noise displaced them
toward the probe; under isotropic noise that preference has no net
direction and cancels, but under anisotropic noise it does not, and
Proposition~2 fails in both directions depending on where the anisotropy
sits (Table~S10). Noise along the radial direction erases the sagitta,
since a radially thickened shell allows selection to pull the neighbors'
mean radius toward the probe's own; noise concentrated tangentially
inflates it, because tangentially displaced points from further around
the manifold become selectable. Noise in the residual normal space is
inert. Two further channels do not act here: a probe's own displacement
cannot move the anchor within the validation construction, by design
(Section~\protect\hyperlink{the-anchor-is-selected-by-the-test-point-not-by-the-base-point}{3.1}),
so its absence is not evidence about other constructions; and under the
isotropic reference-noise model tested here, reference noise enters as
fluctuation and by inflating the \emph{measured} \(\mathcal{l}\) rather
than as a bias channel of its own (Table S9).

\hypertarget{gate-4-does-the-gain-axis-exist}{%
\subsection{\texorpdfstring{\emph{Gate 4: does the gain axis
exist?}}{Gate 4: does the gain axis exist?}}\label{gate-4-does-the-gain-axis-exist}}

The alignment \(a = \parallel P_{N}\widehat{\rho} \parallel\) governs
this gate and, unlike Gate 5, is computable from \(\mu\) and \(V_{d}\)
alone. The validation manifolds are origin-centered spheres, where the
radial direction is exactly the normal and this gate holds by
construction; the complementary regime is a cone through the origin,
where the alignment is zero in the continuum and small under finite
sampling. A torus of revolution provides the intermediate case: its
alignment has a closed form varying over the surface and passing through
zero, so a single manifold contains well-conditioned and degenerate
regions at once. The estimated alignment tracks the analytic value
pointwise and degeneracy is flagged only near the zero crossing
(Table~S8).

\hypertarget{anisotropic-noise}{%
\subsubsection{\texorpdfstring{\emph{Anisotropic
noise}}{Anisotropic noise}}\label{anisotropic-noise}}

Anisotropy in a tangential direction moves the fractions by at most a
few hundredths even when the noise scale along it is about twice the
neighborhood extent (Table S11): the anchor tracks tangential
displacement, and the same tracking that makes \(T\) unidentifiable
absorbs tangential noise. Anisotropy in the residual normal space is
likewise almost inert, entering numerator and denominator of \(O\)
together. Anisotropy along the gain axis is not absorbed. There the
alignment degrades and energy moves out of \(G\) and into \(O\),
monotonically in the ratio of the noise scale along that direction to
the neighborhood extent and with no safe threshold (Table~S11); the
mechanism is the variance route of
Section~\protect\hyperlink{two-mechanisms-by-which-the-frame-rotates}{5.3.1},
and the frame-rotation quantity of Eq.~(17) rises by orders of magnitude
in step with the alignment's fall (Table~S12).

The damage is done while the alignment is still far above any tolerance
one would set: by the time \(a\) would trip a threshold chosen to flag a
non-existent gain axis, most of \(G\) has been converted to \(O\). The
tolerance detects a gain axis that does not exist; it does not detect
one that exists but has been contaminated. Report \(a\) alongside \(G\)
and \(O\) and treat decreases in \emph{a} as places where the \emph{G/O}
split becomes ambiguous.

\hypertarget{gate-5-is-anchor-displacement-in-the-gain-axis-or-residual-normal-space}{%
\subsection{\texorpdfstring{\emph{Gate 5: is anchor displacement in the
gain axis or residual normal
space?}}{Gate 5: is anchor displacement in the gain axis or residual normal space?}}\label{gate-5-is-anchor-displacement-in-the-gain-axis-or-residual-normal-space}}

Decomposing the anchor displacement of Proposition~2 into a component
along the gain axis and a component in the residual normal space,

\(\delta = - \frac{\mathcal{l}^{2}}{2}\,{\overrightarrow{H}}^{\top}{\widehat{n}}_{g},\quad\quad\eta = \frac{\mathcal{l}^{2}}{2}\, \parallel P_{res}\overrightarrow{H} \parallel ,\quad\quad\)
(22)

shows that it enters the denominator of \(O\) as apparent gain through
\(\delta\) and the numerator as spurious novelty through \(\eta\). Only
the first is present when manifold curvature aligns with gain,
\(\overrightarrow{H} \parallel {\widehat{n}}_{g}\). The governing
quantity is therefore a second alignment,

\(c = {\widehat{n}}_{g}^{\top}\overrightarrow{H}/ \parallel \overrightarrow{H} \parallel ,\quad\quad\)
(23)

independent of the \(a\) of Gate~4: at \(|c| = 1\) the displacement is
absorbed entirely into the gain leg and the numerator of \(O\) is
unaffected (Figure 8A), at \(c = 0\) it appears entirely as spurious
novelty (Figure 8B), and in general it is apportioned as \(c^{2}\) and
\(1 - c^{2}\). Unlike \(a\), \(c\) requires an estimate of
\(\mathbb{I}\) and must be argued from the manifold geometry rather than
read off the data.

Table~S13 confirms the case \(|c| = 1\) using pure geodesic probes on
spheres. The absolute off-manifold energy there is flat at the
observation-noise floor across every radius and neighborhood size
tested, while the fraction \(O\) varies from cell to cell through its
denominator alone --- the sagitta and the tangential jitter together ---
not through leakage into the numerator. This is the practical payoff of
building the gain axis inside the normal space: the sagitta is routed
into \(G\) and the numerator of \(O\) is left untouched, whenever
\(|c| = 1\).

That case is not general, and there is a structural reason it is not
merely a property of the sphere. For a \(d\)-manifold whose linear
\emph{span} is \((d + 1)\)-dimensional the normal space inside that span
is one-dimensional, so \(P_{N}\widehat{\rho}\), the surface normal and
\(\overrightarrow{H}\) are forced parallel and \(|c| = 1\) identically
--- which covers the sphere, the spherical cap, every quadric patch in
three-space, the torus of revolution and the ellipsoid. Reaching
\(|c| < 1\) requires the manifold's linear span to be at least
\(d + 2\). The Clifford torus in four dimensions is the simplest such
geometry with alignment exactly one, so it isolates this gate from
Gate~4; Table~S14 shows the absolute novel energy rising from the noise
floor to track \(\eta\) as the torus is made asymmetric, and growing in
proportion to \(\mathcal{l}^{2}\) where on the sphere it was flat.
Because that torus has a parallel second fundamental form (the rate of
change of the curvature is zero across the surface), a circle \(\times\)
ellipse is used to repeat the test where the second fundamental form
varies (Table~S15); the rule survives, with both regimes appearing on a
single manifold.

\hypertarget{gate-6-is-the-ratio-well-conditioned}{%
\subsection{\texorpdfstring{\emph{Gate 6: is the ratio
well-conditioned?}}{Gate 6: is the ratio well-conditioned?}}\label{gate-6-is-the-ratio-well-conditioned}}

O is defined as a ratio (energy along one direction over total energy),
and any ratio estimate is only trustworthy when its denominator is not
itself dominated by noise or fluctuation. On an origin-centered sphere
\(\eta = 0\), and the first-order expansion (13) reduces to

\(\mathbb{E}\lbrack O\rbrack - O_{\star} \simeq - 2\, O_{\star}\sqrt{1 - O_{\star}}\mspace{6mu}\frac{\delta}{\sqrt{\gamma^{2} + \ \beta^{2}}}\, sgn(\gamma) - \frac{O_{\star}\,\tau^{2}}{\gamma^{2} + \ \beta^{2}}\quad\quad\).
(24)

The first term is \emph{signed by the direction of gain}: for a manifold
that curves toward the origin
(\({\overrightarrow{H}}^{\top}{\widehat{n}}_{g}\  < \ 0\), as on an
origin-centered sphere) an amplifying change has \(O\) under-reported
and a suppressive change has it over-reported, so the off-manifold
fraction is a conservative novelty statistic only for gain increases.
The second term is set by the jitter rather than by curvature. At fixed
\(k\) both terms fall together as \(n_{ref}^{- 1}\); they separate in
\(k\), where \(\delta\) grows while \(\tau^{2}\) is \(k\)-invariant at
\(d\  = \ 2\), and the tangential term matters most where
\(\tau^{2}/\delta\) is largest (Table S16).

Because \(O\) is a ratio of quadratic forms rather than a ratio of
expectations, equating the two requires the denominator's fluctuation to
be small relative to its mean. With \(\tau^{2} \sim \mathcal{l}^{2}/k\)
by Proposition~3, a sufficient leading-order condition is
\((\gamma\  + \ \delta)^{2}\  + \ \beta^{2}\  + \ \eta^{2}\  \gg \ \mathcal{l}^{2}/k\).
Its left side is not observable, but \({\| r\|}^{2}\) is: by Eq. (11)
and Proposition 3,
\(E{\| r\|}^{2}\  = \ (\gamma\  + \ \delta)^{2}\  + \ \beta^{2}\  + \ \eta^{2}\  + \ \mathcal{l}^{2}/k\),
so the condition is equivalent to

\(C_{6}\  = \ \frac{k{\| r\|}^{2}}{\mathcal{l}^{2}},\ \ C_{6}\  - \ 1\  \gg \ 1\)
(25)

This fails where a suppressive change is nearly cancelled by the anchor
displacement, and there the leading denominator term is itself a
fluctuation (Figure 9). Under anisotropic noise \(\tau^{2}\) changes,
but it tracks the \(\mathcal{l}^{2}/k\) measured on the same
neighborhoods, so the check remains computable provided \(\mathcal{l}\)
is measured rather than assumed (Table S12). \(C_{6}\) is computed for
each probe from its own \(\| r\|\) and the \(\mathcal{l}\) of its
neighborhood, so this is one of the two gates a reader can act on
directly.

Table~S17 gives the recovered off-manifold fraction against the imposed
value across neighborhood size, imposed magnitude and the sign of gain.
The predicted sign reversal is observed, and the error of the closed
form grows as \(C_{6}\) falls (Table S17 bins it by \(C_{6}\)).
Neighborhood size has no favorable direction here: the jitter that would
improve with larger \(k\) is invariant at \(d = 2\), while the bias
grows regardless.

\hypertarget{what-the-validation-supports}{%
\section{What the validation
supports}\label{what-the-validation-supports}}

On a manifold where the gain axis is defined and the mean curvature
vector is aligned with it, the decomposition recovers the off-manifold
share of a change closely wherever the conditioning of Gate~6 holds.
Biases in this off-manifold estimate can be understood by the geometry
of the local neighborhood: signed by the direction of gain, linear in
the anchor displacement at first order, and computable up to the second
fundamental form.

The cascade defines the scope of the decomposition, and the seven gates
are not equally available to the analyst. Gates 4 and 6 are computable
from the decomposition directly, through \(a\) and \(C_{6}\). Gate
2\textquotesingle s bound is conditional on Gate 1 and on isotropic
noise. The covariance route of Section 5.3.1 means boundary proximity
and curvature gradients remain a per-manifold check. Gates~3 and~5
depend on the mean curvature vector and its alignment with the gain
axis; neither is estimated here, so both must be argued from manifold
geometry. Gate~5's booking rule is validated over a range of \emph{c},
but the rule tells a reader how the displacement is apportioned given
\(c\), not what \(c\) is for their manifold. Gates~0 and~1 are upstream
of everything and produce artifacts shaped like real results.

\hypertarget{practical-interpretation-and-diagnostics}{%
\subsection{\texorpdfstring{\emph{Practical interpretation and
diagnostics}}{Practical interpretation and diagnostics}}\label{practical-interpretation-and-diagnostics}}

Taken together, the results support a cautious interpretation of gain,
reconfiguration, and novelty. The decomposition is most informative when
the reference manifold has a locally valid chart, the intrinsic
dimension is supported by the data, the tangent estimate is stable
across reasonable neighborhood scales, the test state lies within a
plausible projection region, and the radial axis is sufficiently aligned
with the hypothesized gain direction. Ratio-based summaries require an
additional conditioning check because small total or reference-aligned
components can make relative attribution unstable even when the
underlying vector decomposition is well behaved. In general, there is a
trade-off between enforcing large reference manifold sample size and
reducing neighborhood size. For noise along the gain axis the damage is
governed by the noise scale along \(\widehat{\rho}\) relative to the
neighborhood extent, \(\sigma_{w}\mathcal{/l}\) (Table S11, where
\(\sigma_{w}\) is varied at fixed \(\mathcal{l}\)). A larger
neighborhood should therefore reduce it, the opposite of Gates 2 and 3,
where larger neighborhoods increase anchor bias and, near boundaries or
curvature gradients, frame rotation. Where common-mode fluctuation
dominates, that trade-off should be made deliberately rather than by
defaulting to small \emph{k.}

Several tools are available for estimating intrinsic
dimensionality\textsuperscript{11-13}, though neural data provide
several reasons to treat these estimates cautiously. In simulations and
recordings, linear methods can overestimate the dimension of nonlinear
manifolds, noise can create spurious dimensions, and finite sample size
can lead to underestimation when the intrinsic dimension is
large\textsuperscript{14}. Mathematical analyses reach the same
conclusion from a different direction: tangent and dimension estimates
are reliable only when neighborhood scale, sampling density, reach, and
noise are jointly compatible\textsuperscript{15,16}. An apparently
stable estimate of dimension is therefore not sufficient evidence that
the resulting normal space is correctly specified. Over-estimation
biases the off-manifold fraction and, on a manifold whose curvature
exceeds its noise, collapses the alignment of Gate~4 outright.
Under-estimation, not simulated here, leaves off-manifold directions in
the complement but places a true tangent direction there too, so
tangential displacement and jitter are booked as \(O\) and the gain axis
acquires a tangential component. The alignment should therefore be
examined as \emph{d} varies: a sharp drop between \emph{d} and
\emph{d+1} indicates that the normal direction is being absorbed; on an
origin-centered manifold the drop reads the absorbed share directly,
since \(a^{2}\  \approx \ 1\  - \ capture\) (Table S3).

Noise geometry matters as well. Isotropic noise may mainly increase the
variance of the reported components, whereas anisotropic or
state-dependent noise can rotate the estimated tangent space, distort
the radial direction, and create systematic normal displacement. The
local-PCA literature explicitly treats non-uniform distributions and
varying noise as sources of perturbation rather than as harmless
residuals\textsuperscript{16}.The decomposition is robust to anisotropy
in a tangential direction because the anchor tracks it (Table S11); this
matters because correlated variability in population recordings lies
largely within the signal manifold. Anisotropy along the gain axis is
not absorbed, however. In real-world population recordings anisotropic
noise is often driven by shared multiplicative fluctuation. This is a
radial displacement, and on an origin-centered spherical manifold the
radial direction is the gain axis, so shared multiplicative fluctuation
is exactly noise concentrated on \({\widehat{n}}_{g}\)
(Section~\protect\hyperlink{anisotropic-noise}{5.5.1}). Whitening is not
a neutral remedy. Centering, as in z-scoring, moves the origin, which
the construction requires to be zero population activity. A purely
linear rescaling \(\Sigma^{- 1/2}\) keeps the origin and maps rays to
rays, so multiplicative gain stays radial, but it replaces the inner
product: \(P_{N}\), \(a\), \(c\) and the split between \(G\) and \(O\)
are then defined in a different metric, and the decomposition answers a
different question. In practical applications, noise should be assessed
relative to the local signal geometry, not only by its total magnitude.

In an ideal case, the mean curvature is small on the scale of the
neighborhood, or the radial direction is close to the mean curvature
direction, or the analysis is confined to a regime where the
displacement is large against
\(\mathcal{l}^{2} \parallel \overrightarrow{H} \parallel /2\). An
analysis declaring none of these reports \(O\) with a bias of known form
but unknown size: its \(\delta\) part flips sign with the gain, and
where \(|c|\  < \ 1\) its \(\eta^{2}\) part raises \(O\) regardless of
the sign of the gain.

A reported decomposition should therefore be accompanied by diagnostics
for:

\begin{itemize}
\item
  local neighborhood stability and estimated tangent-space rotation;
\item
  sensitivity to intrinsic-dimension choice;
\item
  distance to the presumed tubular region and proximity to boundaries or
  bottlenecks;
\item
  radial alignment and radial injectivity over the relevant reference
  neighborhood;
\item
  estimated anchor drift under resampling and under the hypothesized
  transformation;
\item
  anisotropy and scale of measurement noise; and
\item
  conditioning of the ratios (\(C_{6}\), Eq. 25).
\end{itemize}

Analyses that compare sets of test points against a constant manifold
hold the biases of the reference geometry fixed, such that a difference
in G, T, or O across probe sets is interpretable so long as the sets
draw anchors from a comparable distribution over the manifold and have
comparable gain sign and magnitude: by Eq. (24) the bias depends on
both, so two sets with the same anchors but opposite gain acquire
opposite biases (Table S17).

\hypertarget{discussion}{%
\section{Discussion}\label{discussion}}

By decomposing a neural state into tangential reconfiguration, radial
gain, and residual off-manifold displacement, I provide a geometric
framework for asking whether a change is consistent with amplification
of an existing representation or departure from the sampled repertoire.
The cascade of identifiability gates developed here shows that errors in
the reference manifold, intrinsic dimensionality, tangent estimation,
gain-axis alignment, and ratio conditioning can systematically transform
one type of change into another. Thus, the principal contribution is not
simply a new measure of neural novelty or gain, but a framework for
determining when such mechanistic attributions are interpretable and
when the geometry of the measurement itself makes neural state changes
inherently ambiguous.

\hypertarget{neural-distance-compared-to-what}{%
\subsection{\texorpdfstring{\emph{Neural distance compared to
what?}}{Neural distance compared to what?}}\label{neural-distance-compared-to-what}}

The use of a local anchor places the method within the broader theory of
geometric inference from sampled manifolds. For a continuous manifold,
positive reach defines a tubular neighborhood in which the nearest-point
projection is unique. Reach also relates to how rapidly tangent spaces
can vary and captures both high curvature and near self-approaches that
create projection ambiguity\textsuperscript{17}. These results provide a
natural geometric condition for treating a test state as a perturbation
of a particular local reference point.

The empirical anchor used here is not, however, the exact projection
onto the continuous manifold. It is a local statistic computed from a
finite and potentially non-uniform sample. Even when the underlying
manifold has positive reach, a local centroid can be displaced by
curvature, boundaries, density gradients, and sampling variability.
Local tangent estimation faces the same trade-off: small neighborhoods
reduce curvature bias but provide little averaging, whereas large
neighborhoods improve sampling stability while mixing regions with
different tangent spaces \textsuperscript{15,18}. Bounds for local PCA
likewise depend on reach, density, neighborhood radius, and the geometry
of the noise distribution\textsuperscript{16}. Thus, increasing the
number of reference samples does not by itself remove all anchor error.
It can reduce sampling variance while leaving systematic displacement
caused by the chosen neighborhood, with systematic distortions due to
curvature, for example.

The most important limitation is the difference between estimating a
decomposition relative to a selected anchor and recovering the
transformation that generated the test state. Given an estimated
reference manifold and a test point, the method can compute tangential,
gain-axis, and residual components relative to that geometry. What it
cannot generally recover from those observations alone is the unobserved
starting state, the amount of anchor drift, or the causal sequence of
transformations that produced the test point. A state displaced along
the gain axis could reflect true gain, a change in the reference state,
a deformation of the manifold, or a combination of these effects. The
distinction parallels a broader warning in neural-manifold research:
manifold geometry is a useful descriptive summary, but it does not by
itself identify the circuit mechanism that generated that geometry or
its dynamics\textsuperscript{19}.

Correspondence across conditions or sessions can aid in identifying
which state should serve as the reference, making it possible to
distinguish global scale, rotation, and residual deformation.
Low-dimensional alignment methods use stable neural channels or shared
structure to resolve coordinate ambiguity between
sessions\textsuperscript{20,21}. Behavioral variables can serve a
similar role by anchoring latent states to common task conditions and
improving consistency across recordings\textsuperscript{22}. Alignment
of latent dynamics extends this idea across recording sessions by using
the rules governing state evolution as a reference, rather than matching
static point clouds alone\textsuperscript{23-25}.

Even when correspondence may be motivated through estimation of the
initial point of displacement, many of the biases presented here are
still relevant. A starting location resolves the anchor-location
ambiguity of Gate 3 and removes the associated anchor-drift contribution
to Gate 5, but it does not by itself resolve Gate 0 (local chart
validity), Gate 1 (intrinsic dimension), Gate 2 (tangent estimation),
Gate 4 (radial-axis alignment), or Gate 6 (ratio conditioning).

\hypertarget{the-radial-axis-is-a-coding-assumption}{%
\subsection{\texorpdfstring{\emph{The radial axis is a coding
assumption}}{The radial axis is a coding assumption}}\label{the-radial-axis-is-a-coding-assumption}}

The further partition of the normal space into a radial gain direction
and a residual novelty direction introduces a modeling assumption: a
radial axis is appropriate when multiplicative modulation acts on the
population response vector relative to a meaningful origin. Shared
cortical variability has been modeled using multiplicative gain together
with additive offsets, and the two components have different effects on
population covariance and stimulus discriminability\textsuperscript{26}.
Related work has shown that multiplicative modulation can account for
systematic response variability and can be distinguished from private
spike-generation noise\textsuperscript{27}, while different neurons can
express different mixtures of multiplicative and additive
modulation\textsuperscript{26,28,29}. The radial construction should
therefore be presented as a coding hypothesis: it tests whether a
privileged direction in the observed population coordinates behaves like
gain.

Radial injectivity is consequently an identifiability condition for this
particular parameterization. If distinct reference states lie on the
same ray from the origin, radial position cannot uniquely identify the
underlying state. In that setting, a radial displacement may be
compatible with multiple combinations of state change and gain. The
condition is best viewed as a prerequisite for interpretation rather
than as a generic property of neural manifolds. Chung, Lee, and
Sompolinsky (2018) showed that neural manifolds possess meaningful
radial geometry, including characteristic manifold radii and directions
associated with their extent in activity space\textsuperscript{30}.
Their framework therefore provides an important geometric foundation for
treating radial structure as a property of neural representations. Here,
however, I identify a different consequence of radial geometry: radial
scaling can alter the empirical reference anchor itself. When the radial
direction has a tangential component relative to the manifold,
multiplying a state by a gain factor moves the state through the ambient
space and creates ambiguity as to which is the state that is being
gain-modulated. In this sense, the present analysis extends the
geometric perspective of Chung et al. from describing the structure of
neural manifolds to asking whether that structure permits a uniquely and
stably identifiable decomposition of transformed states.

This decomposition also offers a normative argument that situations that
require robust state disambiguation in the face of intense gain
fluctuations should be mapped onto a spherical manifold in ambient state
space (more precisely, the positive orthant, since rates cannot be
negative), while situations in which gain defines the manifold
coordinate should tend towards a cone. Spheres in ambient state space
are compatible with familiar low-D neuronal manifolds. For example, the
coding manifold for a grid cell module is modeled as a two-dimensional
torus\textsuperscript{31}. For the gain axis to be both radial from a
true zero-activity origin and normal to the toroidal surface constrains
its embedding such that the torus lies at constant radius from the
origin (exactly so if \(a\  = \ 1\) everywhere), as in a Clifford torus
embedded in a higher-dimensional sphere. Radial gain then moves the
representation between concentric tori (Figure 10). Although the
toroidal coordinates are preserved, this scaling changes the metric of
the embedded torus\textsuperscript{:} distances between states, and so
the discriminability of nearby positions for a readout with fixed noise,
scale with gain. Grid fields expand in novel
environments\textsuperscript{32}, though the relationship between gain
and grid-field spacing, and more generally how representational geometry
is constrained by the need to disambiguate tangential from gain-related
displacements, remain empirical questions.

\hypertarget{novel-from-whose-point-of-view}{%
\subsection{\texorpdfstring{\emph{Novel from whose point of
view?}}{Novel from whose point of view?}}\label{novel-from-whose-point-of-view}}

The decomposition is defined with reference to a sampled manifold and
the Euclidean geometry of the recording space, but neither is
necessarily what a downstream neural observer sees\textsuperscript{33}.
A linear readout of ensemble activity will impose its own projection,
with a null space in which some displacements are invisible regardless
of their magnitude in neuronal space. This change in reference frame
alters the interpretation of the decomposition in two important ways.
First, for multiplicative gain, the relevant zero is not necessarily a
single point at the origin, but the set of activity patterns that
produce no output in the readout. Second, the reader-centric view
separates two forms of activity that the present geometric construction
treats alike as off-manifold: activity that falls within the
readout\textquotesingle s null space has no downstream consequence,
whereas activity that is visible to the readout but lies outside the
manifold represents a state the population does not normally
produce\textsuperscript{34}. The former is therefore not novelty; the
latter may be. Both appear as off-manifold activity in the present
framework, placing a fundamental limit on what \emph{O} can mean. More
generally, this highlights that novelty is not an intrinsic property of
a neural representation alone, but a property of the relationship
between a representation and the system that reads it.

\hypertarget{acknowledgments}{%
\section{Acknowledgments}\label{acknowledgments}}

This work was funded by NIMH grant R01MH140076. Proofs, software, and
figures were developed with AI assistance from Claude (Opus 5.0/Sonnet
5) and ChatGPT (GPT-5.6 Luna).

Code availability. The MATLAB code that generates every supplement table
is available at
https://github.com/McKenzieNeuro/McKenzieLab/tree/main/GTO.

\hypertarget{references}{%
\section*{References}\label{references}}
\addcontentsline{toc}{section}{References}

1 Baldassano, C. \emph{et al.} Discovering Event Structure in Continuous
Narrative Perception and Memory. \emph{Neuron} \textbf{95}, 709-721
e705, doi:10.1016/j.neuron.2017.06.041 (2017).

2 Ben-Yakov, A. \& Henson, R. N. The Hippocampal Film Editor:
Sensitivity and Specificity to Event Boundaries in Continuous
Experience. \emph{J Neurosci} \textbf{38}, 10057-10068,
doi:10.1523/JNEUROSCI.0524-18.2018 (2018).

3 Wang, Y. C., Adcock, R. A. \& Egner, T. Toward an integrative account
of internal and external determinants of event segmentation.
\emph{Psychon Bull Rev} \textbf{31}, 484-506,
doi:10.3758/s13423-023-02375-2 (2024).

4 Zacks, J. M., Speer, N. K., Swallow, K. M., Braver, T. S. \& Reynolds,
J. R. Event perception: a mind-brain perspective. \emph{Psychol Bull}
\textbf{133}, 273-293, doi:10.1037/0033-2909.133.2.273 (2007).

5 Clewett, D., Huang, R. \& Davachi, L. Locus coeruleus activation
"resets" hippocampal event representations and separates adjacent
memories. \emph{Neuron} \textbf{113}, 2521-2535 e2528,
doi:10.1016/j.neuron.2025.05.013 (2025).

6 McKenzie, S. \emph{et al.} Event boundaries drive norepinephrine
release and distinctive neural representations of space in the rodent
hippocampus. \emph{bioRxiv}, doi:10.1101/2024.07.30.605900 (2024).

7 Buzsaki, G., McKenzie, S. \& Davachi, L. Neurophysiology of
Remembering. \emph{Annu Rev Psychol} \textbf{73}, 187-215,
doi:10.1146/annurev-psych-021721-110002 (2022).

8 Areshenkoff, C. \emph{et al.} Neural excursions from manifold
structure explain patterns of learning during human sensorimotor
adaptation. \emph{Elife} \textbf{11}, doi:10.7554/eLife.74591 (2022).

9 Golub, M. D. \emph{et al.} Learning by neural reassociation. \emph{Nat
Neurosci} \textbf{21}, 607-616, doi:10.1038/s41593-018-0095-3 (2018).

10 Sadtler, P. T. \emph{et al.} Neural constraints on learning.
\emph{Nature} \textbf{512}, 423-426, doi:10.1038/nature13665 (2014).

11 Carter, K. M., Raich, R. \& Hero, A. O. I. On local intrinsic
dimension estimation and its applications. \emph{Transactions on Signal
Processing} \textbf{58}, 650-663 (2009).

12 Mo, D. \& Huang, S. H. Fractal-based intrinsic dimension estimation
and its application in dimensionality reduction. \emph{IEEE Transactions
on Knowledge and Data Engineering} \textbf{24}, 59-71 (2010).

13 Ong, E.-J., Bobrowski, O., Reinert, G. \& Skraba, P. A Universal
Nearest-Neighbor Estimator for Intrinsic Dimensionality. \emph{arXiv},
2603.10493 (2026).

14 Altan, E., Solla, S. A., Miller, L. E. \& Perreault, E. J. Estimating
the dimensionality of the manifold underlying multi-electrode neural
recordings. \emph{PLoS Comput Biol} \textbf{17}, e1008591,
doi:10.1371/journal.pcbi.1008591 (2021).

15 Kaslovsky, D. N. \& Meyer, F. G. Non-Asymptotic Analysis of Tangent
Space Perturbation. \emph{Information and Inference: A Journal of the
IMA} \textbf{3}, 134-187 (2014).

16 Lim, U., Oberhauser, H. \& Nanda, V. Tangent Space and Dimension
Estimation with the Wasserstein Distance. \emph{SIAM Journal on Applied
Algebra and Geometry} \textbf{8}, doi:10.1137/22M1522711 (2024).

17 Boissonnat, J. D., Lieutier, A. \& Wintraecken, M. The reach, metric
distortion, geodesic convexity and the variation of tangent spaces.
\emph{J Appl Comput Topol} \textbf{3}, 29-58,
doi:10.1007/s41468-019-00029-8 (2019).

18 Aamari, E. \& Levrard, C. Nonasymptotic rates for manifold, tangent
space and curvature estimation. \emph{Ann. Statist.} \textbf{47},
177-204 (2019).

19 Langdon, C., Genkin, M. \& Engel, T. A. A unifying perspective on
neural manifolds and circuits for cognition. \emph{Nat Rev Neurosci}
\textbf{24}, 363-377, doi:10.1038/s41583-023-00693-x (2023).

20 Degenhart, A. D. \emph{et al.} Stabilization of a brain-computer
interface via the alignment of low-dimensional spaces of neural
activity. \emph{Nat Biomed Eng} \textbf{4}, 672-685,
doi:10.1038/s41551-020-0542-9 (2020).

21 Haxby, J. V., Guntupalli, J. S., Nastase, S. A. \& Feilong, M.
Hyperalignment: Modeling shared information encoded in idiosyncratic
cortical topographies. \emph{Elife} \textbf{9}, doi:10.7554/eLife.56601
(2020).

22 Schneider, S., Lee, J. H. \& Mathis, M. W. Learnable latent
embeddings for joint behavioural and neural analysis. \emph{Nature}
\textbf{617}, 360-368, doi:10.1038/s41586-023-06031-6 (2023).

23 Karpowicz, B. M. \emph{et al.} Stabilizing brain-computer interfaces
through alignment of latent dynamics. \emph{Nat Commun} \textbf{16},
4662, doi:10.1038/s41467-025-59652-y (2025).

24 Zhou, D. \& Wei, X. X. Learning identifiable and interpretable latent
models of high-dimensional neural activity using pi-VAE. \emph{Advances
in Neural Information Processing Systems} \textbf{33}, 7234-7247 (2020).

25 Gosztolai, A., Peach, R. L., Arnaudon, A., Barahona, M. \&
Vandergheynst, P. MARBLE: interpretable representations of neural
population dynamics using geometric deep learning. \emph{Nat Methods}
\textbf{22}, 612-620, doi:10.1038/s41592-024-02582-2 (2025).

26 Lin, I. C., Okun, M., Carandini, M. \& Harris, K. D. The Nature of
Shared Cortical Variability. \emph{Neuron} \textbf{87}, 644-656,
doi:10.1016/j.neuron.2015.06.035 (2015).

27 Goris, R. L., Movshon, J. A. \& Simoncelli, E. P. Partitioning
neuronal variability. \emph{Nat Neurosci} \textbf{17}, 858-865,
doi:10.1038/nn.3711 (2014).

28 Arandia-Romero, I., Tanabe, S., Drugowitsch, J., Kohn, A. \&
Moreno-Bote, R. Multiplicative and Additive Modulation of Neuronal
Tuning with Population Activity Affects Encoded Information.
\emph{Neuron} \textbf{89}, 1305-1316, doi:10.1016/j.neuron.2016.01.044
(2016).

29 Zhu, R. J. B. \& Wei, X. X. Unsupervised approach to decomposing
neural tuning variability. \emph{Nat Commun} \textbf{14}, 2298,
doi:10.1038/s41467-023-37982-z (2023).

30 Chung, S., Lee, D. D. \& Sompolinsky, H. Classification and geometry
of general perceptual manifolds. \emph{Physical Review X} \textbf{8},
031003 (2018).

31 Gardner, R. J. \emph{et al.} Toroidal topology of population activity
in grid cells. \emph{Nature} \textbf{602}, 123-128,
doi:10.1038/s41586-021-04268-7 (2022).

32 Barry, C., Ginzberg, L. L., O\textquotesingle Keefe, J. \& Burgess,
N. Grid cell firing patterns signal environmental novelty by expansion.
\emph{Proc Natl Acad Sci U S A} \textbf{109}, 17687-17692,
doi:10.1073/pnas.1209918109 (2012).

33 Buzsaki, G. Neural syntax: cell assemblies, synapsembles, and
readers. \emph{Neuron} \textbf{68}, 362-385,
doi:10.1016/j.neuron.2010.09.023 (2010).

34 Vastola, J. J., Cohen, Z. \& Drugowitsch, J. Is the information
geometry of probabilistic population codes learnable? \emph{NeurIPS
Workshop on Symmetry and Geometry in Neural Representations}, 258-277
(2023).

\hypertarget{figures}{%
\section*{Figures}\label{figures}}
\addcontentsline{toc}{section}{Figures}

\includegraphics[width=6.37668in,height=2in]{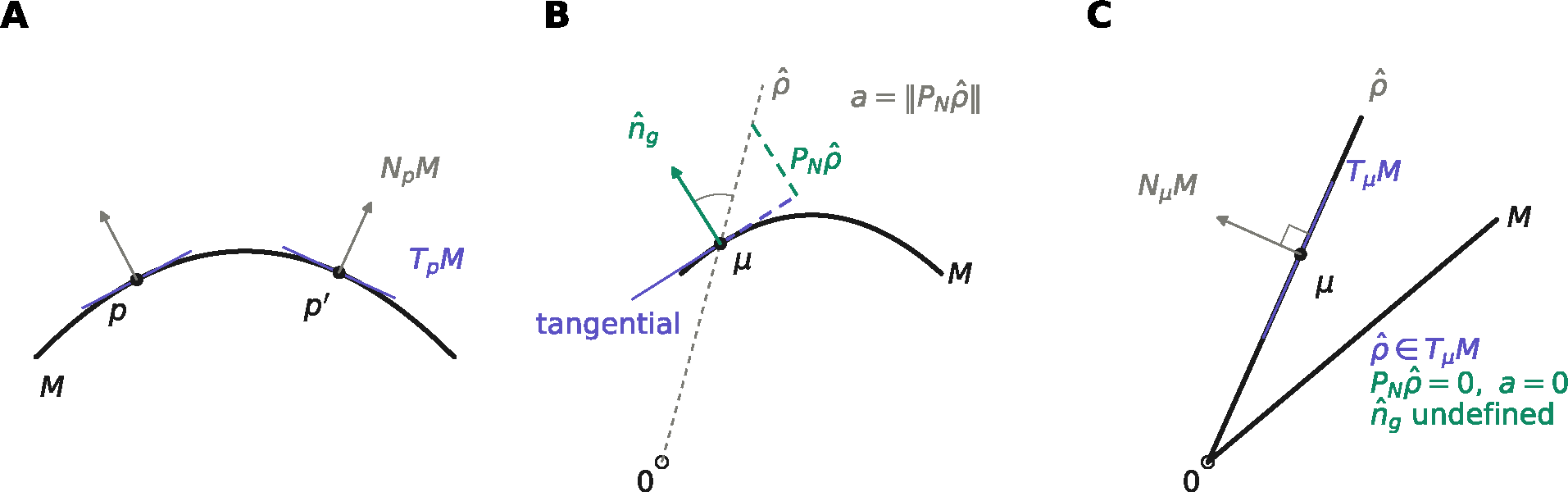}

\textbf{Figure 1. The gain axis and where it fails to exist.}

\textbf{(A)} Tangent and normal bundles at two points of a curved
reference manifold. The partition into tangent and normal directions is
defined pointwise and changes with position, which is what makes it a
bundle rather than a fixed frame.

\textbf{(B)} Construction of the gain axis. The radial direction \ensuremath{\hat{\rho}} from
the zero-activity origin is resolved at \ensuremath{\mu} into a tangential part, which
is discarded, and its projection into the normal space, whose
normalization is the gain axis \({\widehat{n}}_{g}\). The alignment
\(a = \ \left\| P_{N}\rho\hat{} \right\|\) is the cosine of the marked
angle. Defining the axis with \ensuremath{\hat{\rho}} alone would conflate radial change with
tangential movement.

\textbf{(C)} The degenerate case. On a cone through the origin, every
displacement moves equivalently along the radial direction and the
tangent direction and the gain axis is undefined.

\includegraphics[width=6.5in,height=2.13819in]{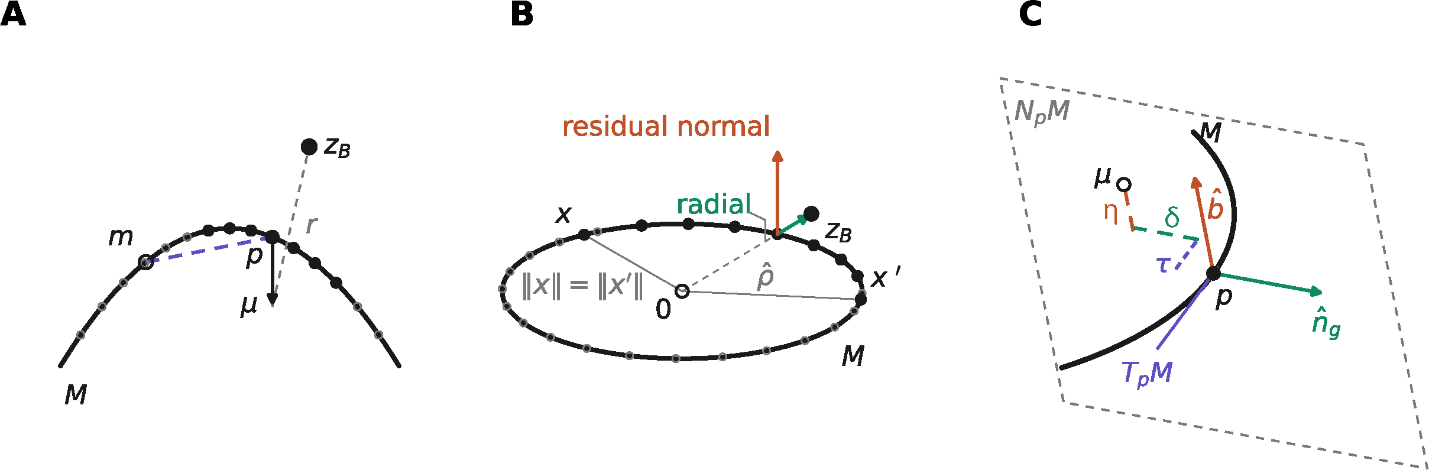}

\textbf{Figure 2. Anchor selection, the anchor frame, and when the
anchor is exact.}

\textbf{(A)} The test point selects its own anchor. \emph{\ensuremath{\mu}} is the mean
of the \emph{k} reference points nearest \(z_{B}\ \), so \emph{\ensuremath{\mu} -- m}
decomposes into a tangential leg \emph{p -- m}, which tracks movement
along the manifold, and a normal leg \emph{\ensuremath{\mu} -- p}, which is the sagitta.

\textbf{(B)} The two cases in which the tangential leg vanishes exactly.
On a circle centered on the origin, a displacement out of the plane of
the reference set is orthogonal to every x, while a radial displacement
also has no tangent component.

\textbf{(C)} The anchor frame. The anchor displacement resolves into \ensuremath{\delta}
along the gain axis, \ensuremath{\eta} in the residual normal space, and \emph{\ensuremath{\tau}} in the
tangent space. Drawn on a curve in three dimensions, where the normal
space is a genuine plane and \ensuremath{\beta} is a real direction.

\emph{\hfill\break
}

\includegraphics[width=4.28001in,height=1.95334in]{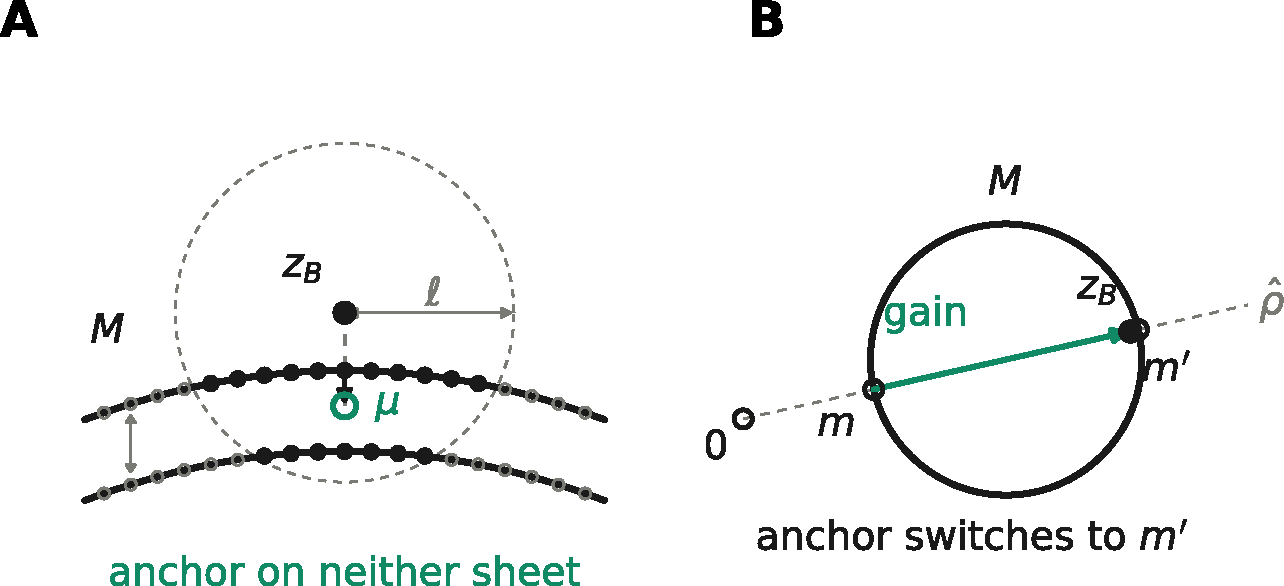}

\textbf{Figure 3. Gate 0: overlapping sheets and radial re-entry.}

\textbf{(A)} The anchor does not uniquely belong to one neighborhood
when two sheets are within the radius \ensuremath{\ell}. The anchor is the mean of
points drawn from both and lands in the gap, on neither sheet. Damage is
greatest when the separation is comparable to \ensuremath{\ell}.

\textbf{(B)} Radial re-entry. The ray from the origin meets M at m and
again at m\ensuremath{\prime}. A large enough gain carries \(z_{B}\) close enough to m\ensuremath{\prime}
that the nearest anchor switches, and the radial map is no longer
injective.

\includegraphics[width=5.17334in,height=2.28in]{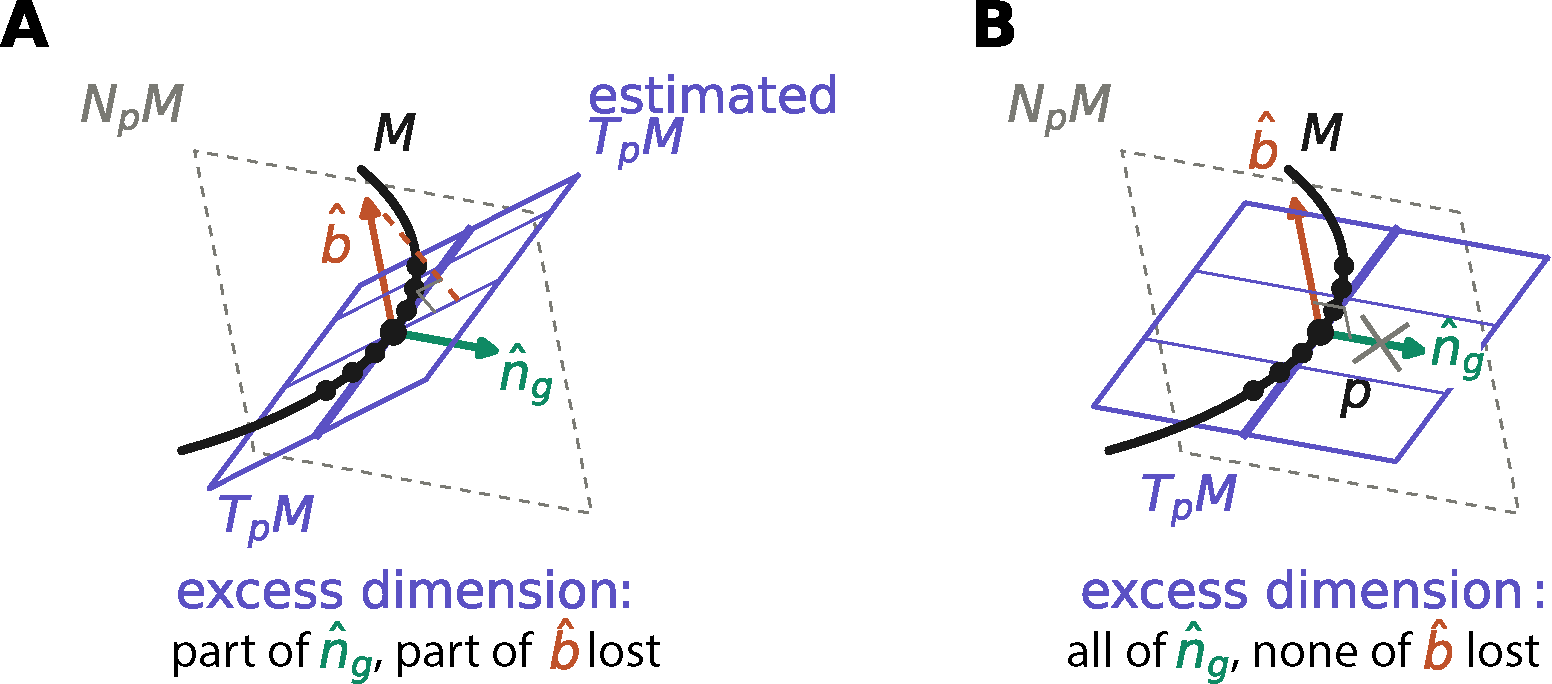}

\textbf{Figure 4. Gate 1: what an over-estimated dimension takes.}

\textbf{(A)} Over-estimating \emph{d} makes the estimated tangent space
a plane where the true tangent space is a line, taking one direction out
of the normal space. A plane that does not perfectly align with
\({\widehat{n}}_{g}\ \)will absorb part of the gain and off-manifold
space.

\textbf{(B)} The same scene with the excess dimension rotated onto the
gain axis. This scenario occurs when the variance around the sagitta
exceeds the noise floor. When the tangent is perfectly aligned with
\({\widehat{n}}_{g}\), \ensuremath{\hat{b}} is perpendicular to the excess direction and
nothing is lost, while the gain axis is absorbed into the tangent
estimate and destroyed.

\emph{\hfill\break
}

\includegraphics[width=6.5in,height=2.3625in]{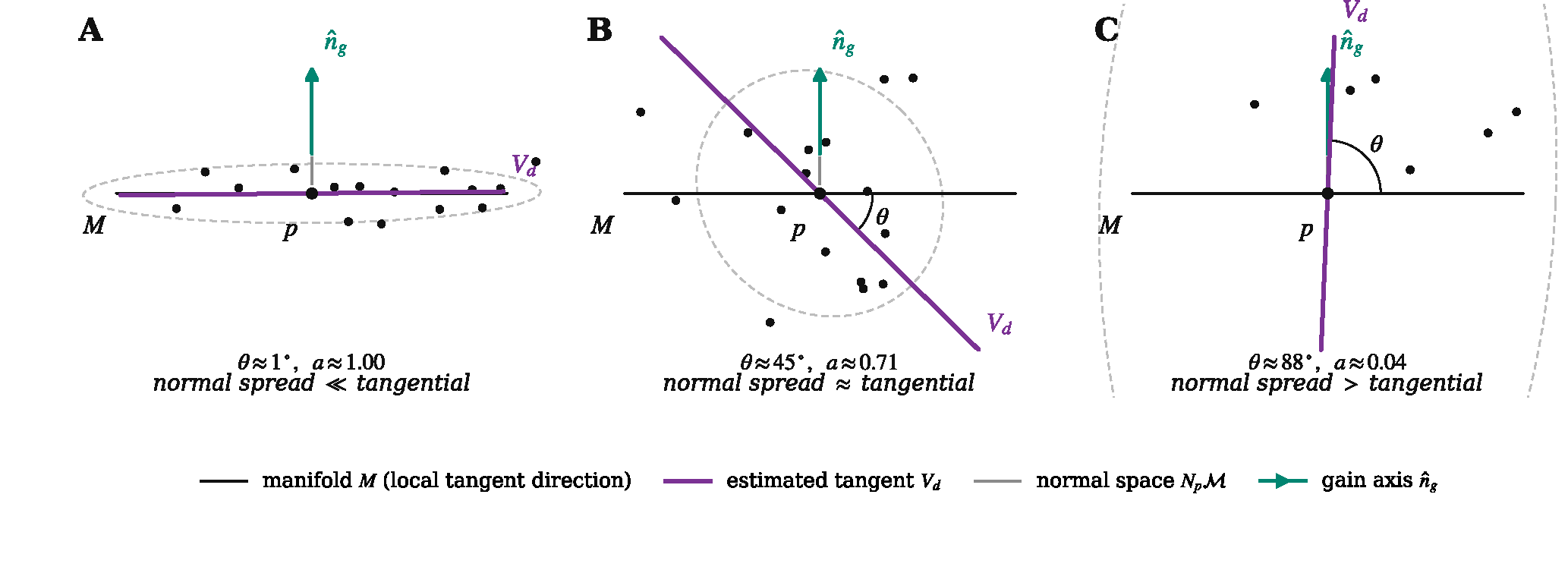}

\textbf{Figure 5. Gate 2, variance route: when normal-direction spread
competes with tangential spread, the fitted plane rotates away from the
true tangent and absorbs the gain axis.}

(A) Normal spread \(\ll\) tangential spread: \(V_{d}\) recovers the true
tangent almost exactly

(B) Normal spread \(\approx\) tangential spread: the local covariance is
close to isotropic, so the leading eigenvector is poorly determined;
\(V_{d}\) rotates by a large, effectively random angle, and roughly a
third of the gain axis's energy has migrated into the estimated tangent
space

(C) Normal spread \(>\) tangential spread: the leading eigenvector of
the neighborhood is the normal direction. \(V_{d}\) has rotated onto
\({\widehat{n}}_{g}\) and the gain axis is destroyed.

\emph{\hfill\break
}

\includegraphics[width=6.5in,height=1.79514in]{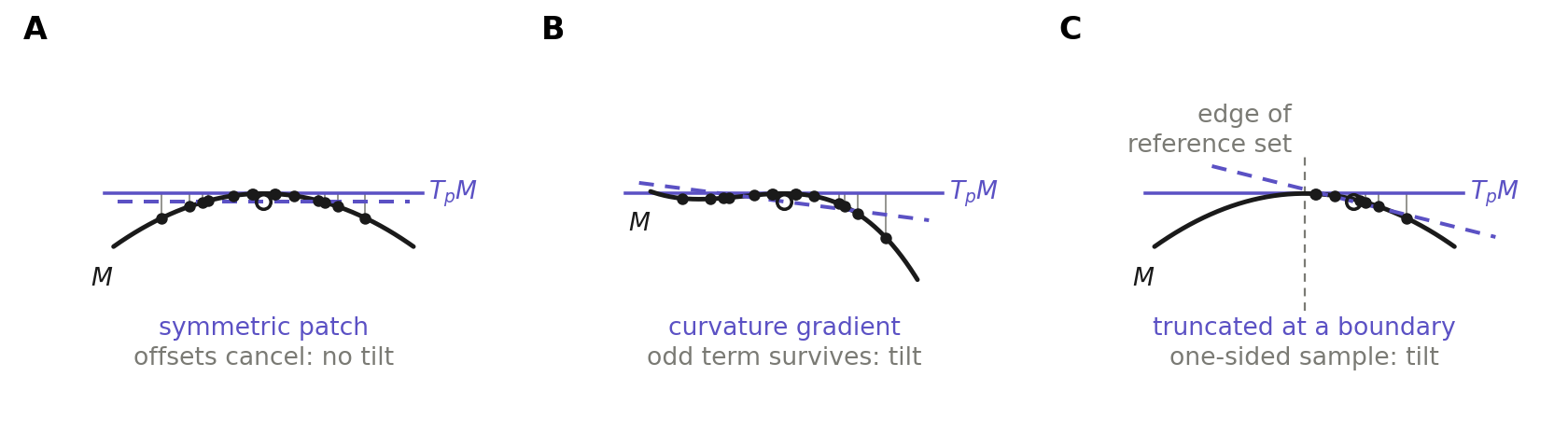}

\textbf{Figure 6. Gate 2, covariance route: what breaks the
cancellation.}

\textbf{(A)} A symmetric patch. The quadratic sagitta is even in the
tangential coordinate, so the normal offsets cancel in pairs and the
fitted direction is parallel to the true tangent. Under symmetric
sampling, curvature alone produces no tilt.

\textbf{(B)} A curvature gradient adds an odd cubic term, the
cancellation fails, and the fitted direction tilts.

\textbf{(C)} The same symmetric curve as A, with the reference set
truncated at its edge. The manifold is unchanged; only the sample is
one-sided, and the fitted direction tilts again.

\emph{\hfill\break
}

\includegraphics[width=6.23335in,height=2.20667in]{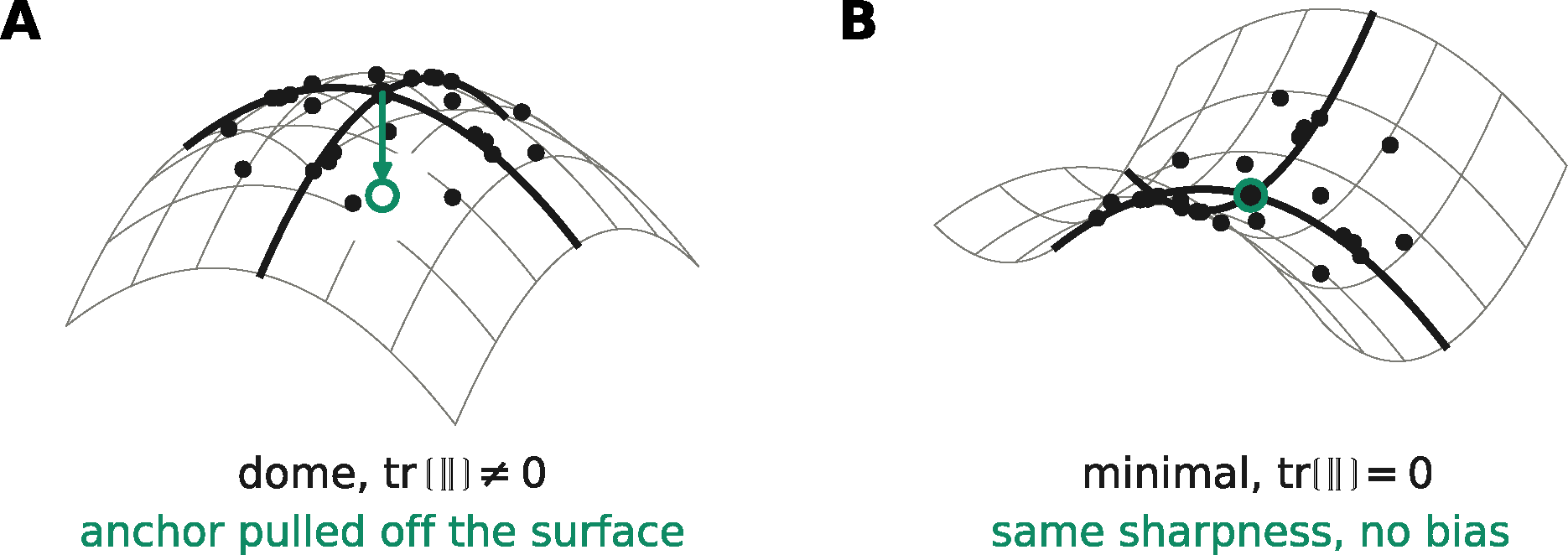}

\textbf{Figure 7. Gate 3: the trace, not the sharpness, biases the
anchor.}

\textbf{(A)} A dome. Both principal curvatures share a sign, the trace
of the second fundamental form \(\mathbb{I}\) is non-zero, and the
neighborhood mean is pulled off the surface along the normal.

\textbf{(B)} A minimal surface with the same principal curvature
magnitude and opposite signs. To leading order, the trace vanishes and
the neighborhood mean remains on the surface. Sharp bending does not
bias the anchor; the trace does.

\emph{\hfill\break
}

\includegraphics[width=6.36001in,height=2.36334in]{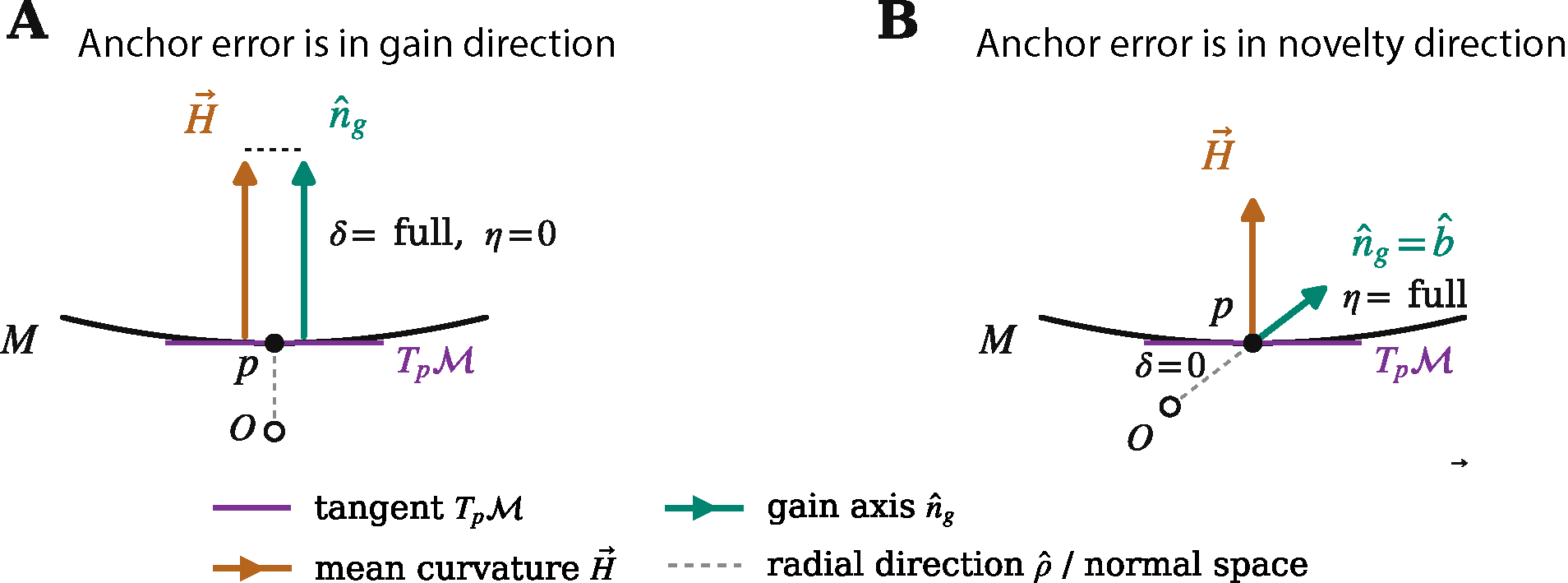}

\textbf{Figure 8. Gate 5: whether anchor-displacement bias is booked as
gain or as novelty depends on where the origin sits relative to the
curve's osculating plane.}

(A) The origin \(O\) lies on the normal line through \(p\), so the
radial direction \(\widehat{\rho} = (p - O)/\| p - O\|\) has no
tangential component and \({\widehat{n}}_{g}\  = \ \widehat{\rho}\) is
parallel to \(\overrightarrow{H}\): \(|c| = 1\), and the anchor's
displacement is absorbed entirely into \(\delta\), read as gain.

(B) The origin leaves the plane the curve locally bends in (the
osculating plane spanned by the tangent and \(\overrightarrow{H}\)),
displaced instead along the binormal \(\widehat{b}\); then
\({\widehat{n}}_{g}\  = \ \widehat{b}\) is orthogonal to
\(\overrightarrow{H}\): \(c = 0\), and the identical displacement is
absorbed entirely into \(\eta\), read as novelty.

\includegraphics[width=6.01668in,height=3.24334in]{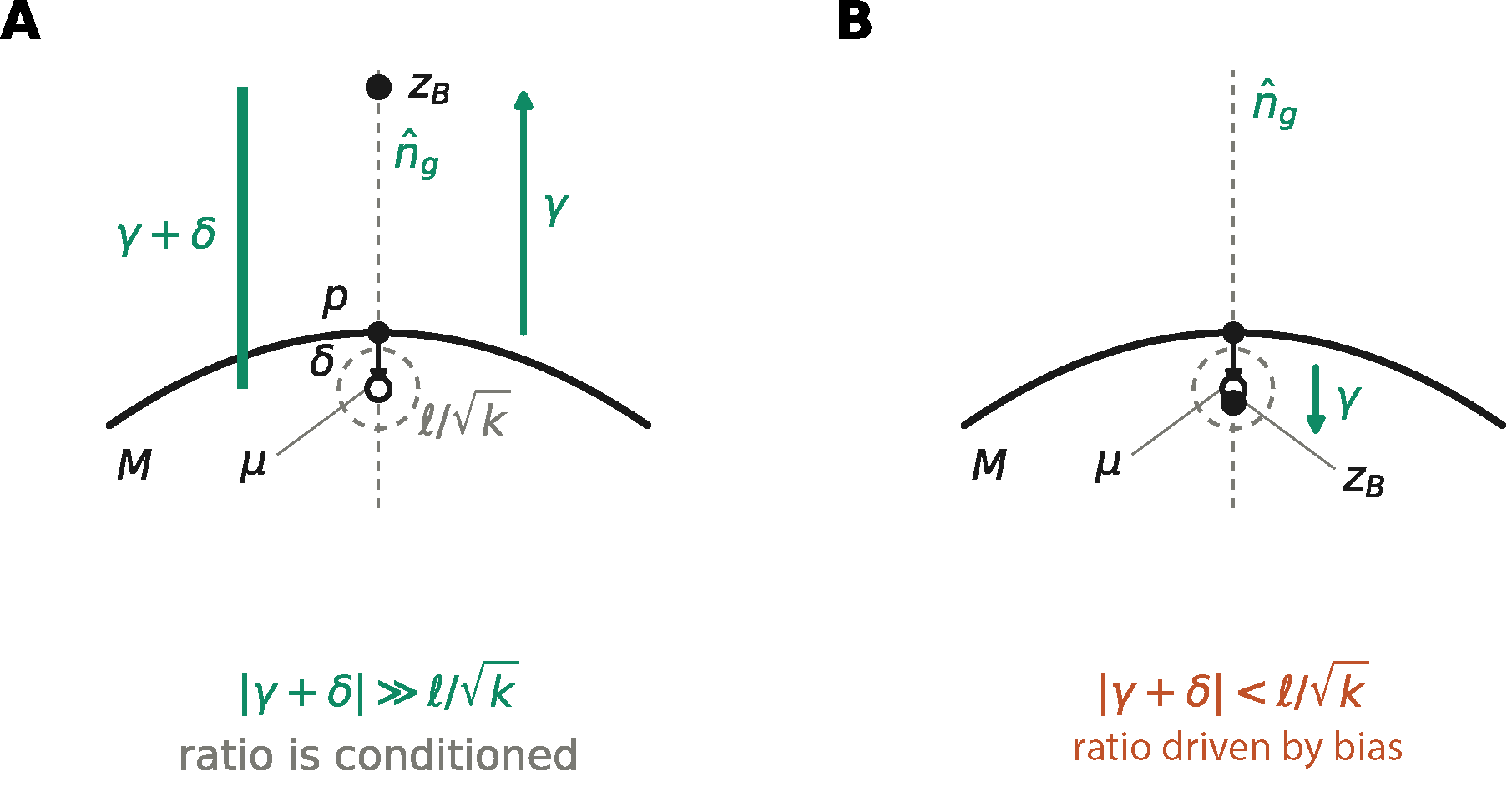}

\textbf{Figure 9. Gate 6: the recovered off-manifold fraction is only as
reliable as the ratio}
\(\mathbf{(}\mathbf{\gamma}\mathbf{+}\mathbf{\delta}\mathbf{)}^{\mathbf{2}}\mathbf{+}\mathbf{\beta}^{\mathbf{2}}\mathbf{+}\mathbf{\eta}^{\mathbf{2}}\mathbf{\  \gg \ }\mathcal{l}^{\mathbf{2}}\mathbf{/}\mathbf{k}\)
\textbf{is large, an observable condition captured by}
\(\mathbf{C}_{\mathbf{6}}\mathbf{\  = \ }\mathbf{k}\mathbf{\|}\mathbf{r}\mathbf{\|}^{\mathbf{2}}\mathbf{/}\mathcal{l}^{\mathbf{2}}\)\textbf{.}

(A) An amplifying change: \(\gamma\) and \(\delta\) add, so the resolved
leg is long compared with the jitter disc and \(C_{6}\  \gg \ 1\) ---
the ratio is conditioned.

(B) A suppressive change on the same manifold, with the same anchor and
the same \(\delta\): \(\gamma\) and \(\delta\) nearly cancel, the
resolved leg falls inside the jitter disc, and \(C_{6}\  \approx \ 1\)
--- the leading term of the ratio is a fluctuation rather than a
measurement.

\includegraphics[width=6.5in,height=5.94028in]{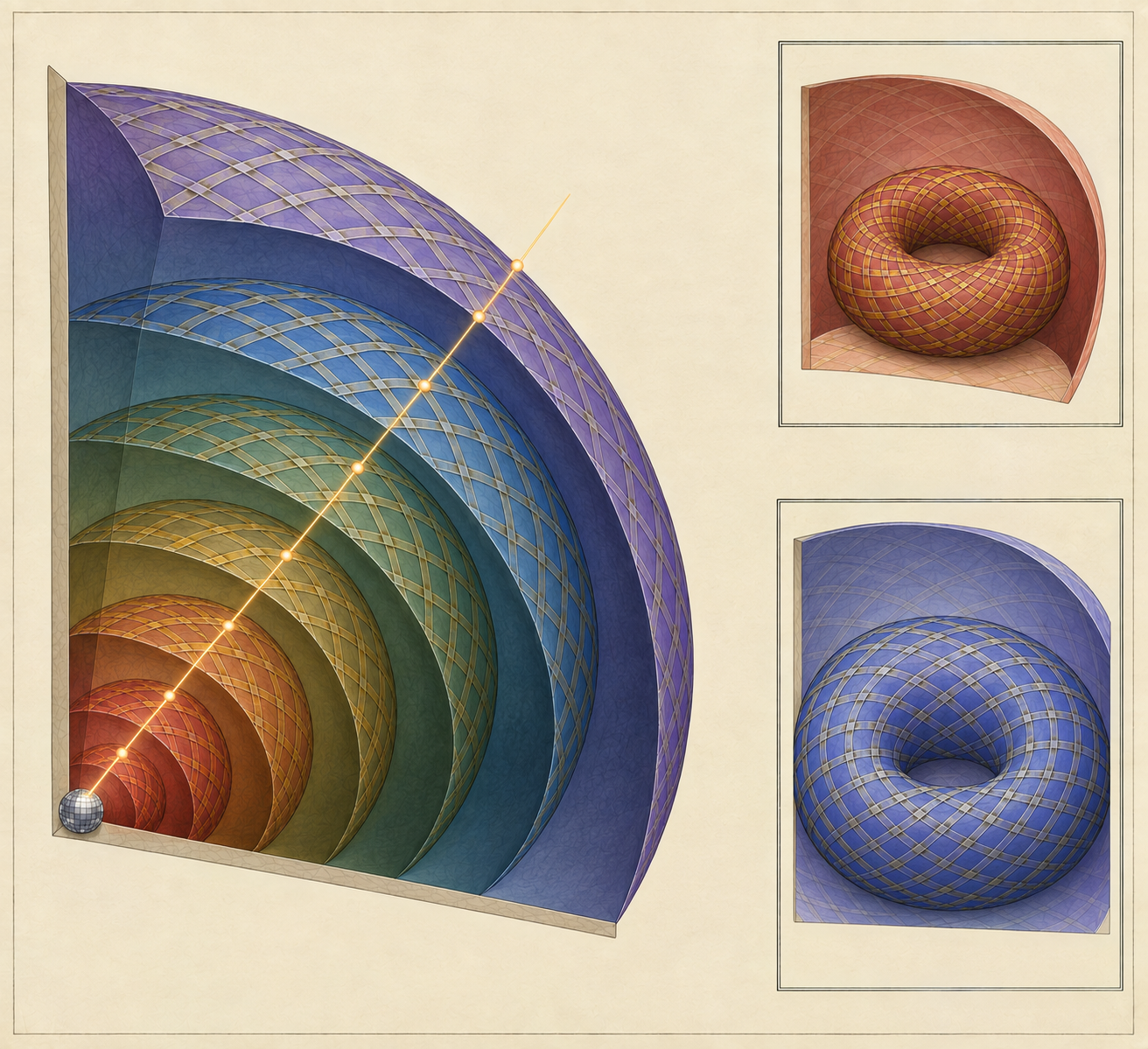}

\textbf{Figure 10. Embedding of a 2-D toroidal manifold on a hypersphere
at multiple gain intensities.} Radial gain moves the coded state between
concentric tori embedded at constant radius from the origin in ambient
state space (main panel; ray marks a trajectory of increasing gain), as
in a Clifford torus embedded in a higher-dimensional sphere. The insets
isolate the smallest (red) and largest (blue) tori from this family,
shown in the torus\textquotesingle s conventional \ensuremath{\mathbb{R}}\ensuremath{^3} embedding for
visualization: the intrinsic angular coordinates
\(\theta_{1},\theta_{2}\) are unchanged, while the embedding radii
\(R_{1},\ R_{2}\) scale with gain, so a fixed angular separation
corresponds to a larger physical distance at higher gain.

\clearpage
\renewcommand{\thetable}{S\arabic{table}}
\renewcommand{\thefigure}{S\arabic{figure}}
\setcounter{table}{0}
\setcounter{figure}{0}
\setcounter{equation}{0}
\setcounter{section}{0}
\renewcommand{\thesection}{S\arabic{section}}
\renewcommand{\thesubsection}{S\arabic{section}.\arabic{subsection}}
\renewcommand{\theequation}{S\arabic{equation}}
\begingroup
\title{Supplement: proofs and simulation tables\\
\large Identifying Neural State Changes due to Gain versus Off-Manifold Displacement}
\author{}
\date{}
\maketitle

\noindent Section~\ref{sec:S-methods} gives the methods common to every table below: reference
sampling, probe construction, the neighborhood extent $\ell$, noise, degeneracy, and aggregation.
Section~\ref{sec:S-proofs} collects the proofs. Everything from Section~\ref{sec:S-gate0} onward is
simulation: what the sweeps measured, on the manifolds they were measured on. Only
Table~\ref{tab:S-exact} checks an exact identity; every other prediction is a local asymptotic
expansion or a model-specific scaling, with scope as stated in the main text.

\paragraph{Typographic convention.} Where a table tests a closed-form prediction, the rightmost
column or columns are headed \textbf{ratio} and give measured divided by predicted, using the
columns named in the caption; a value of $1$ means the prediction is recovered, and departures are
the quantity of interest. Tables with no prediction, or whose prediction can be zero, carry no
ratio column.

\clearpage
\section{Methods}
\label{sec:S-methods}

This section collects the details common to every table: how the reference set and probes are
drawn, how the neighborhood extent is estimated, where noise enters, when a point is flagged
degenerate, and how many trials and seeds each table uses. Sections~\ref{sec:S-proofs} onward
give each table's own parameters in its caption; this section is the reference for what those
parameters mean and is cited from there rather than repeated.

\subsection{Reference sampling}
\label{sec:S-methods-sampling}

Reference sets are drawn per manifold, aiming for uniform area density in every case; the method
differs because the manifolds differ.

\begin{itemize}
\item \textbf{Full sphere.} $v \sim \mathcal{N}(0, I_3)$, normalized to $\|v\|=1$, then scaled by
$R$. Used in Tables~\ref{tab:S-gate0}, \ref{tab:S-gate1} (noise-dominated and curvature-dominated
blocks), \ref{tab:S-anchorbias}, \ref{tab:S-offman}, and the $k$ and $R$ blocks of
Table~\ref{tab:S-rotation}.
\item \textbf{Spherical cap, exact.} Uniform in $\cos\theta$
($\cos\theta = \cos(\mathrm{cap}) + (1-\cos(\mathrm{cap}))\,u$, $u\sim\mathrm{Unif}(0,1)$), which is
uniform in area on the cap exactly, at any cap angle. Used in the cap block of
Table~\ref{tab:S-params} and throughout Table~\ref{tab:S-anisoanchor}.
\item \textbf{Spherical cap, small-angle.} Uniform in the polar angle itself via
$\theta = \mathrm{cap}\cdot\sqrt{u}$ (the construction that gives uniform area on a \emph{flat}
disc), rather than in $\cos\theta$. This is only approximately uniform in area on a sphere, with
relative error $O(\mathrm{cap}^2)$ from $\sin\theta \approx \theta$; at the cap angles used here
(0.25--0.35 rad) the error is a percent or two. Used in Tables~\ref{tab:S-aniso},
\ref{tab:S-anisoframe} and \ref{tab:S-covbound}, where the cap is small enough for the
approximation to be immaterial next to the effects under study.
\item \textbf{Flat disc.} Uniform via $r = a\sqrt{u}$, $\phi \sim \mathrm{Unif}(0,2\pi)$, exact for a
flat patch. Used for the quadric patches of Table~\ref{tab:S-curvlaw} and the curvature-gradient
patch of Table~\ref{tab:S-covgrad}.
\item \textbf{Torus of revolution.} Rejection sampling in $\theta$ against density
$\propto R_c + r\cos\theta$, uniform in $\phi$. Used in Table~\ref{tab:S-torus}.
\item \textbf{Clifford torus.} Uniform in $(u,v)$ directly: the flat product metric makes this
exact. Used in Table~\ref{tab:S-clifford}.
\item \textbf{Circle $\times$ ellipse.} Uniform in the circle coordinate; rejection sampling in the
ellipse coordinate against density $\propto\sqrt{A^2\sin^2v + B^2\cos^2v}$. Used in
Table~\ref{tab:S-ellipse}.
\end{itemize}

\subsection{Probe construction}
\label{sec:S-methods-probes}

Every probe construction falls into one of two families, chosen so that the quantity a table
reports is not confounded with the anchor-tracking behavior of Section~\ref{sec:S-proofs}.

\textbf{Radial and ambient legs on an origin-centered sphere.} A probe is
\begin{equation}
z_B = m + \nu\bigl(\mathrm{sgn}\sqrt{1-O_\star}\;\hat\rho + \sqrt{O_\star}\;\hat o\bigr),
\label{eq:S-probeconstruction}
\end{equation}
with $\hat\rho = m/\|m\|$ the radial direction
and $\hat o$ a direction orthogonal to the sphere's ambient span. Because every reference point has
the same norm $R$, and $\hat o$ is orthogonal to the span both $m$ and every reference point occupy,
the squared distance from $z_B$ to any reference point $x$ changes by the same constant under this
construction; $k$-NN selection is therefore exactly invariant to $\nu$, $O_\star$ and $\mathrm{sgn}$
(the main text's Case 2, Section 3.1.1.1). This isolates whatever the table is testing from the anchor's
own tendency to track the imposed displacement. Used in
Tables~\ref{tab:S-params}, \ref{tab:S-anisoanchor}, \ref{tab:S-aniso}, \ref{tab:S-cor3},
\ref{tab:S-offman}, and the rotation and boundary tests of Tables~\ref{tab:S-rotation},
\ref{tab:S-covgrad} and \ref{tab:S-covbound} (pure off-manifold probes, $O_\star=1$).

\textbf{On-manifold steps, referred to the probe's own projection.} A probe is placed by an exact
step along the manifold (a geodesic on the sphere or Clifford torus, a coordinate step on the torus
of revolution or the circle $\times$ ellipse), and the anchor displacement is measured against the
probe's own base point rather than a point fixed before the step. This is necessary whenever the
manifold is inhomogeneous or curved, where referring to a pre-step point would contaminate the
measurement with the probe's own curvature-driven offset (Table~\ref{tab:S-torus} documents this
confound directly). Used in Tables~\ref{tab:S-gate0}, \ref{tab:S-torus}, \ref{tab:S-anchorbias},
\ref{tab:S-clifford} and \ref{tab:S-ellipse}.

\subsection{Neighborhood extent}
\label{sec:S-methods-ell}

Given the $k$ reference points nearest a test point, with centroid $\mu$ and tangent basis $V_d$
(the leading $d$ right singular vectors of the mean-centered neighbor matrix), two extents are
computed:
\begin{align}
\ell^2 &= \frac{1}{k-1}\sum_{j=1}^k \|V_d^\top(x_j-\mu)\|^2 && \text{tangential, unbiased}
\label{eq:S-elldef}\\
\rho^2 &= \frac{1}{k-1}\sum_{j=1}^k \|x_j-\mu\|^2 && \text{ambient, unbiased}
\notag
\end{align}
$\ell$ is the neighborhood extent of Propositions~2--3 and is used in every closed-form prediction
in this supplement (\texttt{D.ell} in the code); $\rho$ additionally includes normal-direction
spread and any noise not confined to the tangent space, and is not used in predictions
(\texttt{D.rnbr}). Both divide by $k-1$ rather than $k$: since $\mu$ is the sample mean of the same
$k$ points, dividing by $k$ underestimates the population second moment by a factor $(k-1)/k$, which
appears as a systematic $k/(k-1)$ excess in any ratio built from the biased form.

A third value, the areal estimate $k\cdot\mathrm{Area}/(2\pi n_{\mathrm{ref}})$, appears in
Table~\ref{tab:S-torus} and the $k$ sweep of Table~\ref{tab:S-clifford}, where $\ell$ cannot be
measured from a fixed base point beforehand. It must be multiplied by the $k$-NN boundary factor
$(k+1)/k$ to match $\ell^2$: the $k$-th nearest neighbor sits on the boundary of the occupied disc,
so the sampled region is larger by that factor than the area estimate assumes
(Section~\ref{sec:S-proofs}).

On a sphere of radius $R$ with $n_{\mathrm{ref}}$ points uniform over the full sphere,
$\ell^2 \approx 2(k+1)R^2/n_{\mathrm{ref}}$ to leading order (Eq.~21 of the main text); the sagitta
$\delta = \ell^2/2R$ carries a further next-order correction,
$\delta = (\ell^2/2R)(1+\ell^2/3R^2)$, from the exact spherical geometry
(Eq.~\eqref{eq:S-sagitta2}), tested directly by the \emph{ratio2} columns of
Tables~\ref{tab:S-params}, \ref{tab:S-anisoanchor} and \ref{tab:S-anchorbias}.

\subsection{Noise}
\label{sec:S-methods-noise}

Where a table has a noise parameter $\sigma$, it is added as $\mathcal{N}(0,\sigma^2 I_F)$
independently to every one of the $F$ ambient coordinates, applied to the reference set, the
probes, or both as stated in that table's caption; Table~\ref{tab:S-params}'s noise block and
Table~\ref{tab:S-gate1} add it to both. Anisotropic noise
(Tables~\ref{tab:S-anisoanchor}--\ref{tab:S-anisoframe}) instead uses
$\Sigma = \sigma^2(I + \kappa\,ww^\top)$ for a unit vector $w$ placed along a stated direction (the
gain axis, a tangent direction, or an ambient direction), so that $\sigma_w=\sigma\sqrt{1+\kappa}$
is the noise scale along $w$. Tables~\ref{tab:S-curvlaw}, \ref{tab:S-covgrad} and
\ref{tab:S-covbound} are noiseless by construction, since they isolate a purely geometric effect
(curvature, or a sampling boundary) from any noise contribution. Table~\ref{tab:S-clifford} uses a
small isotropic $\sigma=0.001$ so that the noise floor $\sqrt{F-3}\,\sigma$ is available as a
reference scale without swamping the geometric signal; Table~\ref{tab:S-ellipse} and the main block
of Table~\ref{tab:S-torus} are noiseless.

\subsection{Degeneracy}
\label{sec:S-methods-degen}

A point is flagged degenerate when the estimated alignment $a=\|P_N\hat\rho\|$ falls below
\texttt{alignTol}, $0.1$ by default; $G$ and $O$ are then undefined (\texttt{NaN}) and excluded from
every summary in this supplement, while $T$, which does not use the gain axis, remains defined and
included. The threshold is scientific rather than numerical: on a cone, where the gain axis is
undefined in the continuum, the measured alignment is about $0.06$, not machine epsilon, because
finite sampling, noise and curvature keep $\|P_N\hat\rho\|$ off zero even in the fully degenerate
case (\texttt{gto2\_checks}, check C2). Reported degeneracy fractions (columns headed \%degen.\ or
fracDegen) are the share of probes excluded on this basis in that cell.

\subsection{Aggregation and replication}
\label{sec:S-methods-agg}

Summaries are medians over probes unless a table states otherwise; several tables in
Section~\ref{sec:S-gate1} and Section~\ref{sec:S-gate6} report both medians and means explicitly,
because the quantities involved (leakage, the recovered off-manifold fraction) are asymmetrically
distributed and a median-only or mean-only comparison can appear to disagree with a closed-form
prediction that is exact only for one of the two (Table~\ref{tab:S-gate1}'s survival law, checked
against both, is the clearest case: the earlier apparent median/mean discrepancy was the missing
$q=0$ leakage floor, not a choice between summaries). Where a table reports a standard error, it is
a Monte Carlo SE over the stated number of seeds, each seed redrawing the full reference set and
probe set independently; single-seed tables report no SE. Seed counts and trials per cell are
given in each table's caption.

\subsection{Software and code availability}
\label{sec:S-methods-code}

Simulations were run in MATLAB R2026a. The code that generates every table in this supplement is
available at \url{https://github.com/McKenzieNeuro/McKenzieLab/tree/main/GTO}, run via a
single entry point (\texttt{gto2\_run\_all}) that regenerates all tables in the dependency order of
Section~\ref{sec:S-cascade} and logs each step.

\clearpage
\section{Proofs and derivations}
\label{sec:S-proofs}

The main text states Propositions~1--3 and the corollaries drawn from them; the derivations are
collected here.

\subsection*{Definition 1 (orthogonality of the three subspaces)}

$T_p\mathcal{M}$ and $N_p\mathcal{M}$ are orthogonal complements by definition, and
$\hat n_g \in N_p\mathcal{M}$, so $G_p \perp T_p\mathcal{M}$. $R_p$ is defined as
$N_p\mathcal{M}\cap G_p^{\perp}$, so $R_p \perp G_p$ and $R_p \subseteq N_p\mathcal{M}$ gives
$R_p \perp T_p\mathcal{M}$. Within $N_p\mathcal{M}$, $G_p$ is one-dimensional and $R_p$ is its
orthogonal complement, so $G_p \oplus R_p = N_p\mathcal{M}$ and hence
$T_p\mathcal{M}\oplus G_p\oplus R_p = \mathbb{R}^F$. Orthogonal direct sums give unique
resolutions and Pythagorean norms, which is the content of the definition. The construction
requires only $P_N\hat\rho \neq 0$, so that $\hat n_g$ is defined; where that fails, $G_p$ and
$R_p$ are not defined and neither is the decomposition. $\square$

Note that no fourth component exists. Since $R_p$ is the whole orthogonal complement of $G_p$
inside $N_p\mathcal{M}$, the three subspaces already exhaust $\mathbb{R}^F$, and any further
residual is identically zero. Distinguishing residual-normal displacement that the reference
geometry can account for from displacement it cannot would require a further privileged
direction inside $R_p$; the present construction supplies none, and treats all of $R_p$ as
novelty.

\subsection*{Proposition 1 (exact decomposition of the fractions)}

\emph{Proof.} By construction $\hat n_g \perp V_d$, so $\mathrm{span}(V_d)$,
$\mathrm{span}(\hat n_g)$ and the residual normal space are mutually orthogonal, and
$r = (z_B - m) + (m - \mu)$ resolves uniquely across them. The gain-axis component is
$\gamma+\delta$; the residual-normal component is $\beta\hat b + \vec\eta$, of squared norm
$\beta^2+\eta^2+2\beta\eta\cos\psi$; the tangential component has squared norm $\tau^2$.
Pythagoras gives the stated $\|r\|^2$ and the fractions follow by definition. No approximation
is made, so a numerical departure is a coding error rather than a tolerance question. $\square$

The statement assumes that the imposed change $z_B-m$ has no component in the tangent space
\emph{of the anchor frame}. In general the tangential term is $\|P_T(z_B-m)+P_T(m-\mu)\|^2$ rather
than $\tau^2$; Table~\ref{tab:S-exact} checks that general form and reports the share of
$\|r\|^2$ the imposed change actually places there.

\subsection*{Consequences of Proposition 1}

\emph{(ii) First order.} Writing $\nu^2 = \gamma^2+\beta^2$ and $O_\star = \beta^2/\nu^2$, and
ordering terms in the neighborhood extent --- $\delta,\eta = O(\ell^2)$, $\tau^2 = O(\ell^2/k)$,
while $\delta^2$ and $\eta^2$ are $O(\ell^4)$ and dropped ---
\[
O - O_\star = \frac{2\gamma^2\beta\eta\cos\psi}{\nu^4}
- \frac{2\beta^2\gamma\delta}{\nu^4}
- \frac{\beta^2\tau^2}{\nu^4},
\]
which is the form quoted in the main text after collecting $O_\star$.

\emph{(iii) Zero imposed novelty.} Setting $\beta = 0$ in the exact expression leaves
$O = \eta^2/[(\gamma+\delta)^2+\eta^2+\tau^2]$ directly. The first-order form carries $1/\beta$
in its $\eta$ term and is not valid in this limit.

\emph{(iv) Random novel direction.} For $\hat b$ uniform on the unit sphere of a residual
normal space of dimension at least two, $\mathbb{E}[\hat b\cdot\hat\eta] = 0$ by symmetry, so
the cross term has zero mean.

\subsection*{Proposition 2 (anchor displacement)}

Let $e_1,\dots,e_d$ be an orthonormal basis of $T_p\mathcal{M}$ and $\mathrm{I\!I}$ the second
fundamental form, so that a neighbouring manifold point with chart coordinate $u$ is
$x(u) = p + \sum_i u_ie_i + \tfrac12\mathrm{I\!I}(u,u) + O(\|u\|^3)$.

\emph{Proof.} Averaging over the $k$ neighbours about $p$ gives
$\mu - p = \bar u + \tfrac12\mathrm{I\!I}(\overline{uu^\top}) + O(\ell^3)$ with
$\bar u = k^{-1}\sum_j u_j$. Taking expectations, $\mathbb{E}[\bar u] = 0$ and
$\mathbb{E}[\overline{uu^\top}] = (\ell^2/d)I_d$ for a locally isotropic measure, so by
bilinearity $\tfrac12\mathrm{I\!I}(\mathbb{E}[\overline{uu^\top}])
= (\ell^2/2d)\sum_i\mathrm{I\!I}(e_i,e_i) = (\ell^2/2)\vec H$. $\square$

The neglected term is $O(\ell^3)$ in general and $O(\ell^4)$ when the local sampling measure is
symmetric, since the cubic contribution contracts $\nabla\mathrm{I\!I}$ against the third
moment of $u$.

Here $\ell^2 = \mathbb{E}\|u\|^2$ over the $k$ neighbours of the probe. The $k$-th neighbour lies on
the boundary of the selection ball and the other $k-1$ are uniform inside it, so for locally
uniform sampling $\ell^2$ is $(k+1)/k$ times the areal value $k\,\mathrm{Area}/(2\pi
n_{\mathrm{ref}})$. When $\mathbb{E}[u]=0$ the centered estimator
$\sum_j\|u_j-\bar u\|^2/(k-1)$ is unbiased for $\ell^2$.

\subsection*{The spherical case}

For a sphere of radius $R$ with outward unit normal $\hat n$, $\mathrm{I\!I}(e_i,e_i) = -\hat
n/R$ for every $i$, so $\vec H = -\hat n/R$ and $\|\mathbb{E}[\mu]-p\| = \ell^2/2R$ directed
inward, independent of $d$ and of $F$. When the sphere is centered at the origin the normal is
the radial direction, so this displacement is collinear with the gain axis.

With $\ell$ measured in the tangent plane the exact sphere supplies the next order. A neighbour at
tangential radius $\rho$ lies $R-\sqrt{R^2-\rho^2} = \rho^2/2R + \rho^4/8R^3 + O(\rho^6/R^5)$
below the tangent plane, and a uniformly filled disc has
$\mathbb{E}\rho^4 = \tfrac43(\mathbb{E}\rho^2)^2$, so the inward displacement is
\begin{equation}
\delta = \frac{\ell^2}{2R}\Bigl(1+\frac{\ell^2}{3R^2}\Bigr) + O(\ell^6/R^5).
\label{eq:S-sagitta2}
\end{equation}
The columns headed ratio2 in Tables~\ref{tab:S-params}, \ref{tab:S-anisoanchor} and
\ref{tab:S-anchorbias} test this form.

\subsection*{Proposition 3 (anchor jitter)}

$\mathbb{E}\|\bar u\|^2 = k^{-1}\operatorname{tr}\mathbb{E}[uu^\top] = \ell^2/k$ for
independent zero-mean samples, so the root-mean-square tangential fluctuation is
$\ell/\sqrt{k}$. Under locally uniform sampling $\ell \propto k^{1/d}$, hence
$\ell/\sqrt{k} \propto k^{1/d-1/2}$: decreasing in $k$ for $d>2$, exactly $k$-invariant for
$d=2$, increasing for $d=1$, while the bias of Proposition~2 grows as $k^{2/d}$ for every $d$.

\subsection*{The covariance tilt}

The second fundamental form $\mathrm{I\!I}$ is a symmetric bilinear map
$T_p\mathcal{M}\times T_p\mathcal{M}\to N_p\mathcal{M}$, and its derivative
$\nabla\mathrm{I\!I}$ the corresponding symmetric trilinear map. Restricting to a single tangent
direction $e_1$ and a single normal direction $\hat n$ collapses both to scalars: bilinearity
gives $\mathrm{I\!I}(u_1e_1,u_1e_1) = u_1^2\,\mathrm{I\!I}(e_1,e_1)$, and projecting onto
$\hat n$ leaves $\kappa = \hat n^{\top}\mathrm{I\!I}(e_1,e_1)$, the ordinary normal curvature in
direction $e_1$ seen along $\hat n$; the same restrictions give
$g = \hat n^{\top}\nabla\mathrm{I\!I}(e_1,e_1,e_1)$, to leading order the rate of change of
$\kappa$ along $e_1$. The local model is then
\begin{equation}
w(u_1) = \tfrac12\kappa\,u_1^2 + \tfrac16 g\,u_1^3 + O(u_1^4).
\label{eq:S-cubicpatch}
\end{equation}

For this patch the tilt of the fitted plane is the regression slope
$\mathbb{E}[u_1w]/\mathbb{E}[u_1^2]$, valid to leading order while
$\mathbb{E}[ww^\top] \ll \mathbb{E}[uu^\top]$. The quadratic part contributes
$\tfrac{\kappa}{2}\mathbb{E}[u_1^3] = 0$ by parity and the cubic part survives, giving
\[
\theta \simeq \frac{g}{6}\cdot\frac{\mathbb{E}[u_1^4]}{\mathbb{E}[u_1^2]} .
\]
The fourth-moment ratio is a property of the sampling distribution, not of the geometry: for a
uniformly filled disc of radius $a$ it equals $a^2/2 = \ell^2$, giving
$\theta \simeq g\ell^2/6$, but the coefficient would differ for a
Gaussian neighborhood. Since the rotation is $\|V_d^\top\hat n\|^2 = \sin^2\theta \approx \theta^2$, this gives
\begin{equation}
\|V_d^{\top}\hat n\|^{2} \;\propto\; \ell^{4} \;\propto\; k^{2},
\label{eq:S-covtilt}
\end{equation}
using $\ell^2\propto k$ for this sampling regime. The $\ell^4$ scaling is the robust part; the
prefactor is not.

The same calculation shows why a curvature gradient does not disturb Proposition~2: the
\emph{mean} of $w$ retains the quadratic term and loses the cubic to parity, while the
\emph{covariance} loses the quadratic and retains the cubic. Curvature and its gradient are
read by two different statistics of the same neighborhood and do not leak into one another
under symmetric sampling.

\subsection*{The finite-sample tilt floor}

Under symmetric sampling the covariance vanishes in expectation but not in a finite sample, so
the fitted plane tilts at random. Take the normal displacement $w = \|u\|^2/2R$ (a sphere,
$\kappa = 1/R$). The least-squares slope along tangent direction $i$ is
$\widehat{\mathrm{Cov}}(u_i,w)/\widehat{\mathrm{Var}}(u_i)$. For $k$ points uniform in a disc of
radius $a$, $\mathbb{E}u_i^2 = a^2/4$, $\mathbb{E}[u_i^2\|u\|^2] = a^4/6$ and
$\mathbb{E}[u_i^2\|u\|^4] = a^6/8$, so
\[
k\,\mathrm{Var}\,\widehat{\mathrm{Cov}}(u_i,\|u\|^2)
= \mathbb{E}[u_i^2\|u\|^4] + (\mathbb{E}\|u\|^2)^2\,\mathbb{E}u_i^2
- 2\,\mathbb{E}\|u\|^2\,\mathbb{E}[u_i^2\|u\|^2] = \frac{a^6}{48},
\]
and $\mathrm{Var}(\mathrm{slope}_i) = a^2/(12kR^2)$ at leading order in $1/k$. Summing the two
tangent directions and using $a^2 = 2\ell^2$,
\begin{equation}
\mathbb{E}\,\|V_d^\top\hat n\|^2 = \frac{\ell^2\kappa^2}{3k}\,\bigl[1+O(1/k)\bigr].
\label{eq:S-tiltfloor}
\end{equation}
A one-direction patch $w = \tfrac12\kappa u_1^2$ gives the same total ($\kappa^2a^2/8k$ along
$u_1$ plus $\kappa^2a^2/24k$ along $u_2$). For a uniformly sampled 2-sphere,
$\ell^2 = 2kR^2/n_{\mathrm{ref}}$ and the floor is $2/(3n_{\mathrm{ref}})$, independent of $k$ and
$R$. The heuristic that the fitted plane is the tangent plane at the displaced centroid gives
$2/n_{\mathrm{ref}}$; the least-squares plane tilts one third as much. With two approximately
Gaussian slopes the rotation is approximately exponential, so its median is near $\ln 2$ times its
mean. The $O(1/k)$ factor is measured directly in Table~\ref{tab:S-covgrad}.

\subsection*{The survival law}

The survival law of the main text is a standard fact about random subspace projections rather
than a result derived here. By rotational invariance the fixed direction may be held and the
subspace randomised, so the squared projection of a unit vector onto a uniformly random (Haar)
$q$-dimensional subspace of $\mathbb{R}^m$ is
$\sum_{i=1}^{q} g_i^2 \big/ \sum_{i=1}^{m} g_i^2$ for i.i.d.\ standard Gaussians $g_i$: a ratio
of chi-squared variates, hence
$\mathrm{Beta}(q/2,\,(m-q)/2)$, with mean $q/m$. Survival is the complement, giving
$1 - q/(F-d_{\mathrm{true}})$ with $m = F-d_{\mathrm{true}}$.

The Beta median has no closed form but is easily evaluated numerically; at $F = 30$,
$d_{\mathrm{true}} = 2$ the median survival exceeds the mean by $0.017$--$0.020$ for $q = 1$--$4$,
because the leakage distribution is right-skewed. A comparison must therefore be like for like:
median survival against the median law, mean against the mean.

A second correction matters at the same scale. At $q = 0$ the estimated tangent space already
absorbs a small leakage $L_0$ of $\hat b$, since noise tilts it, and the excess directions act on
what remains. Survival is therefore modelled as $(1-L_0)(1-B)$, with
$B\sim\mathrm{Beta}(q/2,(m-q)/2)$ and $L_0$ drawn from the measured per-probe $q = 0$ leakage,
assuming independence. With both corrections the noise-dominated survivals of
Table~\ref{tab:S-gate1} agree with the law to within $0.007$, for medians and means alike; without
the floor the median form misses by up to $0.012$ and the mean form by up to $0.009$. Where the
excess directions are structured rather than noise-driven the law fails outright, which is the
curvature-dominated block of the same table.

\subsection*{The curvature alignment on a Clifford torus}

\begin{sloppypar}\noindent
For $p(u,v) = (R_1\cos u, R_1\sin u, R_2\cos v, R_2\sin v)$ the normal space is spanned by
$n_1 = (\cos u,\sin u,0,0)$ and $n_2 = (0,0,\cos v,\sin v)$, with
$\mathrm{I\!I}(e_u,e_u) = -n_1/R_1$ and $\mathrm{I\!I}(e_v,e_v) = -n_2/R_2$. Hence
$\vec H = -\tfrac12(n_1/R_1 + n_2/R_2)$ and
$\|\vec H\| = \tfrac12\sqrt{R_1^{-2}+R_2^{-2}}$. The position vector
$p = R_1n_1 + R_2n_2$ lies wholly in the normal space, so $\hat n_g = \hat\rho$ and
$\hat n_g^\top\vec H = -\tfrac12(1+1)/\rho = -1/\rho$ with $\rho = \sqrt{R_1^2+R_2^2}$. Then
\[
c = -\frac{1}{\rho\|\vec H\|}
  = -\frac{2}{\sqrt{R_1^2+R_2^2}\;\sqrt{R_1^{-2}+R_2^{-2}}},
\]
which is $-1$ at $R_1 = R_2$ and falls in magnitude as the radii are separated.
\end{sloppypar}

\clearpage
\section{The exact identity}
\label{sec:S-exact}

Proposition~1 is an algebraic identity in the anchor frame, so it admits a check of a different
class from everything else here: a departure is a coding error, not a tolerance question. Every
quantity is computed from the \emph{realized} $\mu$, $V_d$ and $\hat n_g$: $\gamma = \hat
n_g^\top(z_B-m)$, $\delta = \hat n_g^\top(m-\mu)$, $\beta\hat b$ and $\vec\eta$ the residual-normal
parts of $z_B-m$ and $m-\mu$, and $t = P_T(z_B-m)+P_T(m-\mu)$, so that
$\|r\|^2 = (\gamma+\delta)^2 + \|\beta\hat b+\vec\eta\|^2 + \|t\|^2$. Imposed changes have random
composition ($O_\star$ uniform on $[0,1]$), magnitude ($\nu$ uniform on $[0.3,1]$) and gain sign.
The last column is the median share of $\|r\|^2$ that the imposed change places in the anchor
tangent space, which Proposition~1 as stated takes to be zero.

\begin{table}[h]
\centering
\small
\caption{Exact identity, maximum absolute departure over 300 probes per seed. Sphere, $R=5$,
$F=30$, $d=2$, $n_{\mathrm{ref}}=4000$, $k=40$, $\sigma=0.01$.}
\label{tab:S-exact}
\begin{tabular}{lrrrrrr}
\toprule
seed & $\bigl|\,\|r\|^2_{\text{pred}} - \|r\|^2\,\bigr|$ & $|\Delta G|$ & $|\Delta T|$ & $|\Delta O|$ & $|G{+}T{+}O-1|$ & imposed $T$ share \\
\midrule
0 & $6.7\times10^{-16}$ & $2.4\times10^{-15}$ & $4.7\times10^{-16}$ & $1.9\times10^{-15}$ & $3.0\times10^{-15}$ & $4.02\times10^{-4}$ \\
1 & $4.4\times10^{-16}$ & $2.4\times10^{-15}$ & $4.2\times10^{-16}$ & $2.4\times10^{-15}$ & $3.1\times10^{-15}$ & $3.50\times10^{-4}$ \\
2 & $5.6\times10^{-16}$ & $2.2\times10^{-15}$ & $5.6\times10^{-16}$ & $2.9\times10^{-15}$ & $3.2\times10^{-15}$ & $4.36\times10^{-4}$ \\
\bottomrule
\end{tabular}
\end{table}

\clearpage
\section{Gate 0 --- sheet proximity}
\label{sec:S-gate0}

Two concentric spherical shells separated by $h$, both origin-centered, with probes lying
exactly on the inner shell. The imposed change is purely on-manifold, so all reported energy
is artefact. The second block repeats the test with the sheets offset in an ambient direction
instead of radially.

\begin{table}[h]
\centering
\small
\caption{Sheet proximity. $R = 5$, $F = 30$, $d = 2$, $n_{\mathrm{ref}} = 6000$ split evenly
between sheets, $k = 40$, 300 probes, $\sigma = 0.01$; medians over probes, means over 5 seeds.
``wrong'' is the fraction of neighbours drawn from the other sheet. Control is a single sheet.
The noise floor for $\|r_{\mathrm{novel}}\|$ is $\sqrt{F-3}\,\sigma = 0.0520$.}
\label{tab:S-gate0}
\begin{tabular}{lrrrrrr}
\toprule
stacking & $h$ & $h/\ell$ & wrong & $\|r_{\mathrm{rad}}\|$ & $\|r_{\mathrm{novel}}\|$ & $\|r\|$ \\
\midrule
radial & 4.00 & 4.89 & 0.0\% & 0.06600 & 0.05214 & 0.13920 \\
 & 1.00 & 1.36 & 11.4\% & 0.06195 & 0.05219 & 0.14020 \\
 & 0.50 & 0.82 & 38.8\% & \textbf{0.15747} & 0.05231 & 0.19021 \\
 & 0.25 & 0.42 & 46.7\% & 0.08236 & 0.05238 & 0.12934 \\
 & 0.10 & 0.17 & 49.3\% & 0.01608 & 0.05239 & 0.09687 \\
 & 0.05 & 0.09 & 49.8\% & 0.01051 & 0.05240 & 0.09432 \\
 & control & --- & 0.0\% & 0.06600 & 0.05214 & 0.13920 \\
\midrule
ambient & 4.00 & 4.89 & 0.0\% & 0.06600 & 0.05214 & 0.13920 \\
 & 1.00 & 1.38 & 12.6\% & 0.04803 & 0.13661 & 0.18495 \\
 & 0.50 & 0.85 & 40.6\% & 0.02600 & \textbf{0.20959} & 0.23105 \\
 & 0.25 & 0.43 & 47.8\% & 0.03036 & 0.13058 & 0.15932 \\
 & 0.10 & 0.17 & 49.8\% & 0.03305 & 0.07246 & 0.11215 \\
 & 0.05 & 0.09 & 50.0\% & 0.03347 & 0.05807 & 0.10310 \\
 & control & --- & 0.0\% & 0.06600 & 0.05214 & 0.13920 \\
\bottomrule
\end{tabular}
\end{table}

Under radial stacking $\|r_{\mathrm{novel}}\|$ stays at the noise floor across every separation,
$0.0521$--$0.0524$ against $0.0520$, while $\|r_{\mathrm{rad}}\|$ peaks at $2.39\times$ control at
$h/\ell = 0.82$. Under ambient stacking the pattern inverts: $\|r_{\mathrm{novel}}\|$ peaks at
$4.0\times$ the floor at $h/\ell = 0.85$, while $\|r_{\mathrm{rad}}\|$ stays at or below control.
The control's $\|r_{\mathrm{rad}}\|$ is the sagitta itself ($\ell^2/2R = 0.067$). At $h\ll\ell$ the
two sheets act as one denser, thicker sheet and the artefact subsides. The wrong-sheet fraction is
largest ($50\%$) exactly where the artefact is smallest, so it is not a diagnostic.

\clearpage
\section{Gate 1 --- dimension excess}
\label{sec:S-gate1}

Leakage $\|V_d^\top\hat b\|^2$ for an imposed purely ambient $\hat b$, swept over $d$. The upper
table compares survival with the survival law at a noise-dominated and a curvature-dominated noise
level; the lower table sweeps the noise through the transition between them at $q = 1$.
``Capture'' is the median squared overlap of the third singular vector of the neighborhood with
the true normal at the base point: near $1$ when curvature wins that slot, near the chance value
$0.017$ (median of $\mathrm{Beta}(\tfrac12,\tfrac{27}{2})$) when noise does.

\begin{table}[h]
\centering
\footnotesize
\setlength{\tabcolsep}{3pt}
\caption{Dimension excess on a sphere. $R = 5$, $F = 30$, $d_{\mathrm{true}} = 2$,
$n_{\mathrm{ref}} = 4000$, $k = 40$, 2500 probes $\times$ 3 seeds. Survival is $1-\|V_d^\top\hat
b\|^2$, summarized per seed by median or mean and averaged over seeds (sd across seeds in
parentheses). Haar med.\ and Haar mean are the median and mean of
$1-\mathrm{Beta}(q/2,(F-d)/2)$; $\times$floor folds in the per-probe $q=0$ leakage,
$(1-L_0)(1-B)$, by Monte Carlo ($10^5$ draws). Ratio columns are measured over the floor-corrected
prediction of the same summary. $a$ is the median alignment.}
\label{tab:S-gate1}
\begin{tabular}{lrrrrrrrrrr}
\toprule
 & $d$ & med.\ (sd) & mean & Haar med. & $\times$floor & Haar mean & $\times$floor & $a$ & ratio (med.) & ratio (mean) \\
\midrule
$\sigma = 0.30$ & 2 & 0.9925 (0.0002) & 0.9888 & 1.0000 & 0.9925 & 1.0000 & 0.9887 & 0.9959 & 1.000 & 1.000 \\
 & 3 & 0.9708 (0.0007) & 0.9557 & 0.9830 & 0.9698 & 0.9643 & 0.9534 & 0.9827 & 1.001 & 1.002 \\
 & 4 & 0.9390 (0.0007) & 0.9211 & 0.9481 & 0.9365 & 0.9286 & 0.9181 & 0.9629 & 1.003 & 1.003 \\
 & 5 & 0.9049 (0.0005) & 0.8874 & 0.9114 & 0.9006 & 0.8929 & 0.8829 & 0.9422 & 1.005 & 1.005 \\
 & 6 & 0.8695 (0.0010) & 0.8542 & 0.8742 & 0.8638 & 0.8571 & 0.8475 & 0.9212 & 1.007 & 1.008 \\
\midrule
$\sigma = 0.01$ & 2 & 1.0000 (0.0000) & 1.0000 & 1.0000 & 1.0000 & 1.0000 & 1.0000 & 0.9999 & 1.000 & 1.000 \\
 & 3 & 0.9985 (0.0001) & 0.9963 & 0.9830 & 0.9830 & 0.9643 & 0.9643 & 0.2939 & 1.016 & 1.033 \\
 & 4 & 0.9782 (0.0006) & 0.9598 & 0.9481 & 0.9481 & 0.9286 & 0.9286 & 0.2869 & 1.032 & 1.034 \\
 & 5 & 0.9422 (0.0012) & 0.9225 & 0.9114 & 0.9114 & 0.8929 & 0.8930 & 0.2797 & 1.034 & 1.033 \\
 & 6 & 0.9047 (0.0013) & 0.8855 & 0.8742 & 0.8741 & 0.8571 & 0.8572 & 0.2722 & 1.035 & 1.033 \\
\bottomrule
\end{tabular}

\vspace{1em}

\begin{tabular}{rrrrrrr}
\toprule
$\sigma$ & survival ($q{=}1$) & median $O$ & \% degen. & capture & $a$ & ratio $a/\sqrt{1-\text{capture}}$ \\
\midrule
0.005 & 0.9996 & 0.9229 & 5.0 & 0.979 & 0.1452 & 1.002 \\
0.010 & 0.9985 & 0.9225 & 0.0 & 0.913 & 0.2939 & 0.996 \\
0.020 & 0.9936 & 0.9232 & 0.0 & 0.634 & 0.6046 & 0.999 \\
0.030 & 0.9872 & 0.9234 & 0.0 & 0.241 & 0.8707 & 0.999 \\
0.050 & 0.9828 & 0.9235 & 0.0 & 0.050 & 0.9741 & 0.999 \\
0.080 & 0.9816 & 0.9271 & 0.0 & 0.027 & 0.9853 & 0.999 \\
0.120 & 0.9801 & 0.9333 & 0.0 & 0.022 & 0.9874 & 0.998 \\
0.200 & 0.9758 & 0.9438 & 0.0 & 0.020 & 0.9855 & 0.996 \\
0.300 & 0.9708 & 0.9456 & 0.0 & 0.020 & 0.9827 & 0.993 \\
\bottomrule
\end{tabular}
\end{table}

\paragraph{The survival law.} Compared like for like, and with the $q = 0$ floor folded in, the
noise-dominated survivals exceed the law by $0.001$--$0.006$ in medians and $0.002$--$0.007$ in
means for $q = 1$--$4$, against seed-to-seed scatter of at most $0.001$. The residual is
systematic and grows with $q$, so the excess directions of a local PCA are close to Haar-distributed
but not exactly. Without the floor, the median form misses by up to $0.012$ and the mean form by up
to $0.009$ (Section~\ref{sec:S-proofs}).

\paragraph{The curvature-dominated case.} At $\sigma = 0.01$ the median leakage at $q = 1$ is
$0.0015$ against a Haar median of $0.017$ (mean $0.0037$ against $0.036$): the absorbed direction
is the normal, not a novelty-carrying one. The alignment collapses from $1.000$ to $0.294$ in that
single step and stays near $0.28$ as $d$ grows. Dimension over-estimation damages gate~4, not
gate~1's own quantity.

\paragraph{The transition.} Capture falls from $0.98$ at $\sigma = 0.005$ to the chance level by
$\sigma = 0.08$, crossing $0.5$ near $\sigma \approx 0.023$. That is about half the sagitta
$\ell^2/2R = 0.050$, and $0.8$ times its spread along the normal for a uniformly filled disc,
$\ell^2/(2\sqrt3R) = 0.029$. The alignment at $q = 1$ follows $\sqrt{1-\text{capture}}$ to within
$0.7\%$ (capture is printed to three decimals), because on this sphere the gain axis is the normal
and $a^2 = 1 - \|V_d^\top\hat n\|^2$ up to the gate-2 rotation floor. The drop in $a$ between $d$
and $d+1$ is therefore a direct, observable reading of how much of the normal the extra dimension
has captured.

\section{Gate 2 --- tangent rotation}
\label{sec:S-gate2}

Table~\ref{tab:S-rotation} measures the squared component, inside the estimated tangent basis, of
a direction that is truly normal at the base point, on a noiseless uniformly sampled sphere. The
prediction is the finite-sample tilt floor of Eq.~(\ref{eq:S-tiltfloor}): mean
$2/(3n_{\mathrm{ref}})$ and median near $2\ln2/(3n_{\mathrm{ref}})$, independent of $k$ and $R$ at
leading order in $1/k$. Each cell reuses the same draws, so in the $R$ block the construction
differs only by a uniform rescaling.

\begin{table}[h]
\centering
\small
\caption{Tangent rotation, $\|V_d^\top\hat n\|^2\times n_{\mathrm{ref}}$, on a sphere with pure
off-manifold probes. $F = 30$, $d = 2$, 1000 probes per cell, 5 seeds (sd across seeds in
parentheses), noiseless reference. Each block varies one parameter, holding the others at their
default ($n_{\mathrm{ref}}=4000$, $k=40$, $R=5$). Ratio columns divide by $2\ln2/3 = 0.462$
(median) and $2/3$ (mean).}
\label{tab:S-rotation}
\begin{tabular}{llrrrr}
\toprule
block & value & median (sd) & mean (sd) & ratio (median) & ratio (mean) \\
\midrule
$n_{\mathrm{ref}}$ & 1000 & 0.572 (0.025) & 0.862 (0.043) & 1.238 & 1.293 \\
 & 2000 & 0.515 (0.018) & 0.810 (0.043) & 1.114 & 1.215 \\
 & 4000 & 0.499 (0.018) & 0.762 (0.029) & 1.080 & 1.143 \\
 & 8000 & 0.509 (0.033) & 0.786 (0.045) & 1.101 & 1.179 \\
 & 16000 & 0.505 (0.033) & 0.783 (0.010) & 1.093 & 1.175 \\
\midrule
$k$ & 20 & 0.541 (0.011) & 0.911 (0.019) & 1.171 & 1.367 \\
 & 40 & 0.499 (0.018) & 0.762 (0.029) & 1.080 & 1.143 \\
 & 80 & 0.483 (0.040) & 0.715 (0.050) & 1.045 & 1.073 \\
 & 160 & 0.502 (0.032) & 0.730 (0.024) & 1.086 & 1.095 \\
\midrule
$R$ & 3 & 0.499 (0.018) & 0.762 (0.029) & 1.080 & 1.143 \\
 & 5 & 0.499 (0.018) & 0.762 (0.029) & 1.080 & 1.143 \\
 & 12 & 0.499 (0.018) & 0.762 (0.029) & 1.080 & 1.143 \\
 & 20 & 0.499 (0.018) & 0.762 (0.029) & 1.080 & 1.143 \\
\bottomrule
\end{tabular}
\end{table}

The $R$ block returns identical rows at every radius: with the same draws, a uniformly rescaled
configuration selects the same neighbours and the same singular vectors, so the rotation is exactly
scale-invariant, as the prediction requires. Across $n_{\mathrm{ref}}$ the median sits at
$0.50$--$0.57$ and the mean at $0.76$--$0.86$, both close to flat for $n_{\mathrm{ref}}\ge 4000$
and higher at $n_{\mathrm{ref}} = 1000$, where $\ell/R = 0.28$. Across $k$ the mean falls from
$1.37$ times the prediction at $k = 20$ to $1.07$--$1.10$ at $k = 80$--$160$. That is the
$O(1/k)$ correction of Eq.~(\ref{eq:S-tiltfloor}), measured directly on the covariance-route patch
in Table~\ref{tab:S-covgrad}. The $\ell^2 = 2kR^2/n_{\mathrm{ref}}$ used here is the areal value, and
contributes a further $(k+1)/k$. The median is $0.59$--$0.69$ of the mean, at or below the
exponential value $\ln 2 = 0.69$: the tail is at least exponential, and heaviest at small $k$.

\clearpage
\section{Gate 2 --- the covariance route}
\label{sec:S-covroute}

A symmetric neighborhood on a homogeneous manifold has $\mathbb{E}[uw^\top] = 0$ by parity, so
curvature displaces the anchor without tilting the frame. These two sweeps break that parity
deliberately.

\begin{table}[h]
\centering
\small
\setlength{\tabcolsep}{4pt}
\caption{Curvature gradient. Patch $w = \tfrac12\kappa u_1^2 + \tfrac16 g u_1^3$ with
$\kappa = 0.2$, uniform disc of radius 1, $F = 30$, $n_{\mathrm{ref}} = 6000$, 2000 trials per cell,
noiseless; means. Every cell uses the same draws, so the $g=0$ rotation at the same $k$ is a paired
measurement of the finite-sample floor. Ratio columns: measured floor over $\ell^2\kappa^2/3k$
(Eq.~\ref{eq:S-tiltfloor}); raw rotation over $(g\ell^2/6)^2$; rotation minus the measured floor over
$(g\ell^2/6)^2$ (subtr.).}
\label{tab:S-covgrad}
\begin{tabular}{rrrrrrrr}
\toprule
\multicolumn{8}{l}{\emph{$g$ sweep at $k = 40$}} \\
$g$ & $\ell$ & $\|V_d^\top\hat n\|^2$ & $(g\ell^2/6)^2$ & floor & ratio (floor) & ratio (raw) & ratio (subtr.) \\
\midrule
0.00 & 0.0585 & $1.273\times10^{-6}$ & 0 & $1.273\times10^{-6}$ & 1.115 & --- & --- \\
0.50 & 0.0585 & $1.349\times10^{-6}$ & $8.142\times10^{-8}$ & $1.273\times10^{-6}$ & 1.115 & 16.572 & 0.938 \\
1.00 & 0.0585 & $1.591\times10^{-6}$ & $3.257\times10^{-7}$ & $1.273\times10^{-6}$ & 1.115 & 4.885 & 0.977 \\
2.00 & 0.0585 & $2.571\times10^{-6}$ & $1.303\times10^{-6}$ & $1.273\times10^{-6}$ & 1.115 & 1.974 & 0.997 \\
4.00 & 0.0585 & $6.518\times10^{-6}$ & $5.211\times10^{-6}$ & $1.273\times10^{-6}$ & 1.115 & 1.251 & 1.007 \\
\midrule
\multicolumn{8}{l}{\emph{$k$ sweep at $g = 2$}} \\
$k$ & $\ell$ & $\|V_d^\top\hat n\|^2$ & $(g\ell^2/6)^2$ & floor & ratio (floor) & ratio (raw) & ratio (subtr.) \\
\midrule
20 & 0.0419 & $1.757\times10^{-6}$ & $3.439\times10^{-7}$ & $1.395\times10^{-6}$ & 1.190 & 5.108 & 1.050 \\
40 & 0.0585 & $2.571\times10^{-6}$ & $1.303\times10^{-6}$ & $1.273\times10^{-6}$ & 1.115 & 1.974 & 0.997 \\
80 & 0.0821 & $6.359\times10^{-6}$ & $5.045\times10^{-6}$ & $1.208\times10^{-6}$ & 1.076 & 1.261 & 1.021 \\
160 & 0.1157 & $2.098\times10^{-5}$ & $1.991\times10^{-5}$ & $1.164\times10^{-6}$ & 1.043 & 1.054 & 0.995 \\
320 & 0.1634 & $8.062\times10^{-5}$ & $7.916\times10^{-5}$ & $1.162\times10^{-6}$ & 1.045 & 1.018 & 1.004 \\
\bottomrule
\end{tabular}
\end{table}

Three things are read from the three ratio columns. First, the finite-sample floor: at $g = 0$ the
rotation exceeds the leading-order $\ell^2\kappa^2/3k$ by $1.19$, $1.12$, $1.08$, $1.04$ and $1.05$
at $k = 20$ to $320$, an $O(1/k)$ excess that is within Monte Carlo error of unity at the largest
$k$. Second, the raw ratio overstates the covariance term wherever the signal is comparable to that
floor ($16.6$ at $g = 0.5$). Third, once the paired floor is subtracted the prediction
$(g\ell^2/6)^2$ holds in every cell: $0.94$--$1.01$ across the $g$ sweep, the lowest where the
signal is only $6\%$ of the floor, and $0.995$--$1.050$ across a sixteenfold range of $k$.

The rotation rises $46$-fold from $k = 20$ to $k = 320$. It is the only frame term in this
supplement that grows with the neighborhood, and it grows as $\ell^4$.

\begin{table}[h]
\centering
\footnotesize
\setlength{\tabcolsep}{4pt}
\caption{Boundary. Spherical cap of half-angle $0.35$, $R = 5$, $F = 30$,
$n_{\mathrm{ref}} = 8000$, noiseless, $d = d_{\mathrm{true}}$. Each cell places 200 probes at a
fixed fraction of the cap half-angle from the pole and random azimuth, over 3 reference draws;
``edge dist'' is arc length to the cap edge. SE is that of the median. ``interior'' is the
Eq.~(\ref{eq:S-tiltfloor}) floor $\ell^2/(3kR^2)$ at the measured $\ell$; the ratio is the mean
rotation over it.}
\label{tab:S-covbound}
\begin{tabular}{rrrrrrrr}
\toprule
$k$ & ang/cap & edge dist & edge/$\ell$ & median & mean (SE of median) & interior & ratio (mean) \\
\midrule
20 & 0.30 & 1.2250 & 20.25 & $2.288\times10^{-6}$ & $3.582\times10^{-6}$ ($2.1\times10^{-7}$) & $2.440\times10^{-6}$ & 1.47 \\
 & 0.60 & 0.7000 & 11.03 & $1.956\times10^{-6}$ & $3.370\times10^{-6}$ ($2.1\times10^{-7}$) & $2.684\times10^{-6}$ & 1.26 \\
 & 0.80 & 0.3500 & 5.56 & $2.047\times10^{-6}$ & $3.617\times10^{-6}$ ($2.5\times10^{-7}$) & $2.646\times10^{-6}$ & 1.37 \\
 & 0.90 & 0.1750 & 2.79 & $2.200\times10^{-6}$ & $3.476\times10^{-6}$ ($2.0\times10^{-7}$) & $2.622\times10^{-6}$ & 1.33 \\
 & 0.95 & 0.0875 & 1.38 & $2.295\times10^{-6}$ & $3.778\times10^{-6}$ ($2.2\times10^{-7}$) & $2.690\times10^{-6}$ & 1.40 \\
 & 0.99 & 0.0175 & 0.25 & $4.306\times10^{-5}$ & $4.945\times10^{-5}$ ($1.7\times10^{-6}$) & $3.169\times10^{-6}$ & 15.60 \\
\midrule
40 & 0.30 & 1.2250 & 14.17 & $1.628\times10^{-6}$ & $2.865\times10^{-6}$ ($1.8\times10^{-7}$) & $2.490\times10^{-6}$ & 1.15 \\
 & 0.60 & 0.7000 & 7.81 & $1.726\times10^{-6}$ & $2.660\times10^{-6}$ ($1.4\times10^{-7}$) & $2.679\times10^{-6}$ & 0.99 \\
 & 0.80 & 0.3500 & 4.00 & $1.854\times10^{-6}$ & $2.919\times10^{-6}$ ($1.6\times10^{-7}$) & $2.553\times10^{-6}$ & 1.14 \\
 & 0.90 & 0.1750 & 1.98 & $1.842\times10^{-6}$ & $3.039\times10^{-6}$ ($1.8\times10^{-7}$) & $2.595\times10^{-6}$ & 1.17 \\
 & 0.95 & 0.0875 & 0.98 & $7.701\times10^{-6}$ & $1.012\times10^{-5}$ ($4.7\times10^{-7}$) & $2.655\times10^{-6}$ & 3.81 \\
 & 0.99 & 0.0175 & 0.18 & $9.415\times10^{-5}$ & $1.032\times10^{-4}$ ($2.4\times10^{-6}$) & $3.153\times10^{-6}$ & 32.73 \\
\midrule
80 & 0.30 & 1.2250 & 9.99 & $1.844\times10^{-6}$ & $2.847\times10^{-6}$ ($1.5\times10^{-7}$) & $2.505\times10^{-6}$ & 1.14 \\
 & 0.60 & 0.7000 & 5.59 & $1.765\times10^{-6}$ & $2.661\times10^{-6}$ ($1.4\times10^{-7}$) & $2.617\times10^{-6}$ & 1.02 \\
 & 0.80 & 0.3500 & 2.82 & $1.680\times10^{-6}$ & $2.596\times10^{-6}$ ($1.4\times10^{-7}$) & $2.564\times10^{-6}$ & 1.01 \\
 & 0.90 & 0.1750 & 1.41 & $2.126\times10^{-6}$ & $3.142\times10^{-6}$ ($1.7\times10^{-7}$) & $2.583\times10^{-6}$ & 1.22 \\
 & 0.95 & 0.0875 & 0.68 & $4.645\times10^{-5}$ & $4.936\times10^{-5}$ ($1.1\times10^{-6}$) & $2.782\times10^{-6}$ & 17.74 \\
 & 0.99 & 0.0175 & 0.13 & $2.195\times10^{-4}$ & $2.249\times10^{-4}$ ($3.5\times10^{-6}$) & $3.128\times10^{-6}$ & 71.90 \\
\midrule
160 & 0.30 & 1.2250 & 7.02 & $2.024\times10^{-6}$ & $2.855\times10^{-6}$ ($1.5\times10^{-7}$) & $2.539\times10^{-6}$ & 1.12 \\
 & 0.60 & 0.7000 & 4.02 & $1.830\times10^{-6}$ & $2.574\times10^{-6}$ ($1.3\times10^{-7}$) & $2.521\times10^{-6}$ & 1.02 \\
 & 0.80 & 0.3500 & 2.01 & $1.700\times10^{-6}$ & $2.468\times10^{-6}$ ($1.2\times10^{-7}$) & $2.534\times10^{-6}$ & 0.97 \\
 & 0.90 & 0.1750 & 0.99 & $2.363\times10^{-5}$ & $2.538\times10^{-5}$ ($7.0\times10^{-7}$) & $2.621\times10^{-6}$ & 9.68 \\
 & 0.95 & 0.0875 & 0.47 & $1.931\times10^{-4}$ & $1.956\times10^{-4}$ ($2.6\times10^{-6}$) & $2.877\times10^{-6}$ & 67.99 \\
 & 0.99 & 0.0175 & 0.09 & $5.115\times10^{-4}$ & $5.183\times10^{-4}$ ($5.1\times10^{-6}$) & $3.130\times10^{-6}$ & 165.59 \\
\bottomrule
\end{tabular}
\end{table}

Away from the edge the mean rotation matches the interior floor: $0.97$--$1.22$ at $k\ge40$ and
$1.26$--$1.47$ at $k = 20$, the same finite-$k$ excess as Table~\ref{tab:S-covgrad}. (A probe at the
pole, ang/cap $= 0$, has the same neighborhood at every azimuth, so that row carries only 3 effective
samples and is omitted.) The rotation leaves the floor between edge/$\ell\approx1.4$ and $1.0$: it
is $3.8$ times the floor at edge/$\ell = 0.98$ ($k = 40$) and $9.7$ times at $0.99$ ($k = 160$), and
$16$--$166$ times at the ring nearest the edge. At edge/$\ell\approx1$ the excess over the floor rises
$3.1$-fold from $k = 40$ to $k = 160$ for a $3.9$-fold increase in $\ell^2$, so the growth with
neighborhood size is close to, but slightly slower than, $\ell^2$.

\clearpage
\section{Gate 3 --- the anchor displacement law}
\label{sec:S-gate3}

\subsection{Quadric patches: the sagitta and jitter laws}

Table~\ref{tab:S-curvlaw} is the direct test of Propositions~2--3 on quadric patches: the
anchor displacement measured against the mean-curvature prediction, and the tangential jitter
against $\ell/\sqrt{k}$.

\begin{table}[h]
\centering
\footnotesize
\caption{Proposition~2 (anchor displacement) and Proposition~3 (tangential jitter) on
quadric patches, $F=30$, $n_{\mathrm{ref}}=4000$, chart radius $1.00$. Upper block:
curvature sweep at $k=40$, 400 trials/cell. Lower block: $k$ sweep at fixed isotropic
curvature $\kappa_1=\kappa_2=0.20$. Noiseless; $\ell^2$ is the mean of $\|u\|^2$ about the base point in the
true chart. Ratio columns are measured/predicted, for $n_3$ in the upper block; \emph{---} marks a
prediction of zero, where a ratio is not a measurement.}
\label{tab:S-curvlaw}

\begin{tabular}{lrrrrrr}
\toprule
surface & $\ell^2$ & $\langle\mu,n_3\rangle$ & pred. & $\langle\mu,n_4\rangle$ & pred. & ratio \\
\midrule
isotropic $\kappa_1{=}\kappa_2{=}0.2$, one normal & 0.00517 & 0.000517 & 0.000517 & $-$0.000000 & 0.000000 & 1.000 \\
minimal $\kappa_1{=}{-}\kappa_2{=}0.2$, one normal ($\vec H=0$) & 0.00515 & 0.000002 & 0.000000 & 0.000000 & 0.000000 & --- \\
anisotropic $\kappa_1{=}0.4$, $\kappa_2{=}0$, one normal & 0.00517 & 0.000519 & 0.000517 & $-$0.000000 & 0.000000 & 1.004 \\
isotropic $\kappa_1{=}\kappa_2{=}0.5$, one normal & 0.00507 & 0.001267 & 0.001267 & 0.000000 & 0.000000 & 1.000 \\
biaxial $\kappa_1{=}0.2$ on $n_3$, $\kappa_2{=}{-}0.2$ on $n_4$ & 0.00523 & 0.000261 & 0.000262 & $-$0.000262 & $-$0.000262 & 0.998 \\
biaxial $\kappa_1{=}\kappa_2{=}0.2$, two normals & 0.00500 & 0.000249 & 0.000250 & 0.000251 & 0.000250 & 0.995 \\
\bottomrule
\end{tabular}

\vspace{1em}

\begin{tabular}{rrrrrrrr}
\toprule
$k$ & $\ell^2$ & $\langle\mu,n_3\rangle$ & pred. & rms jitter & $\ell/\sqrt{k}$ & ratio (sagitta) & ratio (jitter) \\
\midrule
20  & 0.00265 & 0.000265 & 0.000265 & 0.01127 & 0.01150 & 1.000 & 0.979 \\
40  & 0.00510 & 0.000510 & 0.000510 & 0.01155 & 0.01130 & 1.000 & 1.022 \\
80  & 0.01008 & 0.001008 & 0.001008 & 0.01107 & 0.01122 & 1.000 & 0.986 \\
160 & 0.02006 & 0.002006 & 0.002006 & 0.01132 & 0.01120 & 1.000 & 1.011 \\
320 & 0.04016 & 0.004016 & 0.004016 & 0.01132 & 0.01120 & 1.000 & 1.010 \\
\bottomrule
\end{tabular}
\end{table}

Read the ratio columns with the construction in mind. On a one-normal patch with
$\kappa_1 = \kappa_2$ the displacement along $n_3$ is $\tfrac{\kappa}{2}$ times the mean of
$\|u\|^2$ over the same neighbours that define $\ell^2$, so the isotropic rows and the whole
sagitta column of the lower block are $1$ by construction and test only the arithmetic. The
informative rows are the others. The anisotropic row ($1.004$) and the two biaxial rows ($0.998$,
$0.995$) require the neighborhood's second moment to be isotropic, $\mathbb{E}[u_1^2] =
\mathbb{E}[u_2^2] = \ell^2/2$, and it is. The lower block's jitter column tests Proposition~3
directly, and sits at $0.979$--$1.022$ at every $k$ from 20 to 320.

The upper block's standout row is \emph{minimal}, meaning zero mean curvature at the vertex. There
$\kappa_1=-\kappa_2=0.2$ --- the same curvature magnitude as the isotropic row directly above it ---
yet the predicted displacement is exactly zero and the measured one, $2\times10^{-6}$, is the
finite-sample difference between $\mathbb{E}[u_1^2]$ and $\mathbb{E}[u_2^2]$. This is the
trace-of-$\mathrm{I\!I}$ argument from Gate~3 of the main text made concrete: the two principal
curvatures cancel in the sum that governs $\vec H$, so a surface that is locally just as curved as
the isotropic case produces no anchor bias at all. The two biaxial rows make the complementary
point: with curvature split across two normal directions $n_3,n_4$, each direction's displacement
matches \emph{its own} curvature component independently --- $\langle\mu,n_3\rangle$ against
$\ell^2\kappa_1/4$ and $\langle\mu,n_4\rangle$ against $\ell^2\kappa_2/4$ --- which is the
mechanism Gate~5's booking rule depends on.

In the lower block $\langle\mu,n_3\rangle$ grows sixteenfold with $k$, tracking $\ell^2$
(Proposition~2), while the rms jitter stays at $0.0111$--$0.0116$, matching the $k$-invariant
$\ell/\sqrt{k}$ of Proposition~3. With $\ell\propto k^{1/d}$ the jitter scales as $k^{1/d-1/2}$,
an exponent that is negative for $d>2$, positive at $d=1$ and exactly zero at $d=2$; these are
two-dimensional patches, and the flat column is that exponent vanishing.

\subsection{An inhomogeneous manifold: torus of revolution}

On the sphere and the Clifford torus $\nabla\mathrm{I\!I} = 0$ everywhere, and a quadric patch has
$\nabla\mathrm{I\!I} = 0$ at its vertex, so the term Proposition~2 neglects is structurally zero in
all three. A torus of revolution removes that. With $R_c < 2r$ the mean curvature changes sign on
the inner region, so the anchor displacement should \emph{reverse} there. The same table carries
gate~4: the alignment $a = |R_c\cos\theta + r|/\|p\|$ varies over the surface and passes through
zero.

\begin{table}[h]
\centering
\footnotesize
\setlength{\tabcolsep}{4pt}
\caption{Torus of revolution, $R_c = 3$, $r = 2$, $F = 30$, $k = 40$, 4000 probes, noiseless.
Probes sit \emph{at} their base points. The prediction is $-(\ell^2/4)(\kappa_1+\kappa_2)$ along
the outward normal, with $\ell^2$ the measured mean of the tangential extent over probes; this is
$1.010$, $1.016$, $1.019$ and $1.024$ times the areal value at $n_{\mathrm{ref}} = 5000$, $10000$,
$20000$, $40000$, against the $k$-NN boundary factor $(k+1)/k = 1.025$. The ratio is the
least-squares slope of per-probe displacement on per-probe prediction within the bin; its SE
treats probes as independent and is a lower bound, since probes share reference points. Upper
table: $n_{\mathrm{ref}} = 40000$, bin medians. Lower table: the slope at each $n_{\mathrm{ref}}$.}
\label{tab:S-torus}
\begin{tabular}{lrrrrrrrr}
\toprule
$\theta$ bin & $\kappa_1{+}\kappa_2$ & disp.\ meas. & pred. & $a$ meas. & $a$ true & max $|\Delta a|$ & \% degen. & ratio (SE) \\
\midrule
0.00--0.79 & 0.6919 & $-0.00649$ & $-0.00668$ & 0.976 & 0.976 & 0.009 & 0.0 & 0.988 (0.007) \\
0.79--1.57 & 0.6064 & $-0.00569$ & $-0.00585$ & 0.761 & 0.761 & 0.016 & 0.0 & 0.996 (0.008) \\
1.57--2.36 & 0.3470 & $-0.00339$ & $-0.00335$ & 0.321 & 0.319 & 0.035 & 13.7 & 0.999 (0.017) \\
2.36--3.14 & $-0.3129$ & $+0.00251$ & $+0.00302$ & 0.559 & 0.577 & 0.067 & 4.7 & 0.974 (0.036) \\
3.14--3.93 & $-0.2394$ & $+0.00210$ & $+0.00231$ & 0.459 & 0.455 & 0.052 & 7.7 & 0.973 (0.040) \\
3.93--4.71 & 0.3436 & $-0.00310$ & $-0.00332$ & 0.317 & 0.314 & 0.024 & 15.4 & 0.975 (0.016) \\
4.71--5.50 & 0.6075 & $-0.00580$ & $-0.00587$ & 0.763 & 0.763 & 0.015 & 0.0 & 1.014 (0.008) \\
5.50--6.28 & 0.6916 & $-0.00678$ & $-0.00668$ & 0.975 & 0.975 & 0.006 & 0.0 & 1.022 (0.007) \\
\bottomrule
\end{tabular}

\vspace{1em}

\begin{tabular}{lrrrr}
\toprule
$\theta$ bin & ratio, $n_{\mathrm{ref}}=5000$ & $10000$ & $20000$ & $40000$ \\
\midrule
0.00--0.79 & 0.994 (0.006) & 1.006 (0.007) & 1.006 (0.007) & 0.988 (0.007) \\
0.79--1.57 & 1.065 (0.009) & 1.021 (0.007) & 1.037 (0.008) & 0.996 (0.008) \\
1.57--2.36 & 1.064 (0.014) & 1.003 (0.016) & 1.006 (0.016) & 0.999 (0.017) \\
2.36--3.14 & 0.909 (0.041) & 1.099 (0.041) & 0.995 (0.046) & 0.974 (0.036) \\
3.14--3.93 & 1.095 (0.036) & 1.068 (0.038) & 0.966 (0.036) & 0.973 (0.040) \\
3.93--4.71 & 1.015 (0.014) & 0.998 (0.015) & 0.965 (0.014) & 0.975 (0.016) \\
4.71--5.50 & 1.024 (0.008) & 1.018 (0.008) & 1.005 (0.007) & 1.014 (0.008) \\
5.50--6.28 & 1.002 (0.006) & 0.995 (0.007) & 1.006 (0.007) & 1.022 (0.007) \\
\bottomrule
\end{tabular}
\end{table}

The sign of the displacement reverses with $\kappa_1+\kappa_2$ in the inner bins at every
$n_{\mathrm{ref}}$, as Proposition~2 requires. At $n_{\mathrm{ref}} = 40000$ the slopes are
$0.973$--$1.022$; the four outer bins average $1.005$, and the two inner bins, where
$\ell|\kappa|$ is largest, sit within one SE of $1$. Bins at $\theta$ and $2\pi-\theta$ are
geometrically equivalent, yet the $0$--$0.79$ and $5.50$--$6.28$ pair differs by $0.034$ at
$n_{\mathrm{ref}} = 40000$, several printed SE; the SE is a lower bound, and pooling mirror bins is
the better summary ($1.005$, $1.005$, $0.986$, $0.974$ from outer to inner, inverse-variance
weighted). The largest departures at $n_{\mathrm{ref}} = 5000$ ($1.065$ and $1.064$ in the bins
from $0.79$ to $2.36$) have fallen to $1.021$ and $1.003$ by $10000$: a neglected higher-order term
shrinking with $\ell$, as it should. With the areal $\ell^2$ in place of the measured one the outer
bins sit $2.5$--$3\%$ high at every $n_{\mathrm{ref}}$ instead --- an $\ell$-independent offset,
which is the boundary factor $(k+1)/k$, and which the measured-to-areal ratio in the caption
approaches as $\ell$ falls and the tangent-projection shrinkage vanishes.

The median measured alignment matches the analytic value to within $0.003$ in every bin outside the
inner region and within $0.02$ inside it; per-probe errors reach $0.07$ in the inner bins.
Degeneracy is flagged in the bins containing the zeros of $a$ ($\cos\theta = -2/3$), at $14$--$15\%$
of probes, and at $5$--$8\%$ in the inner bins beside them.

\paragraph{A measurement confound, recorded because it is easy to reproduce.} If the probes are
stepped an arc length $s$ along the manifold and the anchor displacement is referred to the
\emph{pre-step} base point rather than to the probe's own position, the probe's own normal offset
$\kappa_n s^2/2$ contaminates the measurement. At $\theta = \pi$ and $s = 0.10$ that offset is about
$-0.005$, against a predicted inner-bin sagitta of $+0.002$ to $+0.003$ at $n_{\mathrm{ref}} =
40000$. The contaminant does not shrink with $\ell$, so the apparent ratio \emph{diverges} as
$n_{\mathrm{ref}}$ rises --- the signature of an $\ell$-independent contaminant rather than of a
neglected higher-order term. Referring the displacement to the probe's own position, as here,
removes it.

\subsection{Free-parameter dependence}

Table~\ref{tab:S-params} tests the closed-form scalings of the main text on a sphere, and tests two
further channels --- the magnitude of a probe's own displacement, and reference noise --- that
might be expected to drive the anchor displacement but do not.

\begin{table}[h]
\centering
\footnotesize
\setlength{\tabcolsep}{4pt}
\caption{Anchor bias against the free parameters. Defaults $F = 30$, $R = 5$,
$n_{\mathrm{ref}} = 4000$, $k = 40$, full sphere, noiseless; 250 trials per cell, the same draws in
every cell. $\ell$ is measured tangentially about the neighborhood centroid with the $(k-1)$
normalization. ratio $= \delta/(\ell^2/2R)$; ratio2 $= \delta/[(\ell^2/2R)(1+\ell^2/3R^2)]$
(Eq.~\ref{eq:S-sagitta2}), with its Monte Carlo SE.}
\label{tab:S-params}
\begin{tabular}{llrrrrrrr}
\toprule
sweep & value & $\ell$ & $\delta$ & $\ell^2/2R$ & jitter & $\ell/\sqrt{k}$ & ratio & ratio2 (SE) \\
\midrule
$n_{\mathrm{ref}}$ & 500 & 1.9446 & 0.39997 & 0.37816 & 0.30657 & 0.30747 & 1.058 & 1.007 (0.011) \\
 & 1000 & 1.4250 & 0.20852 & 0.20305 & 0.21440 & 0.22531 & 1.027 & 1.000 (0.012) \\
 & 2000 & 1.0065 & 0.10276 & 0.10130 & 0.15948 & 0.15914 & 1.014 & 1.001 (0.011) \\
 & 4000 & 0.7116 & 0.05088 & 0.05064 & 0.10712 & 0.11252 & 1.005 & 0.998 (0.011) \\
 & 8000 & 0.4996 & 0.02507 & 0.02496 & 0.08038 & 0.07899 & 1.004 & 1.001 (0.011) \\
 & 16000 & 0.3590 & 0.01292 & 0.01289 & 0.05742 & 0.05676 & 1.002 & 1.001 (0.012) \\
\midrule
$R$ & 3.0 & 0.4270 & 0.03053 & 0.03038 & 0.06427 & 0.06751 & 1.005 & 0.998 (0.011) \\
 & 5.0 & 0.7116 & 0.05088 & 0.05064 & 0.10712 & 0.11252 & 1.005 & 0.998 (0.011) \\
 & 8.0 & 1.1386 & 0.08140 & 0.08102 & 0.17140 & 0.18003 & 1.005 & 0.998 (0.011) \\
 & 12.0 & 1.7079 & 0.12210 & 0.12154 & 0.25709 & 0.27004 & 1.005 & 0.998 (0.011) \\
 & 20.0 & 2.8465 & 0.20350 & 0.20256 & 0.42849 & 0.45006 & 1.005 & 0.998 (0.011) \\
\midrule
cap angle & 0.30 & 0.1065 & 0.00113 & 0.00113 & 0.01691 & 0.01683 & 1.000 & 1.000 (0.011) \\
 & 0.60 & 0.2105 & 0.00443 & 0.00443 & 0.03343 & 0.03328 & 1.001 & 1.000 (0.011) \\
 & 1.00 & 0.3413 & 0.01167 & 0.01165 & 0.05421 & 0.05396 & 1.002 & 1.000 (0.011) \\
 & 1.50 & 0.4848 & 0.02359 & 0.02350 & 0.07701 & 0.07666 & 1.003 & 1.000 (0.011) \\
 & $\pi$ & 0.7116 & 0.05088 & 0.05064 & 0.10712 & 0.11252 & 1.005 & 0.998 (0.011) \\
\midrule
gain offset & $-1.0$ to $+2.0$ (8) & 0.7116 & 0.05088 & 0.05064 & 0.10712 & 0.11252 & 1.005 & 0.998 (0.011) \\
\midrule
noise $\sigma$ & 0.00 & 0.7116 & 0.05088 & 0.05064 & 0.10712 & 0.11252 & 1.005 & 0.998 (0.011) \\
 & 0.01 & 0.7032 & 0.04975 & 0.04945 & 0.11072 & 0.11118 & 1.006 & 1.000 (0.011) \\
 & 0.05 & 0.7041 & 0.04969 & 0.04957 & 0.11059 & 0.11132 & 1.002 & 0.996 (0.015) \\
 & 0.10 & 0.7066 & 0.04978 & 0.04992 & 0.11346 & 0.11172 & 0.997 & 0.990 (0.023) \\
 & 0.20 & 0.7368 & 0.05414 & 0.05428 & 0.11854 & 0.11649 & 0.997 & 0.990 (0.037) \\
\bottomrule
\end{tabular}
\end{table}

With the tangential term of Eq.~(\ref{eq:S-sagitta2}) included, ratio2 is $0.998$--$1.007$ in every
noiseless cell, within one SE. The uncorrected ratio runs from $1.058$ at $n_{\mathrm{ref}} = 500$
to $1.002$ at $16000$, shrinking with $\ell/R$, so the excess is that next-order term and not a bias.
The $n_{\mathrm{ref}}$ block gives $\delta\propto n_{\mathrm{ref}}^{-1}$ (a $31$-fold fall over a
$32$-fold range) and jitter $\propto n_{\mathrm{ref}}^{-1/2}$, within $5\%$ of $\ell/\sqrt k$ in every
row.

The $R$ block repeats one row at every radius up to scale: $\delta/R = 0.0102$ and both ratios are
identical. On a closed sphere at fixed $n_{\mathrm{ref}}$ and $k$ the whole configuration scales with
$R$, so $\delta\propto R$ is dimensional. A flatter reference manifold is worse only against a fixed
absolute probe magnitude or noise; on a patch of fixed area $\ell$ would not grow with $R$ and
$\delta$ would fall. Cap angle acts through sampled area: ratio2 is $1.000$ across a sevenfold range
in $\ell$. All eight gain offsets give the default cell to every printed digit, because radial
probes on an origin-centered sphere leave the $k$-NN ordering exactly unchanged. Reference noise up
to $\sigma = 0.2$ inflates the measured $\ell$ but leaves ratio2 at $0.990$--$1.000$, within its SE:
noise is a fluctuation, not a bias channel.

\subsection{Anisotropy and the anchor}

Proposition~2 assumes isotropic noise. This sweep tests it directly.

\begin{table}[h]
\centering
\small
\caption{Proposition~2 under anisotropy, $\Sigma = \sigma^2(I+\kappa\,ww^\top)$. Anchor displacement
measured directly at a single base point on a sphere, $R = 5$, $F = 30$, $n_{\mathrm{ref}} = 4000$,
$k = 40$, $\sigma = 0.05$, 2000 trials per cell, the same draws for every role and $\kappa$. Ratio
columns as in Table~\ref{tab:S-params}.}
\label{tab:S-anisoanchor}
\begin{tabular}{lrrrrrrr}
\toprule
$w$ role & $\kappa$ & $\sigma_w$ & $\ell$ & $\delta$ & $\ell^2/2R$ & ratio & ratio2 (SE) \\
\midrule
isotropic & 0 & 0.050 & 0.7130 & 0.05118 & 0.05084 & 1.007 & 1.000 (0.005) \\
\midrule
radial & 25 & 0.255 & 0.7164 & 0.03871 & 0.05132 & 0.754 & 0.749 (0.017) \\
 & 100 & 0.502 & 0.7442 & 0.02399 & 0.05538 & 0.433 & 0.430 (0.024) \\
 & 400 & 1.001 & 0.8437 & 0.01720 & 0.07118 & 0.242 & 0.239 (0.026) \\
\midrule
tangent & 25 & 0.255 & 0.7119 & 0.05746 & 0.05068 & 1.134 & 1.126 (0.006) \\
 & 100 & 0.502 & 0.7097 & 0.07508 & 0.05037 & 1.491 & 1.481 (0.007) \\
 & 400 & 1.001 & 0.7132 & 0.13234 & 0.05086 & 2.602 & 2.585 (0.011) \\
\midrule
ambient & 25 & 0.255 & 0.7159 & 0.05160 & 0.05125 & 1.007 & 1.000 (0.005) \\
 & 100 & 0.502 & 0.7439 & 0.05584 & 0.05534 & 1.009 & 1.002 (0.005) \\
 & 400 & 1.001 & 0.8417 & 0.07162 & 0.07085 & 1.011 & 1.002 (0.004) \\
\bottomrule
\end{tabular}
\end{table}

Isotropic noise leaves Proposition~2 exact to its next order (ratio2 $1.000\pm0.005$). Radial
anisotropy erases the sagitta: $\delta$ falls threefold, from $0.0512$ to $0.0172$, \emph{while
$\ell$ rises}, so the effect is in the displacement itself and not in the convention for measuring
$\ell$; ratio2 falls to $0.749$, $0.430$ and $0.239$. Tangential anisotropy inflates it $2.6$-fold
with $\ell$ flat: tangentially displaced points from further around the sphere become selectable, so
the selected set spans more arc than the measured $\ell$ reports. Ambient anisotropy is inert
(ratio2 $1.000$--$1.002$) even as it inflates $\ell$ by $18\%$. Proposition~2's insensitivity to the
noise scale, established under isotropy, does not extend to noise concentrated along the radial or
tangential directions.

\clearpage
\section{Gate 4 --- anisotropic noise}
\label{sec:S-aniso}

Anisotropy imposed as $\Sigma = \sigma^2(I + \kappa\,ww^\top)$ on an origin-centered spherical
cap small enough that the local frame is near-constant, with $w$ placed in each of three
geometric roles. The imposed change is $(G,T,O) = (0.5,\,0,\,0.5)$ throughout, so departures
from those values are the effect of the noise geometry alone.

\begin{table}[h]
\centering
\small
\caption{Anisotropic noise by direction. Cap half-angle $0.30$, $R = 5$, $F = 30$, $d = 2$,
$n_{\mathrm{ref}} = 1500$, $k = 100$, 300 probes, $\sigma = 0.01$, imposed $\nu = 1$,
$O_\star = 0.5$; medians, the same draws in every cell. $\sigma_w = \sigma\sqrt{1+\kappa}$ is the
noise scale along $w$ and $\ell$ the tangential neighborhood extent.}
\label{tab:S-aniso}
\begin{tabular}{lrrrrrrr}
\toprule
$w$ role & $\kappa$ & $\sigma_w/\ell$ & $G$ & $T$ & $O$ & alignment & \% degen. \\
\midrule
gain axis & 0 & 0.035 & 0.4987 & 0.0014 & 0.4925 & 1.0000 & 0.0 \\
 & 100 & 0.302 & 0.4732 & 0.0017 & 0.5246 & 0.9960 & 0.0 \\
 & 400 & 0.529 & 0.4095 & 0.0038 & 0.5849 & 0.9748 & 0.0 \\
 & 900 & 0.729 & 0.3355 & 0.0086 & 0.6519 & 0.9183 & 0.0 \\
 & 1600 & 0.924 & 0.2681 & 0.0156 & 0.7125 & 0.8052 & 0.3 \\
 & 6400 & 1.557 & \textbf{0.1876} & 0.0168 & \textbf{0.7992} & \textbf{0.4782} & 3.7 \\
\midrule
tangent & 0 & 0.035 & 0.4987 & 0.0014 & 0.4925 & 1.0000 & 0.0 \\
 & 100 & 0.351 & 0.4966 & 0.0017 & 0.4935 & 0.9999 & 0.0 \\
 & 400 & 0.685 & 0.4919 & 0.0018 & 0.4953 & 0.9993 & 0.0 \\
 & 900 & 0.994 & 0.4900 & 0.0022 & 0.4942 & 0.9980 & 0.0 \\
 & 1600 & 1.293 & 0.4864 & 0.0025 & 0.4952 & 0.9969 & 0.0 \\
 & 6400 & 2.272 & 0.4760 & 0.0046 & 0.4924 & 0.9908 & 0.0 \\
\midrule
ambient & 0 & 0.035 & 0.4987 & 0.0014 & 0.4925 & 1.0000 & 0.0 \\
 & 100 & 0.347 & 0.4968 & 0.0014 & 0.4946 & 1.0000 & 0.0 \\
 & 400 & 0.649 & 0.4980 & 0.0014 & 0.4982 & 1.0000 & 0.0 \\
 & 900 & 0.890 & 0.4998 & 0.0023 & 0.4963 & 1.0000 & 0.0 \\
 & 1600 & 1.080 & 0.5005 & 0.0036 & 0.4924 & 0.9998 & 0.0 \\
 & 6400 & 1.679 & 0.5011 & 0.0067 & 0.4856 & 0.9972 & 0.0 \\
\bottomrule
\end{tabular}
\end{table}

Tangential anisotropy leaves $O$ within $0.003$ and $G$ within $0.023$ of their isotropic values out to
$\sigma_w/\ell = 2.272$, with $T\le0.005$; ambient anisotropy leaves both within $0.007$ out to
$\sigma_w/\ell = 1.679$. Along the gain axis the same ratios produce a monotone transfer of energy from
$G$ to $O$ with no threshold. At $\sigma_w/\ell = 0.529$ the recovered $G$ has already fallen by
$18\%$; at $0.924$ it is $0.268$ with $O = 0.713$; at $1.557$ it retains $38\%$ of its isotropic value, $O$ has
risen to $0.799$, the alignment is $0.478$, and $3.7\%$ of probes are flagged degenerate.

\begin{sloppypar}
\paragraph{Why the directional quadratic forms are not the correction.} Substituting
$\hat n_g^\top\Sigma\hat n_g$, $\operatorname{tr}(P_{\mathrm{res}}\Sigma P_{\mathrm{res}})$ and
$\operatorname{tr}(V_d^\top\Sigma V_d)$ into the noise-augmented form of Proposition~1 predicts, at
$\kappa = 6400$, $O \to 0.30$ for tangential anisotropy and $O \to 0.68$ for ambient. Neither
occurs: the measured values are $0.492$ and $0.486$, against the isotropic $0.492$. The substitution treats
the noise as added to the probe against a fixed anchor and frame, whereas both are estimated from
the same noisy cloud and absorb the shared component.
\end{sloppypar}

\paragraph{The alignment tolerance does not protect.} At $\kappa = 1600$ the alignment is still
$0.81$ --- far above any threshold one would set to flag a non-existent gain axis --- while $G$ has
already lost $46\%$ of its isotropic value. Alignment must be read as a graded quantity for this
failure mode to be visible at all.

\clearpage
\subsection{Anisotropy and the frame (gates 2 and 6)}

\begin{table}[h]
\centering
\small
\caption{Gates 2 and 6 under anisotropy. Spherical cap of half-angle $0.25$, $R = 5$,
$F = 30$, $n_{\mathrm{ref}} = 4000$, $k = 40$, $\sigma = 0.02$, 200 probes; medians, the same draws
in every cell. $\tau^2$ is the tangential energy of $r$; the ratio is $\tau^2$ over $\ell^2/k$
measured on the same neighborhoods.}
\label{tab:S-anisoframe}
\begin{tabular}{lrrrrrrr}
\toprule
$w$ role & $\kappa$ & $\sigma_w$ & $\|V_d^\top\hat n\|^2$ & vs.\ isotropic & $\tau^2$ & $\ell^2/k$ & ratio \\
\midrule
isotropic & 0 & 0.020 & $2.175\times10^{-3}$ & 1.00 & 0.00031 & 0.00038 & 0.817 \\
\midrule
radial & 25 & 0.102 & $9.850\times10^{-2}$ & 45.3 & 0.00041 & 0.00045 & 0.891 \\
 & 100 & 0.201 & $5.381\times10^{-1}$ & 247.4 & 0.00066 & 0.00059 & 1.127 \\
 & 400 & 0.400 & $7.225\times10^{-1}$ & 332.2 & 0.00086 & 0.00085 & 1.020 \\
\midrule
tangent & 25 & 0.102 & $3.587\times10^{-3}$ & 1.65 & 0.00038 & 0.00039 & 0.961 \\
 & 100 & 0.201 & $5.992\times10^{-3}$ & 2.75 & 0.00038 & 0.00041 & 0.918 \\
 & 400 & 0.400 & $1.222\times10^{-2}$ & 5.62 & 0.00048 & 0.00045 & 1.083 \\
\midrule
ambient & 25 & 0.102 & $1.933\times10^{-3}$ & 0.89 & 0.00047 & 0.00047 & 1.009 \\
 & 100 & 0.201 & $1.340\times10^{-3}$ & 0.62 & 0.00085 & 0.00061 & 1.395 \\
 & 400 & 0.400 & $9.675\times10^{-4}$ & 0.44 & 0.00102 & 0.00084 & 1.216 \\
\bottomrule
\end{tabular}
\end{table}

The gate-2 rotation term is strongly direction-sensitive. Radial anisotropy raises it $332$-fold, to
$7.225\times10^{-1}$, so the normal direction lies mostly \emph{inside} the estimated tangent space. Tangential
anisotropy raises it $5.6$-fold. Ambient anisotropy lowers it to $0.44$ of the isotropic value,
since a captured ambient direction occupies a basis slot that would otherwise tilt toward the
normal. The gate-6 term $\tau^2$ also moves with the anisotropy --- $2.8$-fold radially,
$1.5$-fold tangentially, $3.3$-fold in the ambient role --- but it moves with the $\ell^2/k$ of the
same neighborhoods: the ratio is $0.82$--$1.13$ in the radial and tangential roles, and reaches
$1.40$ only where ambient noise competes for a tangent slot. The observable conditioning check of
gate~6 therefore survives anisotropy, provided $\ell$ is measured on the data rather than assumed.

\clearpage
\section{Gate 5 --- where the anchor displacement is booked}
\label{sec:S-gate5}

Table~\ref{tab:S-anchorbias} is the pure on-manifold (geodesic) sweep on spheres of five radii and
four neighborhood sizes. It isolates the absolute novel energy from the $O$ fraction, and compares
the radial displacement with the sagitta.

\begin{table}[h]
\centering
\footnotesize
\setlength{\tabcolsep}{4pt}
\caption{Pure on-manifold (geodesic) sweep on spheres of five radii and four neighborhood sizes,
$F=30$, $n_{\mathrm{ref}}=6000$, 300 probes/cell, arc length $0.30$, ambient noise $\sigma=0.01$
(noise floor $\sqrt{F-3}\,\sigma=0.0520$). Every probe is a geodesic step, so the imposed $(G,T,O)$
is $(0,1,0)$ exactly and any radial or off-manifold energy in $r$ is the anchor's, not the
probe's. Ratio columns, left to right: $\|r_{\mathrm{rad}}\|$ over $\ell^2/2R$; the same over
$(\ell^2/2R)(1+\ell^2/3R^2)$, with the SE of the median; $\|r_{\mathrm{novel}}\|$ over the noise
floor.}
\label{tab:S-anchorbias}
\begin{tabular}{rrrrrrrrrr}
\toprule
$R$ & $k$ & $\ell$ & $\|r\|$ & $\|r_{\mathrm{rad}}\|$ & $\ell^2/2R$ & $\|r_{\mathrm{novel}}\|$ & ratio & ratio2 (SE) & ratio (novel) \\
\midrule
3.0 & 20 & 0.247 & 0.0732 & 0.0107 & 0.0102 & 0.0525 & 1.053 & 1.050 (0.060) & 1.011 \\
3.0 & 40 & 0.346 & 0.0735 & 0.0195 & 0.0200 & 0.0523 & 0.973 & 0.969 (0.040) & 1.007 \\
3.0 & 80 & 0.492 & 0.0820 & 0.0405 & 0.0403 & 0.0520 & 1.006 & 0.997 (0.020) & 1.001 \\
3.0 & 160 & 0.690 & 0.1090 & 0.0799 & 0.0793 & 0.0511 & 1.007 & 0.990 (0.011) & 0.983 \\
5.0 & 20 & 0.413 & 0.0973 & 0.0159 & 0.0171 & 0.0525 & 0.929 & 0.927 (0.041) & 1.010 \\
5.0 & 40 & 0.576 & 0.0954 & 0.0323 & 0.0332 & 0.0519 & 0.972 & 0.968 (0.025) & 0.999 \\
5.0 & 80 & 0.811 & 0.1128 & 0.0657 & 0.0657 & 0.0519 & 1.000 & 0.991 (0.014) & 0.998 \\
5.0 & 160 & 1.149 & 0.1641 & 0.1341 & 0.1321 & 0.0516 & 1.015 & 0.997 (0.008) & 0.992 \\
8.0 & 20 & 0.662 & 0.1411 & 0.0268 & 0.0274 & 0.0523 & 0.981 & 0.979 (0.032) & 1.007 \\
8.0 & 40 & 0.927 & 0.1414 & 0.0522 & 0.0538 & 0.0521 & 0.972 & 0.967 (0.020) & 1.002 \\
8.0 & 80 & 1.301 & 0.1682 & 0.1070 & 0.1057 & 0.0511 & 1.012 & 1.003 (0.012) & 0.983 \\
8.0 & 160 & 1.834 & 0.2576 & 0.2150 & 0.2102 & 0.0508 & 1.023 & 1.005 (0.007) & 0.978 \\
12.0 & 20 & 0.975 & 0.1954 & 0.0393 & 0.0396 & 0.0528 & 0.991 & 0.989 (0.023) & 1.015 \\
12.0 & 40 & 1.393 & 0.2089 & 0.0815 & 0.0808 & 0.0514 & 1.008 & 1.003 (0.016) & 0.989 \\
12.0 & 80 & 1.967 & 0.2455 & 0.1605 & 0.1613 & 0.0515 & 0.995 & 0.987 (0.010) & 0.992 \\
12.0 & 160 & 2.739 & 0.3708 & 0.3172 & 0.3126 & 0.0505 & 1.015 & 0.997 (0.006) & 0.972 \\
20.0 & 20 & 1.646 & 0.2862 & 0.0662 & 0.0677 & 0.0533 & 0.977 & 0.975 (0.023) & 1.026 \\
20.0 & 40 & 2.322 & 0.3471 & 0.1349 & 0.1348 & 0.0517 & 1.001 & 0.996 (0.013) & 0.994 \\
20.0 & 80 & 3.256 & 0.4233 & 0.2650 & 0.2650 & 0.0515 & 1.000 & 0.991 (0.010) & 0.992 \\
20.0 & 160 & 4.605 & 0.6241 & 0.5376 & 0.5301 & 0.0521 & 1.014 & 0.996 (0.007) & 1.002 \\
\bottomrule
\end{tabular}
\end{table}

Against Eq.~(\ref{eq:S-sagitta2}) the radial displacement is recovered in every cell: ratio2 averages
$0.984$, $0.981$, $0.994$ and $0.997$ at $k = 20$, $40$, $80$, $160$, and no cell departs from $1$ by
more than $1.8$ SE. The scatter at small $k$ is sampling error, since the bias is smallest there and
Proposition~3's jitter is largest relative to it. At $k = 160$, where $\ell/R\approx0.23$ at every
radius, the uncorrected ratio runs $1.007$--$1.023$; that excess is the tangent-projection term
$\ell^2/3R^2\approx1.8\%$.

The result the table exists to establish is $\|r_{\mathrm{novel}}\|$: $0.0505$--$0.0533$ against a
noise floor of $0.0520$, with no trend in either $R$ or $k$. Curvature on a sphere produces real,
radial anchor displacement but no off-manifold energy at all; the entire sagitta is absorbed by the
gain axis, exactly as gate~5's booking rule requires when $|c|=1$. The fraction $O$ nonetheless
varies from cell to cell, entirely through its denominator: $\|r\|^2$ is the sagitta, the noise, and
the tangential jitter $\tau^2$, which carries roughly $25$--$90\%$ of $\|r\|^2$ (from the medians of the
components). This is the noise floor the
Clifford torus below is measured against.

\subsection{The $|c| < 1$ case: Clifford torus}

For a $d$-manifold whose linear span is $(d+1)$-dimensional the normal space inside the span is
one-dimensional, forcing $|c| = 1$; this covers the sphere, all quadrics in three-space, the torus
of revolution and the ellipsoid. Reaching $|c| < 1$ requires a linear span of dimension at least
$d+2$. The Clifford torus $p(u,v) = (R_1\cos u, R_1\sin u, R_2\cos v, R_2\sin v)$ is the simplest
such geometry with alignment exactly $1$ (so gate~4 is clean). It is intrinsically flat (so probes
can be stepped exactly along it), and gives $c = -(1/\rho)\|\vec H\|^{-1}$,
$\rho = \sqrt{R_1^2+R_2^2}$.

\begin{table}[h]
\centering
\small
\setlength{\tabcolsep}{4pt}
\caption{Clifford torus, $R_1 = 1$, $F = 30$, $n_{\mathrm{ref}} = 8000$, 300 probes,
$\sigma = 0.001$ (noise floor $\sqrt{F-3}\,\sigma = 0.0052$), arc $0.15$; medians. Upper block:
curvature alignment at $k = 40$, $R_2 = 1, 2, 3, 5$. Lower block: $k$ sweep at $R_2 = 3$. $\eta$ and
$\delta$ use the areal $\ell^2$. Ratio columns: $\|r_{\mathrm{novel}}\|$ over $\eta$, and over
$\sqrt{\eta^2+\text{floor}^2}$, since the noise adds in quadrature.}
\label{tab:S-clifford}
\begin{tabular}{lrrrrrrrr}
\toprule
$R_1{:}R_2$ & $c$ & $a$ & $\|r_{\mathrm{rad}}\|$ & $\delta$ pred. & $\|r_{\mathrm{novel}}\|$ & $\eta$ pred. & ratio & ratio (quad.) \\
\midrule
1:1 & $-1.000$ & 1.000 & 0.01097 & 0.01111 & 0.00530 & 0 (exactly) & --- & 1.019 \\
1:2 & $-0.800$ & 1.000 & 0.01379 & 0.01405 & 0.01197 & 0.01054 & 1.136 & 1.019 \\
1:3 & $-0.600$ & 1.000 & 0.01463 & 0.01490 & 0.02028 & 0.01987 & 1.021 & 0.987 \\
1:5 & $-0.385$ & 1.000 & 0.01501 & 0.01540 & 0.03758 & 0.03697 & 1.016 & 1.007 \\
\midrule
\multicolumn{9}{l}{\emph{$k$ sweep at $R_1{:}R_2 = 1{:}3$; $\ell^2$ areal. On the sphere this quantity was flat.}} \\
$k$ & $\ell^2$ & & & & $\|r_{\mathrm{novel}}\|$ & $\eta$ pred. & ratio & ratio (quad.) \\
\midrule
20 & 0.04712 &  &  &  & 0.01113 & 0.00993 & 1.121 & 0.993 \\
40 & 0.09425 &  &  &  & 0.02028 & 0.01987 & 1.021 & 0.987 \\
80 & 0.18850 &  &  &  & 0.04080 & 0.03974 & 1.027 & 1.018 \\
160 & 0.37699 &  &  &  & 0.08118 & 0.07948 & 1.021 & 1.019 \\
\bottomrule
\end{tabular}
\end{table}

At $|c| = 1$ the novel energy sits at the noise floor ($1.019$), reproducing
Table~\ref{tab:S-anchorbias}. As the torus is made asymmetric it rises to track $\eta$, and in the
$k$ sweep it doubles with each doubling of $k$, as $\eta\propto\ell^2\propto k$ requires --- against
the sphere, where the same quantity was flat in $k$ at the noise floor. With the floor added in
quadrature the ratio is $0.987$--$1.019$ in every row. The radial component tracks $\delta$ to
within $3\%$. The $k$ sweep should be read with the areal $\ell^2$ in mind: it is low by the $k$-NN
boundary factor $(k+1)/k$ (Table~\ref{tab:S-torus}) and, because $R_1 = 1$, at large $k$ it also
exceeds the tangentially projected extent. The two effects partly offset across the sweep, so the
$c$ sweep at fixed $k = 40$ is the cleaner test of the booking rule.

\subsection{Varying second fundamental form: circle $\times$ ellipse}

\begin{sloppypar}\noindent
The Clifford torus is intrinsically flat, so $\nabla\mathrm{I\!I} = 0$ and Proposition~2's
neglected term is structurally zero there. Replacing the second circle by an ellipse,
$p(u,v) = (R_1\cos u,\,R_1\sin u,\,A\cos v,\,B\sin v)$, keeps the linear span at $d+2$ while making
the normal curvature $\kappa_e = AB/q^3$, $q = \sqrt{A^2\sin^2 v + B^2\cos^2 v}$, vary around the
ellipse --- and with it $\|\vec H\|$, the alignment, and $c$.
\end{sloppypar}

\begin{table}[h]
\centering
\small
\caption{Circle $\times$ ellipse, $R_1 = 1$, $A = 1$, $B = 3$, $F = 30$, $d = 2$, $k = 60$,
1500 probes, noiseless; probes with $|c| < 0.95$. $\eta$ is predicted \emph{per probe} from that
probe's own $v$, with the areal $\ell^2$. The middle columns are medians over probes; the ratio is
the median of the per-probe ratio, not the ratio of those medians.}
\label{tab:S-ellipse}
\begin{tabular}{rrrrr}
\toprule
$n_{\mathrm{ref}}$ & $\ell^2$ & $\|r_{\mathrm{novel}}\|$ & $\eta$ pred. & ratio \\
\midrule
5000 & 0.16038 & 0.02545 & 0.02540 & 1.005 \\
10000 & 0.08019 & 0.01254 & 0.01270 & 1.004 \\
20000 & 0.04009 & 0.00616 & 0.00635 & 0.979 \\
40000 & 0.02005 & 0.00310 & 0.00318 & 0.987 \\
\bottomrule
\end{tabular}
\end{table}

The ratio is $0.979$--$1.005$ across an eightfold range in $\ell^2$, with no systematic drift, so
the booking rule is not an artefact of the Clifford torus's flatness. At the latitudes where
$|c|\to1$ (excluded above) the prediction $\eta\to0$.

\paragraph{A binning artefact worth avoiding.} Because $\eta$ varies strongly with $v$, evaluating
the prediction at a bin's median coordinate rather than per probe biases the ratio by tens of
percent --- an artefact of aggregating before predicting, not a property of the rule. Predict per
probe and aggregate afterwards.

\clearpage
\section{Gate 6 --- recovery of the off-manifold fraction}
\label{sec:S-gate6}

\subsection{The first-order form and its tangential term}

On a noiseless origin-centered sphere $\eta = 0$ identically, so the first-order form of the main
text reduces to a $\delta$ term plus a $\tau^2$ term. Table~\ref{tab:S-cor3} compares the measured
error with that form, with and without its tangential term, with the exact Proposition~1 form at
the predicted $\delta$ and $\tau^2$, and with an anchor model: for each probe the anchor is placed at
$(R-\delta)\hat e + \bar u$, with $\bar u\sim N(0,\tfrac{\tau^2}{2}I_2)$ in the tangent plane
(Propositions~2--3 and nothing else), and $O$ is computed with the exact sphere frame there.

\begin{table}[h]
\centering
\footnotesize
\setlength{\tabcolsep}{3.5pt}
\caption{Sphere, $R = 5$, $F = 30$, $d = 2$, noiseless, $\nu = 1$, $O_\star = 0.5$, 400 probes
$\times$ 5 seeds. $\delta = \ell^2/2R$ and $\tau^2 = \ell^2/k$ from the measured $\ell$; the exact
form and the model use $\delta(1+\ell^2/3R^2)$. $O-O_\star$ is the median and the mean over probes,
averaged over seeds. The first four ratio columns divide the \emph{mean} error by: the first-order
$\delta$ term alone; the first-order form with its tangential term; the exact form; the model's
mean (sd across seeds). The last divides the median error by the model's median.}
\label{tab:S-cor3}
\begin{tabular}{rrrrrrrrrrr}
\toprule
 & & & & \multicolumn{2}{c}{$O-O_\star$} & \multicolumn{4}{c}{ratio: mean error over} & ratio \\
\cmidrule(lr){5-6}\cmidrule(lr){7-10}
$n_{\mathrm{ref}}$ & $k$ & $\delta$ & $\tau^2$ & median & mean & $\delta$ only & 1st order & exact & model (sd) & med./model \\
\midrule
4000 & 40 & 0.05010 & 0.01252 & $-0.039236$ & $-0.040127$ & 1.133 & 0.963 & 1.009 & 1.012 (0.025) & 1.032 \\
8000 & 40 & 0.02503 & 0.00626 & $-0.020023$ & $-0.020568$ & 1.162 & 0.988 & 1.011 & 1.009 (0.011) & 1.027 \\
16000 & 40 & 0.01266 & 0.00317 & $-0.010216$ & $-0.010509$ & 1.174 & 0.997 & 1.009 & 1.009 (0.013) & 1.024 \\
16000 & 20 & 0.00648 & 0.00324 & $-0.005793$ & $-0.006201$ & 1.353 & 1.000 & 1.008 & 1.007 (0.017) & 1.016 \\
\bottomrule
\end{tabular}
\end{table}

Compared mean to mean, the first-order form with its tangential term converges to the measured
error as $\delta$ shrinks: $0.963$, $0.988$, $0.997$, $1.000$ down the table. Without the tangential term the
ratio stays at $1.133$--$1.174$ at $k = 40$ and reaches $1.353$ at $k = 20$. The $n_{\mathrm{ref}}$ rows cannot
by themselves show that the omitted term is the tangential one, because $\tau^2/\delta = 0.25$ in all
three --- both scale as $1/n_{\mathrm{ref}}$ at fixed $k$ --- so the evidence is the $k = 20$ row,
where $\tau^2/\delta$ doubles to $0.50$ and so does the shortfall. The exact form and the anchor
model account for the mean error to about $1\%$ in every cell ($1.008$--$1.011$ and $1.007$--$1.012$): Propositions~2
and~3 are sufficient. Medians compared with the model's medians agree less well ($1.016$--$1.032$), because
the model's per-probe jitter distribution only approximates the real one.

\subsection{The full recovery sweep}

Table~\ref{tab:S-offman} is the full sweep underlying the gate~6 discussion: recovered $\widehat O$
against imposed $O_\star$, swept over neighborhood size, gain sign and imposed magnitude, with a
final block on a finer $O_\star$ grid in the one ill-conditioned condition. The oracle arm supplies
the gate~2 estimator bound quoted in the main text.

$O_\star = 0$ occurs throughout, so no ratio column is given --- a ratio against zero is not a
number. Read the columns as comparisons instead, each isolating a different source of error.

\begin{itemize}
\item $O_\star$ is the imposed, ground-truth off-manifold fraction.
\item $\widehat O$ is what the full estimator reports, with anchor, frame and gain axis estimated
from a finite, noisy neighborhood --- the number an analysis of real data would produce.
\item $O_{\mathrm{orac}}$ replaces the estimated frame with the generating manifold's exact geometry
at the same anchor. $\widehat O$ against $O_{\mathrm{orac}}$ isolates frame-estimation error, gate~2's
contribution.
\item $O_{\mathrm{pred}} = (b^2+(F-3)\sigma^2)/[(g+\delta)^2+b^2+(F-3)\sigma^2+\ell^2/k+\sigma^2]$,
with $g = \pm\nu\sqrt{1-O_\star}$, $b = \nu\sqrt{O_\star}$ and
$\delta = (\ell^2/2R)(1+\ell^2/3R^2)$, is the closed form evaluated at the \emph{mean} jitter.
$O_{\mathrm{pred}}$ against $O_\star$ is anchor bias with a perfect estimator.
\item $O_{\mathrm{model}}$ is the median of the per-probe anchor model of Table~\ref{tab:S-cor3},
with anchor noise $\sigma/\sqrt k$ added. Unlike $O_{\mathrm{pred}}$ it has the full per-probe
distribution, so it is the like-for-like comparison for a median.
\item $C_6 = \mathrm{median}\,(k\|r\|^2/\ell^2)$ is the observable conditioning index of gate~6:
$\|r\|^2 \approx (g+\delta)^2 + b^2 + \ell^2/k$ plus noise, so $C_6 - 1\gg1$ is Eq.~(25) of the
main text.
\end{itemize}

\begingroup
\footnotesize
\setlength{\tabcolsep}{4.5pt}
\begin{longtable}{rrrrrrrrrrr}
\caption{Recovered $\widehat O$ against imposed $O_\star$. Sphere, $F=30$, $R=5$, $d=2$,
$n_{\mathrm{ref}} = 4000$, $\sigma = 0.01$ in all $F$ dimensions, 300 probes per cell; medians.
Swept over $k\in\{40,160\}$, gain sign, and imposed displacement magnitude $\nu\in\{0.30,1.00\}$.
$\ell$ is the median tangential extent, $\delta = (\ell^2/2R)(1+\ell^2/3R^2)$.
$G_{\mathrm{rec}}+T_{\mathrm{rec}}+\widehat O = 1$ per probe; medians need not sum to one. The final
block repeats one condition on a finer $O_\star$ grid with independent draws.}
\label{tab:S-offman} \\
\toprule
$O_\star$ & $\widehat O$ & $O_{\mathrm{orac}}$ & $O_{\mathrm{pred}}$ & $O_{\mathrm{model}}$ & $G_{\mathrm{rec}}$ & $T_{\mathrm{rec}}$ & $\|r\|$ & $\ell$ & $\delta$ & $C_6$ \\
\midrule
\endfirsthead
\multicolumn{11}{l}{\emph{Table~\ref{tab:S-offman} continued}} \\
\toprule
$O_\star$ & $\widehat O$ & $O_{\mathrm{orac}}$ & $O_{\mathrm{pred}}$ & $O_{\mathrm{model}}$ & $G_{\mathrm{rec}}$ & $T_{\mathrm{rec}}$ & $\|r\|$ & $\ell$ & $\delta$ & $C_6$ \\
\midrule
\endhead
\bottomrule
\multicolumn{11}{r}{\emph{continued on next page}} \\
\endfoot
\bottomrule
\endlastfoot
\multicolumn{11}{l}{\emph{gain amplifying, $k=40$, $\nu=0.30$}} \\
0.00 & 0.020 & 0.020 & 0.020 & 0.020 & 0.905 & 0.076 & 0.371 & 0.706 & 0.050 & 11.0 \\
0.25 & 0.190 & 0.190 & 0.188 & 0.191 & 0.738 & 0.068 & 0.363 & 0.710 & 0.051 & 10.6 \\
0.50 & 0.374 & 0.374 & 0.371 & 0.372 & 0.541 & 0.071 & 0.357 & 0.701 & 0.049 & 10.5 \\
0.75 & 0.580 & 0.580 & 0.573 & 0.580 & 0.329 & 0.076 & 0.350 & 0.703 & 0.050 & 9.9 \\
1.00 & 0.871 & 0.872 & 0.860 & 0.891 & 0.023 & 0.104 & 0.327 & 0.706 & 0.050 & 8.8 \\
\multicolumn{11}{l}{\emph{gain amplifying, $k=40$, $\nu=1.00$}} \\
0.00 & 0.003 & 0.002 & 0.002 & 0.003 & 0.988 & 0.010 & 1.058 & 0.709 & 0.051 & 89.1 \\
0.25 & 0.226 & 0.227 & 0.229 & 0.227 & 0.762 & 0.009 & 1.049 & 0.709 & 0.051 & 87.7 \\
0.50 & 0.461 & 0.461 & 0.463 & 0.463 & 0.525 & 0.011 & 1.045 & 0.699 & 0.049 & 89.0 \\
0.75 & 0.705 & 0.705 & 0.705 & 0.706 & 0.282 & 0.010 & 1.034 & 0.706 & 0.050 & 85.9 \\
1.00 & 0.987 & 0.987 & 0.985 & 0.989 & 0.003 & 0.010 & 1.011 & 0.703 & 0.050 & 82.4 \\
\multicolumn{11}{l}{\emph{gain amplifying, $k=160$, $\nu=0.30$}} \\
0.00 & 0.010 & 0.010 & 0.010 & 0.010 & 0.951 & 0.038 & 0.517 & 1.405 & 0.203 & 21.8 \\
0.25 & 0.099 & 0.099 & 0.101 & 0.101 & 0.855 & 0.042 & 0.503 & 1.401 & 0.201 & 20.5 \\
0.50 & 0.205 & 0.205 & 0.207 & 0.209 & 0.742 & 0.049 & 0.484 & 1.398 & 0.200 & 19.0 \\
0.75 & 0.335 & 0.335 & 0.339 & 0.341 & 0.598 & 0.056 & 0.457 & 1.405 & 0.203 & 16.9 \\
1.00 & 0.642 & 0.642 & 0.639 & 0.651 & 0.270 & 0.068 & 0.381 & 1.397 & 0.200 & 11.9 \\
\multicolumn{11}{l}{\emph{gain amplifying, $k=160$, $\nu=1.00$}} \\
0.00 & 0.002 & 0.002 & 0.002 & 0.002 & 0.990 & 0.008 & 1.209 & 1.394 & 0.199 & 120.1 \\
0.25 & 0.179 & 0.179 & 0.181 & 0.180 & 0.811 & 0.008 & 1.185 & 1.393 & 0.199 & 116.0 \\
0.50 & 0.376 & 0.376 & 0.376 & 0.377 & 0.613 & 0.008 & 1.156 & 1.392 & 0.199 & 110.5 \\
0.75 & 0.599 & 0.599 & 0.599 & 0.599 & 0.389 & 0.008 & 1.122 & 1.400 & 0.201 & 102.4 \\
1.00 & 0.951 & 0.951 & 0.950 & 0.953 & 0.038 & 0.009 & 1.029 & 1.399 & 0.201 & 86.4 \\
\multicolumn{11}{l}{\emph{gain suppressive, $k=40$, $\nu=0.30$}} \\
0.00 & 0.034 & 0.034 & 0.035 & 0.034 & 0.839 & 0.122 & 0.274 & 0.713 & 0.051 & 6.1 \\
0.25 & 0.312 & 0.312 & 0.307 & 0.319 & 0.559 & 0.114 & 0.282 & 0.694 & 0.048 & 6.9 \\
0.50 & 0.565 & 0.565 & 0.551 & 0.576 & 0.309 & 0.111 & 0.289 & 0.703 & 0.050 & 7.0 \\
0.75 & 0.770 & 0.770 & 0.757 & 0.792 & 0.108 & 0.123 & 0.304 & 0.703 & 0.050 & 7.7 \\
1.00 & 0.874 & 0.875 & 0.861 & 0.899 & 0.021 & 0.101 & 0.326 & 0.705 & 0.050 & 8.6 \\
\multicolumn{11}{l}{\emph{gain suppressive, $k=40$, $\nu=1.00$}} \\
0.00 & 0.003 & 0.003 & 0.003 & 0.003 & 0.988 & 0.010 & 0.958 & 0.704 & 0.050 & 73.9 \\
0.25 & 0.272 & 0.272 & 0.271 & 0.271 & 0.715 & 0.011 & 0.965 & 0.701 & 0.050 & 76.5 \\
0.50 & 0.530 & 0.530 & 0.531 & 0.531 & 0.456 & 0.010 & 0.973 & 0.703 & 0.050 & 77.3 \\
0.75 & 0.779 & 0.779 & 0.778 & 0.781 & 0.207 & 0.010 & 0.982 & 0.706 & 0.050 & 77.5 \\
1.00 & 0.990 & 0.990 & 0.986 & 0.989 & 0.002 & 0.007 & 1.009 & 0.696 & 0.049 & 83.4 \\
\multicolumn{11}{l}{\emph{gain suppressive, $k=160$, $\nu=0.30$}} \\
0.00 & 0.110 & 0.109 & 0.108 & 0.119 & 0.413 & 0.445 & 0.153 & 1.394 & 0.199 & 1.9 \\
0.25 & 0.645 & 0.645 & 0.614 & 0.666 & 0.088 & 0.245 & 0.199 & 1.396 & 0.200 & 3.3 \\
0.50 & 0.826 & 0.825 & 0.794 & 0.843 & 0.007 & 0.166 & 0.240 & 1.388 & 0.198 & 4.9 \\
0.75 & 0.843 & 0.843 & 0.826 & 0.862 & 0.029 & 0.115 & 0.289 & 1.397 & 0.200 & 6.9 \\
1.00 & 0.648 & 0.648 & 0.638 & 0.656 & 0.273 & 0.063 & 0.381 & 1.398 & 0.200 & 11.9 \\
\multicolumn{11}{l}{\emph{gain suppressive, $k=160$, $\nu=1.00$}} \\
0.00 & 0.004 & 0.004 & 0.004 & 0.004 & 0.984 & 0.012 & 0.812 & 1.402 & 0.202 & 53.4 \\
0.25 & 0.356 & 0.356 & 0.357 & 0.358 & 0.625 & 0.013 & 0.841 & 1.399 & 0.201 & 58.1 \\
0.50 & 0.650 & 0.650 & 0.651 & 0.651 & 0.334 & 0.011 & 0.881 & 1.396 & 0.200 & 64.2 \\
0.75 & 0.880 & 0.881 & 0.880 & 0.881 & 0.106 & 0.010 & 0.924 & 1.392 & 0.199 & 70.4 \\
1.00 & 0.951 & 0.951 & 0.950 & 0.953 & 0.038 & 0.008 & 1.029 & 1.399 & 0.201 & 86.5 \\
\midrule
\multicolumn{11}{l}{\emph{gain suppressive, $k=160$, $\nu=0.30$, fine grid}} \\
0.00 & 0.117 & 0.116 & 0.107 & 0.120 & 0.467 & 0.392 & 0.150 & 1.391 & 0.199 & 1.9 \\
0.05 & 0.260 & 0.261 & 0.254 & 0.270 & 0.343 & 0.371 & 0.164 & 1.387 & 0.197 & 2.2 \\
0.10 & 0.391 & 0.391 & 0.373 & 0.410 & 0.254 & 0.323 & 0.174 & 1.389 & 0.198 & 2.6 \\
0.15 & 0.504 & 0.505 & 0.469 & 0.510 & 0.186 & 0.283 & 0.180 & 1.389 & 0.198 & 2.7 \\
0.20 & 0.597 & 0.597 & 0.548 & 0.604 & 0.133 & 0.245 & 0.191 & 1.392 & 0.199 & 3.0 \\
0.25 & 0.646 & 0.645 & 0.614 & 0.660 & 0.086 & 0.237 & 0.197 & 1.397 & 0.200 & 3.2 \\
0.30 & 0.712 & 0.713 & 0.668 & 0.727 & 0.058 & 0.217 & 0.202 & 1.404 & 0.202 & 3.4 \\
0.35 & 0.734 & 0.735 & 0.709 & 0.747 & 0.037 & 0.207 & 0.219 & 1.395 & 0.200 & 4.0 \\
0.40 & 0.767 & 0.767 & 0.744 & 0.799 & 0.022 & 0.194 & 0.224 & 1.391 & 0.199 & 4.2 \\
0.45 & 0.795 & 0.796 & 0.772 & 0.823 & 0.010 & 0.183 & 0.234 & 1.393 & 0.199 & 4.4 \\
0.50 & 0.830 & 0.831 & 0.792 & 0.846 & 0.005 & 0.161 & 0.242 & 1.405 & 0.203 & 4.8 \\
0.55 & 0.834 & 0.834 & 0.809 & 0.850 & 0.004 & 0.153 & 0.252 & 1.401 & 0.201 & 5.2 \\
0.60 & 0.830 & 0.831 & 0.820 & 0.880 & 0.004 & 0.156 & 0.262 & 1.398 & 0.200 & 5.7 \\
0.65 & 0.844 & 0.843 & 0.824 & 0.874 & 0.009 & 0.144 & 0.273 & 1.405 & 0.203 & 6.1 \\
0.70 & 0.848 & 0.847 & 0.828 & 0.860 & 0.015 & 0.124 & 0.280 & 1.397 & 0.200 & 6.5 \\
0.75 & 0.837 & 0.837 & 0.827 & 0.865 & 0.028 & 0.130 & 0.290 & 1.394 & 0.199 & 6.9 \\
0.80 & 0.837 & 0.838 & 0.820 & 0.856 & 0.047 & 0.103 & 0.300 & 1.393 & 0.199 & 7.3 \\
0.85 & 0.826 & 0.826 & 0.806 & 0.841 & 0.066 & 0.105 & 0.312 & 1.392 & 0.199 & 8.1 \\
0.90 & 0.792 & 0.793 & 0.780 & 0.806 & 0.099 & 0.095 & 0.326 & 1.400 & 0.201 & 8.8 \\
0.95 & 0.753 & 0.752 & 0.745 & 0.770 & 0.148 & 0.085 & 0.344 & 1.399 & 0.201 & 9.7 \\
1.00 & 0.628 & 0.628 & 0.633 & 0.647 & 0.278 & 0.073 & 0.384 & 1.407 & 0.203 & 11.9 \\
\end{longtable}
\endgroup

\noindent\small
Main grid (40 cells): $\max|\widehat O - O_{\mathrm{orac}}| = 0.0009$;
$\max|\widehat O - O_{\mathrm{pred}}| = 0.0316$; $\max|\widehat O - O_{\mathrm{model}}| = 0.0251$
(medians), $0.0228$ (means). For $O_\star < 1$, mean signed $(\widehat O - O_\star)$ is $-0.102$
under amplifying gain and $+0.099$ under suppressive gain, and mean $|\widehat O - O_\star|$ is
$0.107$ and $0.099$. Fine grid (21 cells): $\max|\widehat O - O_{\mathrm{orac}}| = 0.0010$;
$\max|\widehat O - O_{\mathrm{pred}}| = 0.049$; $\max|\widehat O - O_{\mathrm{model}}| = 0.050$
(medians), $0.043$ (means).
\normalsize

\begin{center}
\small
\begin{tabular}{lrrrr}
\toprule
$C_6$ (main grid) & cells & mean $|\widehat O - O_\star|$ & mean $|\widehat O - O_{\mathrm{pred}}|$ & mean $|\widehat O - O_{\mathrm{model}}|$ \\
\midrule
$<2$     &  1 & 0.110 & 0.0018 & 0.0096 \\
$2$--$5$ &  2 & 0.361 & 0.0316 & 0.0190 \\
$5$--$20$ & 15 & 0.155 & 0.0071 & 0.0089 \\
$\ge 20$ & 22 & 0.055 & 0.0010 & 0.0010 \\
\bottomrule
\end{tabular}
\end{center}

The oracle comparison settles the estimator question on its own: across all forty conditions,
recovery with the estimated frame is indistinguishable from recovery with the generating manifold's
geometry at the same anchor. Whatever error remains is not gate~2's.

That error is anchor bias, and it is signed by the gain direction: $\widehat O$ underestimates
$O_\star$ under amplifying gain and overestimates it under suppressive gain, by nearly equal amounts
once the $O_\star = 1$ rows are set aside. Those rows have no gain leg, so the sign of the gain is
undefined there and the amplifying and suppressive rows coincide.

The conditioning index decides how well the closed form predicts. With $C_6\ge20$ the recovered
fraction matches both $O_{\mathrm{pred}}$ and the model to $0.001$ on average; between $5$ and $20$ to
about $0.01$; below $5$ the errors reach several hundredths.

\paragraph{The non-monotone condition.} Under suppressive gain at $k = 160$ and $\nu = 0.30$ the
recovered fraction does not rise monotonically with $O_\star$. On the fine grid it climbs from $0.117$
at $O_\star = 0$ to a plateau of $0.83$--$0.85$ over $O_\star = 0.50$--$0.80$, then falls to $0.628$ at
$O_\star = 1$. Every cell of that block has $C_6$ between $1.9$ and $11.9$. The shape follows from the
gain leg and the sagitta having opposite signs: $g+\delta$ crosses zero at
$O_\star = 1-(\delta/\nu)^2 \approx 0.56$ ($\delta\approx0.20$), where the gain energy in the
denominator vanishes; beyond it the uncancelled sagitta re-enters the denominator, fully so at
$O_\star = 1$, where $\delta^2 = 0.04$ stands against $b^2 = 0.09$. The closed form and the anchor
model both reproduce the rise, the plateau and the fall.

\clearpage
\section{Cascade summary, with numbers}
\label{sec:S-cascade}

\begin{table}[h]
\centering
\small
\caption{The cascade of the main text, with the quantitative outcome of each validated gate.
Derived from the tables above.}
\label{tab:S-cascade}
\begin{tabular}{clp{8.2cm}}
\toprule
gate & quantity & measured outcome \\
\midrule
0 & chart existence & artefact peaks at $h\approx0.8\ell$ and is booked by stacking direction: radial
stacking reads as gain ($2.4\times$ control), ambient as novelty ($4.0\times$ the floor)
(Table~\ref{tab:S-gate0}) \\
1 & $q = d - d_{\mathrm{true}}$, leakage & noise-dominated: survival within $0.007$ of the
floor-corrected Haar law; the normal captures the first excess direction below
$\sigma\approx\tfrac12\,\ell^2/2R$, where $a$ collapses ($a^2\approx1-$capture)
(Table~\ref{tab:S-gate1}) \\
2 & $\|\hat V_d - V_d\|$; rotation & $\max|\widehat O - O_{\mathrm{orac}}|\le0.0010$
(Table~\ref{tab:S-offman}); rotation $\approx\ell^2\kappa^2/3k$, i.e.\ $2/(3n_{\mathrm{ref}})$ on
a sphere, times a finite-$k$ factor: $1.04$--$1.19$ on the patch (Table~\ref{tab:S-covgrad}),
$1.07$--$1.37$ on the sphere with the areal $\ell^2$ (Table~\ref{tab:S-rotation}), largest at
$k = 20$ \\
3 & $(\ell^2/2)\vec H$, jitter $\ell/\sqrt k$ & ratio2 $0.998$--$1.007$ (Table~\ref{tab:S-params}),
$1.000\pm0.005$ (Table~\ref{tab:S-anisoanchor}); torus slopes $0.97$--$1.02$
(Table~\ref{tab:S-torus}); jitter $0.98$--$1.02$ (Table~\ref{tab:S-curvlaw}) \\
4 & $a = \|P_N\hat\rho\|$ & $a = 1.00$ on the sphere; gain-aligned anisotropy moves $G$ from
$0.499$ to $0.188$ and $a$ to $0.48$ at $\sigma_w/\ell = 1.56$ (Table~\ref{tab:S-aniso}) \\
5 & $c = \hat n_g^\top\hat H$ & booking $\propto c^2 : 1-c^2$ confirmed over $|c| = 1\to0.385$,
ratios $0.98$--$1.02$ (Tables~\ref{tab:S-clifford}, \ref{tab:S-ellipse}) \\
6 & $C_6 = k\|r\|^2/\ell^2$ & mean $|\widehat O - O_{\mathrm{pred}}|$: $0.001$ at $C_6\ge20$,
$0.007$ at $5$--$20$, $0.03$ below $5$ (Table~\ref{tab:S-offman}) \\
\bottomrule
\end{tabular}
\end{table}

\endgroup

\end{document}